\documentclass[preprint,nofootinbib,amsmath,prd,aps,superscriptaddress,tightenlines,12pt]{revtex4}
\usepackage[bookmarks=true]{hyperref}
\usepackage{amsmath}
\usepackage{graphicx}
\usepackage{mathtools}
\usepackage{subfigure}

\begin{document}


\title{Disentangling radiative corrections using high-mass Drell-Yan at the LHC}

\vspace*{1cm}

\author{Radja Boughezal}
\email[]{rboughezal@anl.gov}
\affiliation{High Energy Physics Division, Argonne National Laboratory, Argonne, IL 60439, USA} 
\author{Ye Li}
\email[]{yli@slac.stanford.edu}
\affiliation{SLAC National Accelerator Laboratory, Stanford University, Stanford, CA 94309, USA}
\author{Frank Petriello}
\email[]{f-petriello@northwestern.edu}
\affiliation{High Energy Physics Division, Argonne National Laboratory, Argonne, IL 60439, USA} 
\affiliation{Department of Physics \& Astronomy, Northwestern University, Evanston, IL 60208, USA}


\begin{abstract}

  \vspace{0.5cm}

We present a detailed numerical study of lepton-pair production via the Drell-Yan process above the $Z$-peak at the LHC.  Our results consistently combine next-to-next-to-leading order QCD corrections and next-to-leading order electroweak effects, and include the leading photon-initiated processes using a recent extraction of the photon distribution function. We focus on the effects of electroweak corrections and of photon-photon scattering contributions, and demonstrate which kinematic distributions exhibit sensitivity to these corrections.  We show that a combination of measurements allows them to be disentangled and separately determined.

\end{abstract}

\maketitle

\section{Introduction}

The ATLAS and CMS experiments are producing measurements of kinematic distributions with cross sections spanning orders of magnitude, and in which all bins have experimental errors approaching the percent level.  These results allow for unprecedented detail in comparisons of Standard Model (SM) theory with experiment.  Perhaps the most striking examples of these measurements are those of the Drell-Yan spectra from 7 TeV $pp$ collisions at both ATLAS and CMS~\cite{Aad:2013iua,Chatrchyan:2013tia}.  Both experiments determine the invariant-mass spectrum of lepton pairs from 15 GeV (CMS) or 116 GeV (ATLAS) through 1.5 TeV.  The CMS collaboration further bins their measurement according to the dilepton rapidity.  The extraordinary lever arms provided by these data sets, and the degree to which systematic errors can be controlled, open unique windows onto radiative corrections in the SM, and into the structure of the proton.  Higher-order QCD corrections are crucial in bringing SM into agreement with these measurements.  The Drell-Yan distribution has also been helpful in determining the parton distribution function (PDF) of the photon inside the proton~\cite{Ball:2013hta}.

Future measurements of the Drell-Yan spectrum at a 14 TeV LHC will access an even larger kinematic range.  Along with this larger phase space comes an increased sensitivity to a host of effects. In addition to QCD corrections through next-to-next-to-leading order (NNLO), electroweak Sudakov corrections~\cite{sudakov,Fadin:1999bq,Kuhn:1999nn,Denner:2000jv,Kuhn:2001hz,Denner:2003wi} become increasingly more important at high-energy colliders, as has been emphasized recently in the literature~\cite{EWsummary}.  Photon-initiated corrections also increase in importance at high energies, as has been recently emphasized for both Drell-Yan and $W$-pair production processes~\cite{Dittmaier:2009cr,Bierweiler:2012kw}.  The ATLAS collaboration has performed detailed studies showing the importance of these effects for a host of measurements in the Drell-Yan channel~\cite{uta}.  One view of these corrections is that they represent additional sources of theoretical uncertainty which must be sufficiently controlled in order to perform interesting measurements.  For example, the measurement of high-mass $WW$ scattering will play a central role in Run II of the LHC, as it directly probes whether the discovered Higgs boson completely unitarizes the SM, or whether additional particles beyond those so far discovered are needed.  Sufficient understanding of electroweak (EW) corrections and photon-initiated processes, in addition to the usual QCD corrections, will be needed to interpret these measurements.  The experimental control over the Drell-Yan channel allows such theoretical effects to be precisely validated before being applied to other processes.  Another viewpoint is that the determination of these corrections is interesting in its own right, and that the experimental control over the Drell-Yan channel at the LHC offers a unique laboratory in which to study in detail the higher-order perturbative structure of the Standard Model.  For both of these reasons, the study of high-mass Drell-Yan production will be of great interest in Run II of the LHC.

A wealth of theoretical information is available for the Drell-Yan process.  QCD corrections up to the next-to-next-to-leading order (NNLO) in the strong coupling constant have been previously calculated, both for the inclusive cross section~\cite{Hamberg:1990np} and for differential quantities~\cite{Anastasiou:2003yy,Anastasiou:2003ds,Melnikov:2006kv,Catani:2009sm,Gavin:2010az,Li:2012wna}.  The NLO EW effects are known~\cite{CarloniCalame:2007cd,Berends:1984xv,Baur:1997wa,Baur:2001ze,Arbuzov:2007db,Dittmaier:2009cr}.  Our goal in this manuscript is to use this knowledge to determine how to separately measure the electroweak and photon-initiated corrections affecting the Drell-Yan process.  We exhaustively study the available kinematic distributions which can be measured, at both a 8 TeV and a 14 TeV LHC.  We show that a combination of several differential measurements in high-mass Drell-Yan lepton pair production at a 14 TeV run of the LHC allows one to disentangle the effect of the various corrections.  Although both the electroweak Sudakov logarithms and the photon-initiated contributions increase with lepton-pair invariant mass, they affect other distributions in distinct ways.  We estimate the observability of these deviations and show the most sensitive phase-space regions by constructing $\chi^2$ distributions that account for both statistical errors and imprecise knowledge of quark and gluon distribution functions.  We also study how different choices of basic acceptance cuts on the leptons affect this analysis.  For our numerical studies we use the latest version of {\tt FEWZ}~\cite{Li:2012wna}, which consistently combines NNLO QCD corrections with both NLO electroweak effects and the leading photon-initiated processes.  For inclusive observables away from phase-space regions in which hadronic radiation is restricted, which form the vast majority of the results presented here, this fixed-order approach represents the appropriate framework in which to perform this study.  Near kinematic boundaries, a combination of EW corrections with a QCD parton shower represents a more appropriate framework~\cite{Balazs:1997xd,Barze:2013yca}.  We point out where we expect fixed-order perturbation theory to break down when presenting our results.  We summarize the main conclusions of our study below.
\begin{itemize}

\item The most sensitive observable to the photon distribution function is the low end of the lepton transverse momentum ($p_{Tl}$) distribution, due to the underlying $t$-channel singularity of the corresponding matrix elements.  At lower invariant masses the EW corrections must be under good control in order to extract the photon PDF, as the two effects strongly cancel.  The photon-initiated processes increase more quickly with invariant mass, reducing the effect of this cancellation and making the high-mass, low-$p_{Tl}$ region an ideal place from which to determine the photon PDF.  This distribution is also expected to be less sensitive to potential new physics affecting the high-mass Drell-Yan tail, as the decay of a heavy object will generally populate higher $p_{Tl}$ bins.  Use of $p_{Tl}$ will therefore also help disentangle new physics from the Standard Model.

\item Although the low-$p_{Tl}$ region is most sensitive to the photon PDF, there is no benefit to reducing the experimental lepton $p_{Tl}$ requirement unless the pseudorapidity ($\eta_l$) constraints can also be loosened, due to the kinematics of the underlying process.  Relaxing the pseudorapidity cut on the leptons does enhance the observable deviations from photon-initiated processes.

\item The central region of dilepton rapidity ($Y_{ll}$) is sensitive to the photon PDF, indicating that a measurement of the three-dimensional distribution in $Y_{ll}$, $p_{Tl}$ and invariant mass should be a goal for the 14 TeV run.

\item The electroweak corrections are largest in the high-mass, central $\eta_l$ region of phase space, due to the underlying angular dependence of the Sudakov logarithms.  Once enough integrated luminosity is collected this distribution offers a window into the structure of the electroweak radiative corrections.

\end{itemize}
We view our results as an atlas of radiative corrections that can help guide the 14 TeV experimental study of the Drell-Yan process.  Other detailed studies of Drell-Yan production at the LHC exist~\cite{Dittmaier:2009cr}.  We extend upon this important work in several ways: we more exhaustively study the invariant-mass dependence of the available kinematic distributions; we control higher-order QCD corrections and uncertainties through the use of the full NNLO QCD corrections, which we show to be crucial in determining the other effects of interest; we estimate the observability of the various corrections using detailed and up-to-date error estimates; we use the latest results on the photon PDF that are informed by Run I LHC measurements; finally, we study the effect of varying the experimental acceptance cuts on the leptons.

Our paper is organized as follows.  We present our notation and setup in Section~\ref{sec:setup}.  Numerical results for an 8 TeV and a 14 TeV LHC are presented in Section~\ref{sec:numerics}.  Finally, we conclude in Section~\ref{sec:conc}.

\section{Setup}
\label{sec:setup}

We describe here the parameters and framework we employ in our study.  All numerical results presented are obtained with the program {\tt FEWZ}~\cite{Melnikov:2006kv,Gavin:2010az,Li:2012wna}, which consistently combines NNLO QCD corrections with NLO electroweak corrections and the leading photon-initiated processes.  The EW corrections and photon-initiated contributions, which are the focus of our study, have been extensively validated against the literature in Ref.~\cite{Li:2012wna}, and we do not repeat that comparison here.  We use the recent NNPDF 2.3 PDFs~\cite{Ball:2013hta}, which consistently include QED corrections and allow for an initial-state photon distribution function, at NNLO in QCD perturbation theory.  This is the most recent PDF set which allows for an initial-state photon, and is the only one in which the photon PDF has been constrained by data within a global fit.
The MRST 2004 PDF set~\cite{Martin:2004dh} also allows for a photon PDF, with its form given primarily by a model parameterization.  Another possible approach to including the photon PDF would be to use the MRST 2004 set for the photon, while using a more up-to-date set for the quark and gluon PDFs~\cite{dittmaiertalk}.  We use the $G_{\mu}$ scheme as our electroweak input scheme, which is known to reduce the size of higher-order electroweak corrections~\cite{Dittmaier:2009cr}.  In a fixed-order calculation, the same input scheme must be chosen for real and virtual corrections to preserve the cancellation of infrared singularities, meaning that we use this same scheme for the calculation of the real photonic corrections.  Another possibility is the $\alpha(0)$ scheme.  These two choices differ by uncalculated, and most likely small, ${\cal O}(\alpha^2)$ terms.  Both are of course completely consistent to the order we are working.

We write the cross section in the schematic form
\begin{equation}
\sigma_{\text{full}} = \sigma_{\text{NNLO QCD}}+\Delta \sigma_{\text{NLO EW}}+\Delta \sigma_{\gamma},
\label{eq:crdef}
\end{equation}
where $\sigma_{\text{NNLO QCD}}$ contains the leading-order result together with the higher-order QCD corrections, $\Delta \sigma_{\text{NLO EW}}$ contains the one-loop electroweak radiative corrections, and $\Delta \sigma_{\gamma}$ denotes the contribution from the leading-order $\gamma\gamma \to l^+l^-$ process (we will refer to this contribution as `photon-induced' or `photon-initiated' in our work).  All three pieces are evaluated using the same NNLO PDFs.  For further details on the combination of these contributions, we refer the reader to Ref.~\cite{Li:2012wna}.  We note that we have neglected exclusive and single-dissociative processes, where respectively both or one of the two incoming protons remains intact.  Also, $q\gamma$ scattering processes are known to slightly reduce the size of the photon-initiated contributions~\cite{Dittmaier:2009cr}.  We have neglected processes containing real emission of $W$ or $Z$ bosons, which partially cancel the effects of the EW virtual corrections.  Such corrections lead to diboson contributions to the cross section. How, and whether or not, they should be included in the theoretical prediction of the signal depends on the experimental analysis.  For example, both the ATLAS and CMS Drell-Yan measurements explicitly model such diboson backgrounds, where both bosons decay leptonically, separately~\cite{Aad:2013iua,Chatrchyan:2013tia}, indicating that this contribution should not be added to the Drell-Yan signal studied here.  The effect of including real-boson emission with subsequent decay to either jets or neutrinos was studied in Ref.~\cite{Baur:2006sn}.  The numerical impact of these additional corrections was to reduce the size of the electroweak corrections by a few percent in the high-invariant mass tail.

We have investigated two different choices of renormalization and factorization scale: a dynamical scale set equal to the invariant mass of the produced lepton pair, and a fixed scale equal to the geometric mean of the upper and lower boundaries of the invariant mass bin under consideration.  The two choices give almost identical results, and are indistinguishable in the plots we present.  We show the dynamic scale choice.  Since the scale uncertainties are in general smaller than the PDF and statistical errors, we do not consider them further.  In later sections we consider the use of the difference between the NLO and NNLO QCD predictions as a conservative estimate of the uncertainties coming from uncalculated QCD corrections.  We compute the $1\sigma$ PDF errors on $\sigma_{full}$ using the procedure suggested by the NNPDF collaboration~\cite{Ball:2011uy}.  When forming our $\chi^2$ function we do not include the uncertainties from the photon distribution function, as one purpose of our study is to identify distributions for which we can control other sources of error reliably enough to extract it.  We display the photon PDF uncertainty separately on our plots.

We impose the following basic acceptance cuts on the final-state leptons:
\begin{equation}
p_{Tl} > 20 \; \text{GeV},\;\;\; |\eta_l| < 2.5.
\label{eq:acccuts}
\end{equation}
In the final section we study the effects of changing these constraints.  Photons satisfying  $\sqrt{(\phi_l - \phi_{\gamma})^2+(\eta_l-\eta_{\gamma})^2} < 0.1$ around a lepton are combined with that lepton by adding together their four-momenta.  The effect of this recombination is very small compared to other corrections in the high invariant mass region.  For 8 TeV collisions we investigate the following three different regions of dilepton invariant mass:
\begin{equation}
M_{ll} \in [0.12,0.2] \; \text{TeV}, \;\;\; M_{ll} \in [0.2,0.5] \; \text{TeV}, \;\;\; M_{ll} \in [0.5,1] \; \text{TeV}.
\label{eq:inv8}
\end{equation}
For 14 TeV collisions, we add on another high-mass bin:
\begin{equation}
M_{ll} \in [0.12,0.2] \; \text{TeV}, \;\;\; M_{ll} \in [0.2,0.5] \; \text{TeV}, \;\;\; M_{ll} \in [0.5,1] \; \text{TeV}, \;\;\; M_{ll} \in [1,3] \; \text{TeV}.
\label{eq:inv14}
\end{equation}
These choices are meant to illustrate the behavior of the various corrections as the invariant mass is changed.  We note that we have also studied the region $M_{ll} \in [3,14] \; \text{TeV}$ at a 14 TeV machine, and have found that the event rates are too small to allow discrimination between different effects until at least $3000 \; \text{fb}^{-1}$ is reached.  

In each invariant mass region we study the following three distributions: the dilepton rapidity $Y_{ll}$, the lepton pseudorapidity $\eta_l$, and the lepton transverse momentum $p_{Tl}$.  We note that the lepton means the negatively-charged lepton; to avoid too great a proliferation of plots we do not show the anti-lepton distributions.  In the final section we also study distributions for the harder and softer leptons, ordered in $p_T$.  We have also studied the dilepton transverse momentum distribution $p_{Tll}$, but have found that this distribution does not help in distinguishing between the various radiative corrections we are considering.  The reason for this is clear: the photon-initiated contributions contain only an $l^+l^-$ pair in the final state, and therefore populate only the $p_{Tll}=0$ bin in our calculation.  Higher-order corrections with an additional photon radiated will not significantly change this conclusion, since these effects are suppressed by $\alpha$, unlike QCD radiation which contributes as $\alpha_s$.  Similarly, QED radiation effects only minimally shift the electroweak corrections away from $p_{Tll}=0$.  Therefore, all deviations in $p_{Tll}$ are concentrated near the origin, and this distribution does not help disentangle the various higher-order effects.  The impact of multiple photon radiation on several distributions was studied in Ref.~\cite{Dittmaier:2009cr}, and found to be much smaller than the effects we are considered here.

To denote a cross section restricted to a given invariant mass bin, and to a bin in another kinematic variable, we will generically use the notation $\sigma_x(i)$, which is shorthand for the following expression where the bin boundaries are explicitly written as arguments of the bin-integrated result:
\begin{equation}
\sigma_x(i) \equiv \sigma_x\left([M_{ll}^{down},M_{ll}^{up}],[v^{down},v^{up}] \right),
\end{equation}
where $x=\text{full},\text{NNLO QCD},\text{NLO EW},\gamma$.  $v$ denotes either the absolute value of the dilepton rapidity or the lepton rapidity, or the lepton transverse momentum.  We only study restrictions in a single variable $v$ at a time in this manuscript, as it makes our results simpler to visualize. 

\section{Numerical Results}
\label{sec:numerics}

We will show numerical results for distributions for both 8 TeV and 14 TeV $pp$ collisions, for the invariant mass regions defined in Eqs.~(\ref{eq:inv8}) and~(\ref{eq:inv14}).  In order to determine the observability of the various deviations induced by photon and electroweak effects, we will compare them to the estimated statistical and PDF errors.  The scale variation coming from missing QCD corrections has been studied previously, including in the course of the Drell-Yan measurements~\cite{Aad:2013iua,Chatrchyan:2013tia}, and has been found to be smaller than the considered error sources.  Later in this manuscript we will consider the difference between NLO QCD and NNLO QCD as a measure of the theoretical uncertainty.  The experimental systematics require detailed investigation by the experimental collaborations, and we therefore do not attempt to include them.  Assuming ${\cal L}$ inverse femtobarns of integrated luminosity, the relative error for a given bin $i$ is
\begin{equation}
\delta_{{\cal L}}(i) = \sqrt{\frac{1}{\sigma_{full}(i) {\cal L}} + \left(\frac{\Delta_{\text{PDF}}(i)}{\sigma_{full}(i)}\right)^2},
\end{equation}
where $\sigma_{full}(i)$ is the cross section in the $i$-th bin (subject to acceptance cuts) expressed in femtobarns.  
From this we can form a $\chi^2$ function to quantitatively determine the significance of the neglect of a particular radiative correction $x=\text{NLO EW},\gamma,\text{NLO EW}+\gamma$:
\begin{equation}
\chi^2_{x,{\cal L}}(i) = {\cal L} \times \frac{[\sigma_{full}(i)-\sigma_x(i)]^2}{\sigma_{full}(i)+{\cal L} \Delta^2_{\text{PDF}}(i)}.
\label{eq:chidef}
\end{equation}
We caution that these $\chi^2$ distributions are only meant to illustrate the most promising regions of phase space in which to pursue the measurements of interest.  In addition to the lack of experimental systematic errors, we have neglected possible bin-to-bin correlations that may appear.  These $\chi^2$ distributions are not meant to serve as a numerical fit of the photon PDF or of other effects.

\begin{figure}
\centering
\includegraphics[width=4.1in]{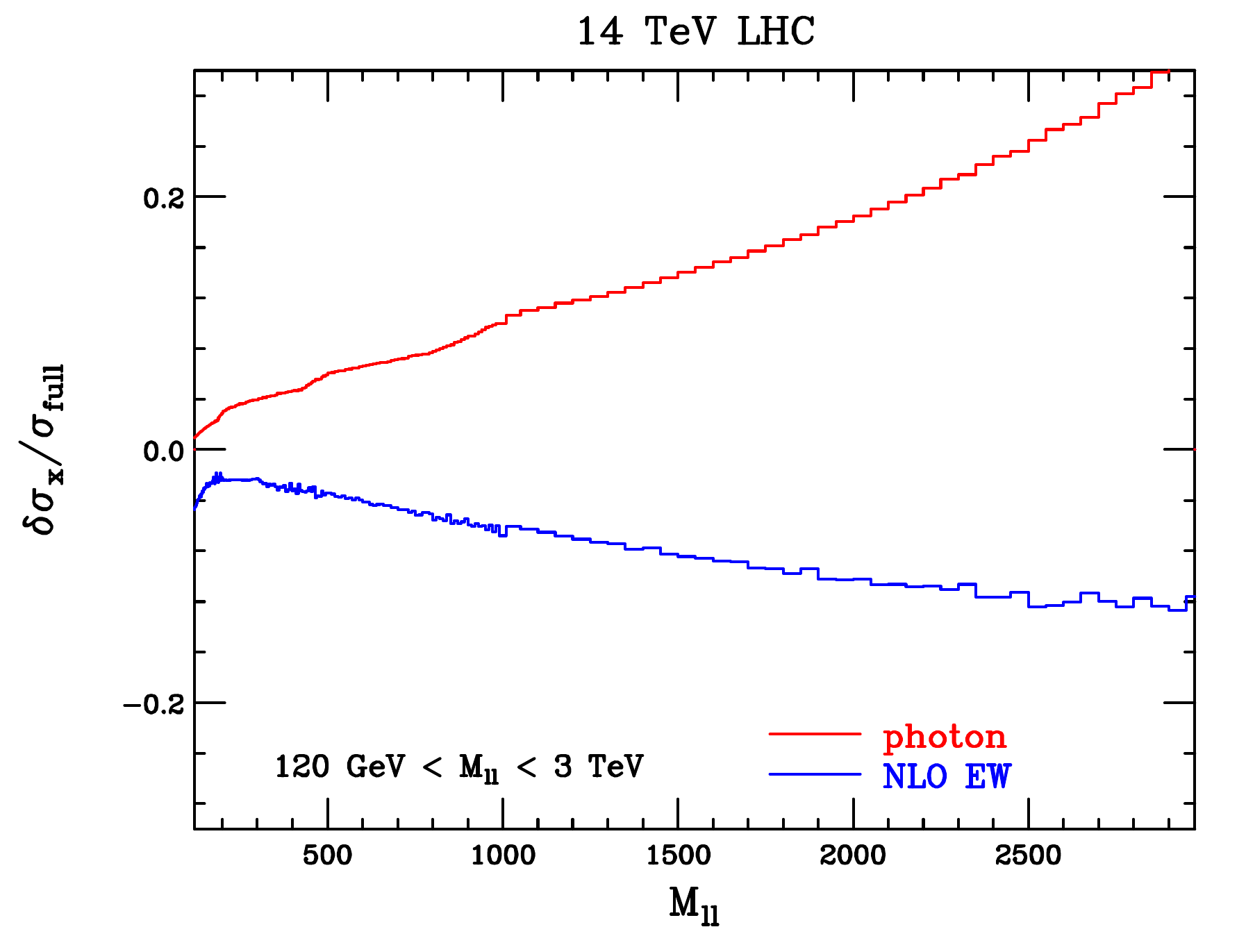}
\caption{Shown are the deviations induced by photon-initiated contributions and electroweak corrections to the dilepton invariant mass distribution at a 14 TeV LHC.}  \label{fig:Qll}
\end{figure}

To begin, we orient the reader by showing in Fig.~\ref{fig:Qll} the relative deviations induced by EW corrections and photon-initiated processes to the invariant mass distribution at a 14 TeV LHC.   Specifically, these are the deviations induced by the $\Delta \sigma_{\text{NLO EW}}$ and $\Delta \sigma_{\gamma}$ contributions in Eq.~(\ref{eq:crdef}), relative to the full cross section $\sigma_{\text{full}}$.  Both corrections grow in magnitude with invariant mass, with the photon corrections reaching $+30\%$ at 3 TeV and the EW contributions reaching $-12\%$.  There is a significant cancellation between the two effects, indicating the importance of simultaneously controlling them, or of finding additional cuts that enhance the effect of one relative to the other.  The goal of the coming sections will be to study additional distributions to determine which are more sensitive to one of these two corrections.

\subsection{Results for an 8 TeV LHC}

\begin{figure}
\centering
\mbox{\subfigure{\includegraphics[width=3.1in]{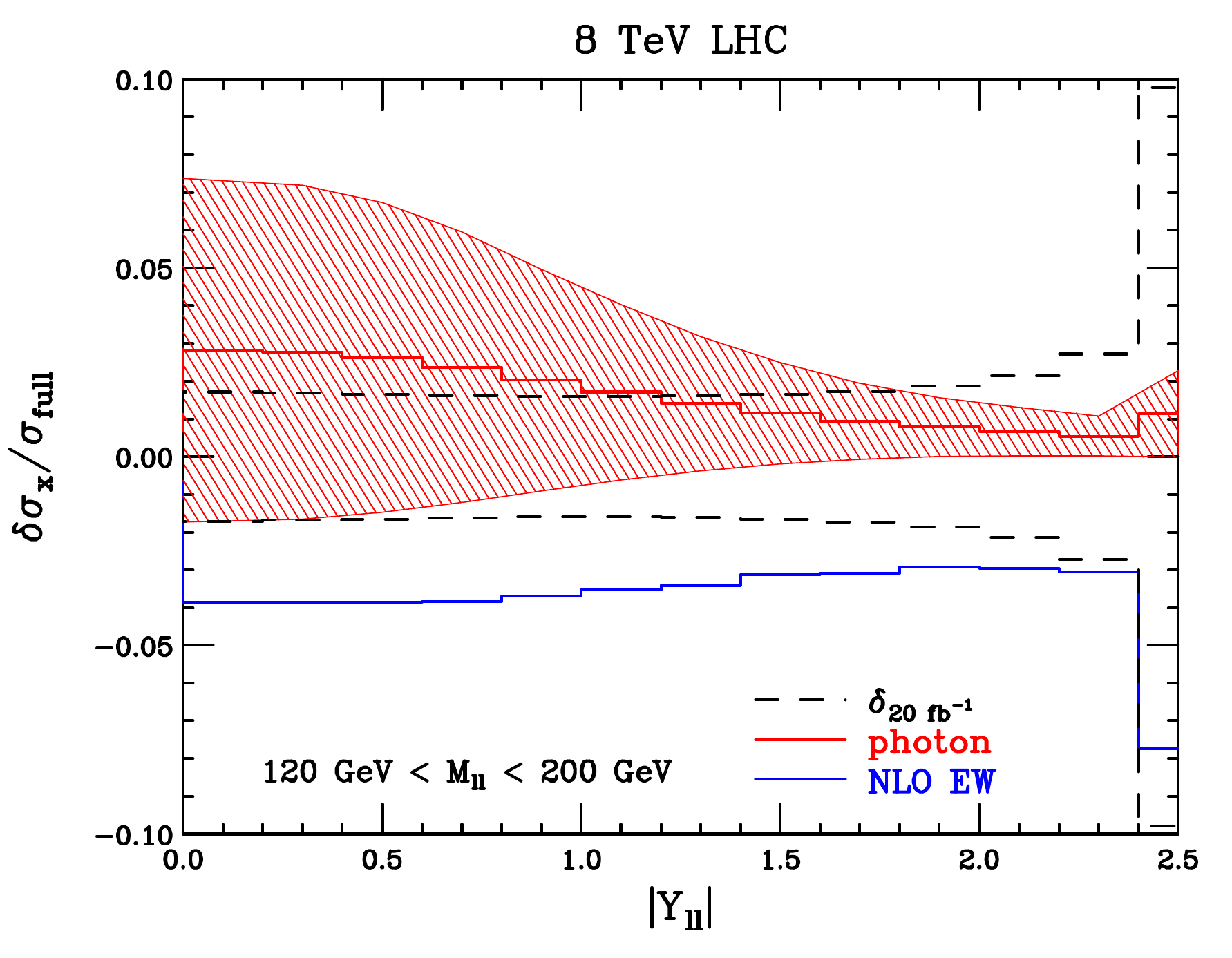}}\quad
\subfigure{\includegraphics[width=3.1in]{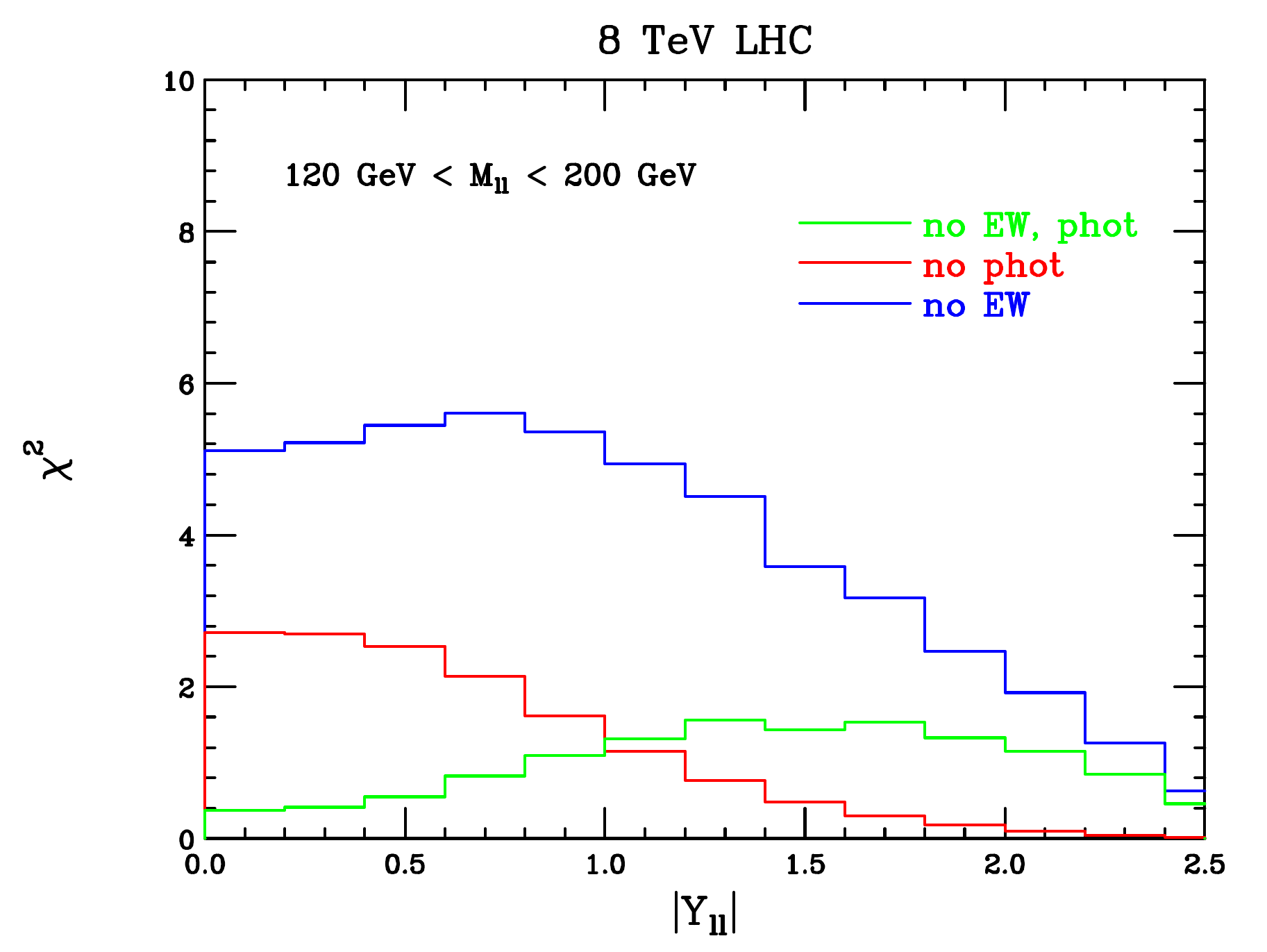}}}
\caption{Shown are the deviations induced by photon-initiated contributions and electroweak corrections to the dilepton rapidity distribution (left panel) at an 8 TeV LHC, for the invariant mass range $M_{ll} \in [120,200]$ GeV.  The band shows the error coming from the photon distribution function.  The dashed lines show the estimated errors coming from statistics and from uncertainties in the quark and gluon distribution functions.  The right panel shows the $\chi^2$ deviation for each bin assuming 20 fb$^{-1}$ of integrated luminosity.}  \label{fig:Zrap8TeV120_200}
\end{figure}

We begin by studying the $|Y_{ll}|$, $|\eta_l|$, and $p_{Tl}$ distributions at an 8 TeV LHC.  In order to estimate statistical errors we assume 20 fb$^{-1}$ of integrated luminosity, consistent with the amount of data collected separately by the ATLAS and CMS experiments.  Our results show the deviations induced by the $\Delta \sigma_{\text{NLO EW}}$ and $\Delta \sigma_{\gamma}$ contributions in Eq.~(\ref{eq:crdef}), relative to the full cross section $\sigma_{\text{full}}$ (with the inclusion of acceptance cuts).  We begin with the dilepton rapidity distribution in the invariant mass range $M_{ll} \in [120,200]$ GeV, shown in Fig.~\ref{fig:Zrap8TeV120_200}.  The uncertainty in the photon PDF is shown as the hatched region.  The $\chi^2$ functions of Eq.~(\ref{eq:chidef}) obtained by turning off the photon-initiated processes, the EW corrections, or both, are shown in the right panel.  The dashed line in the left panel, indicating the estimated errors from statistics and imperfect quark and gluons PDFs, is dominated by the PDF error component for this invariant-mass bin.  Both deviations from photon-induced processes and electroweak corrections are larger than the estimated error, and peak near central rapidity.  The importance of simultaneously controlling both corrections is clear; they almost completely cancel.  Any attempt to extract the photon PDF without accounting for EW corrections, or vice versa, would lead to incorrect results. The error coming from imperfect knowledge of the photon distribution function is large, reaching $\pm 4\%$, twice as large as the estimated error coming from statistics and uncertainties on the quark and gluon PDFs.    The estimated $\chi^2$ values reach five per bin if the EW corrections are neglected, and three per bin if the photon PDF is set to zero.  If other sources of error can be controlled, then measurement of this distribution should provide a handle on the photon content of the proton.  If both corrections are set to zero simultaneously, the $\chi^2$ value remains under two for most of the kinematic range, quantitatively emphasizing the need to control both effects in order to measure either one.

\begin{figure}
\centering
\mbox{\subfigure{\includegraphics[width=3.1in]{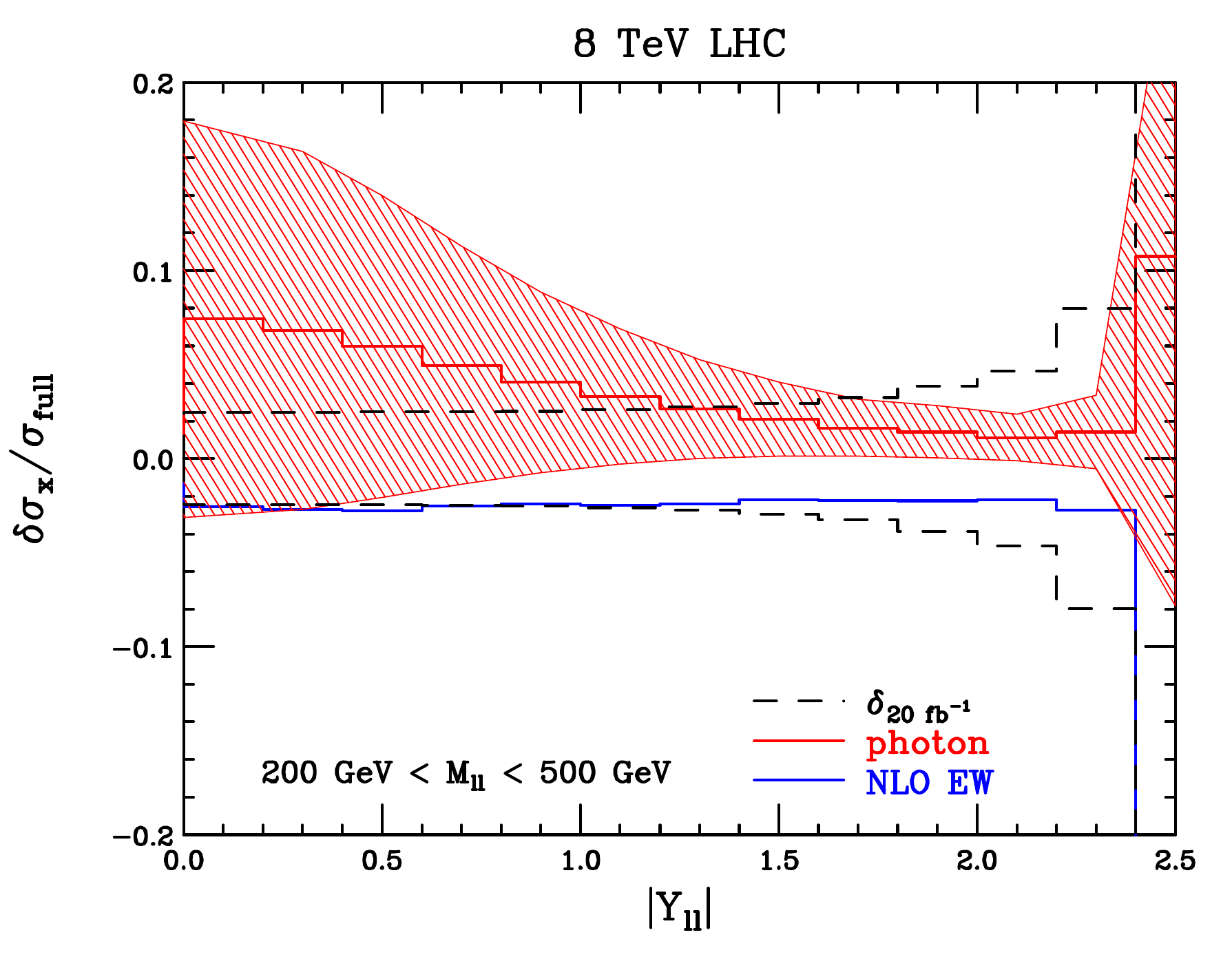}}\quad
\subfigure{\includegraphics[width=3.1in]{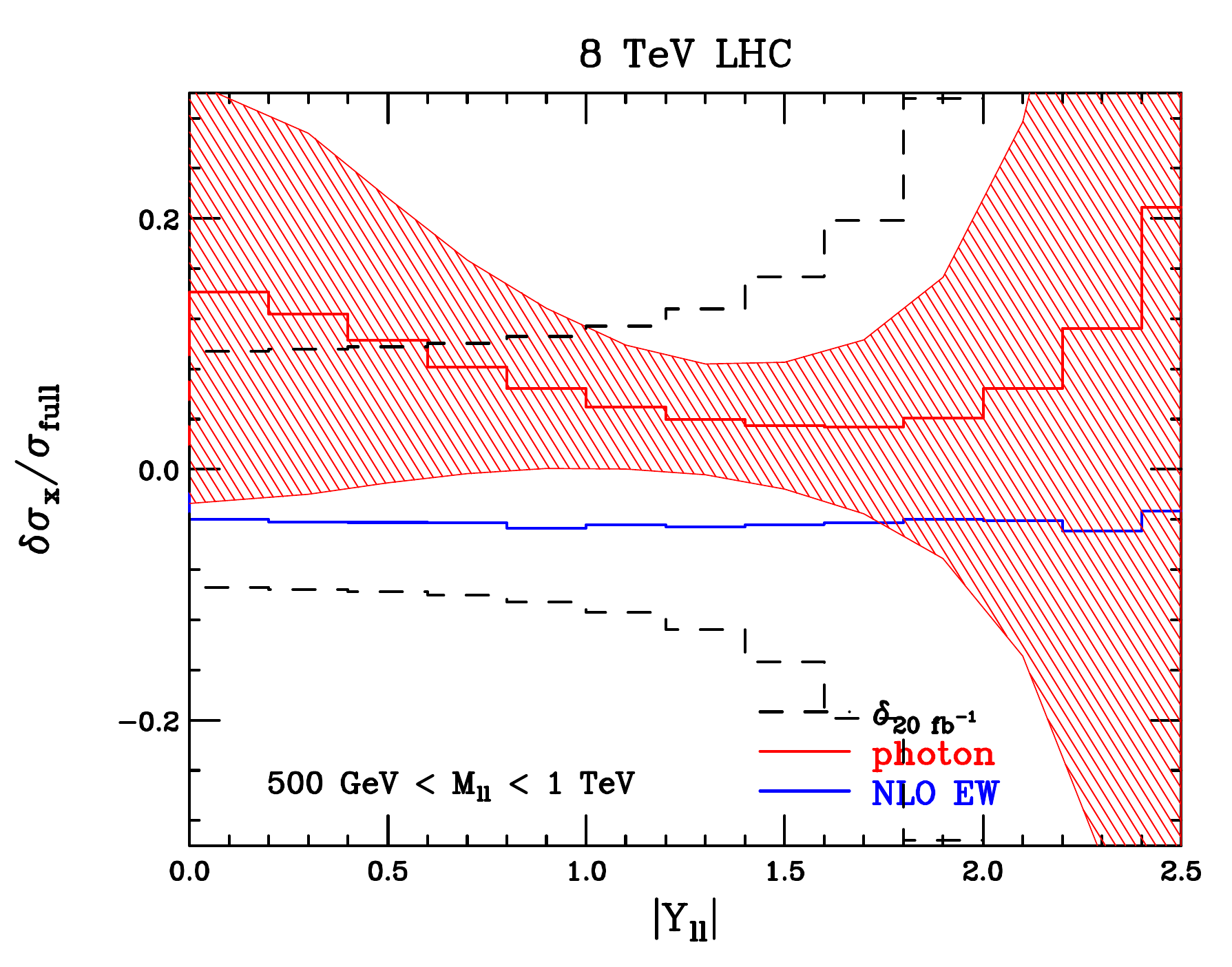}}}
\caption{Shown are the deviations induced by photon-initiated contributions and electroweak corrections to the dilepton rapidity distribution at an 8 TeV LHC, for the invariant mass ranges $M_{ll} \in [200,500]$ GeV (left panel) and $M_{ll} \in [500,1000]$ GeV (left panel).  The bands show the errors coming from the photon distribution function.  The dashed lines show the estimated errors coming from statistics and from uncertainties in the quark and gluon distribution functions. }  \label{fig:Zrap8TeV200_1000}
\end{figure}

In Fig.~\ref{fig:Zrap8TeV200_1000} we show results for the $|Y_{ll}|$ distribution in two higher invariant mass bins, $M_{ll} \in [200,500]$ GeV (left panel) and $M_{ll} \in [500,1000]$ GeV.  The peaking of the photon  corrections near central rapidity is also present in these higher-mass bins.  The photon-induced contributions grow more quickly than the EW corrections as the invariant mass is increased; the cross section contribution from this process depends logarithmically on the lower cut on $p_{Tl}$, and the photon PDF has a smaller downward slope with increasing Bjorken-$x$ compared to the sea-quark distributions, as can be checked using Refs.~\cite{Ball:2013hta} and~\cite{Ball:2011uy}.  This makes the high-mass, central-rapidity phase space region a good place to extract the photon PDF with relatively fewer complications from EW corrections.  The estimated error on the photon PDF also grows rapidly with invariant mass, indicating that any experimental measurement in this region will improve upon the current determination of this quantity.  The estimated error from statistics and uncertainty in the quark and gluon PDFs increases from $\pm 2.5\%$ in the first invariant-mass bin, to $\pm 10\%$ in the second bin.  This is caused primarily by decreased statistics, and not by a change in the PDF errors.  In both bins it is smaller than the expected photon-induced contribution at central rapidities.  The electroweak corrections induce an approximately $-4\%$ correction that is flat over the entire kinematic range.  

\begin{figure}
\centering
\mbox{\subfigure{\includegraphics[width=3.1in]{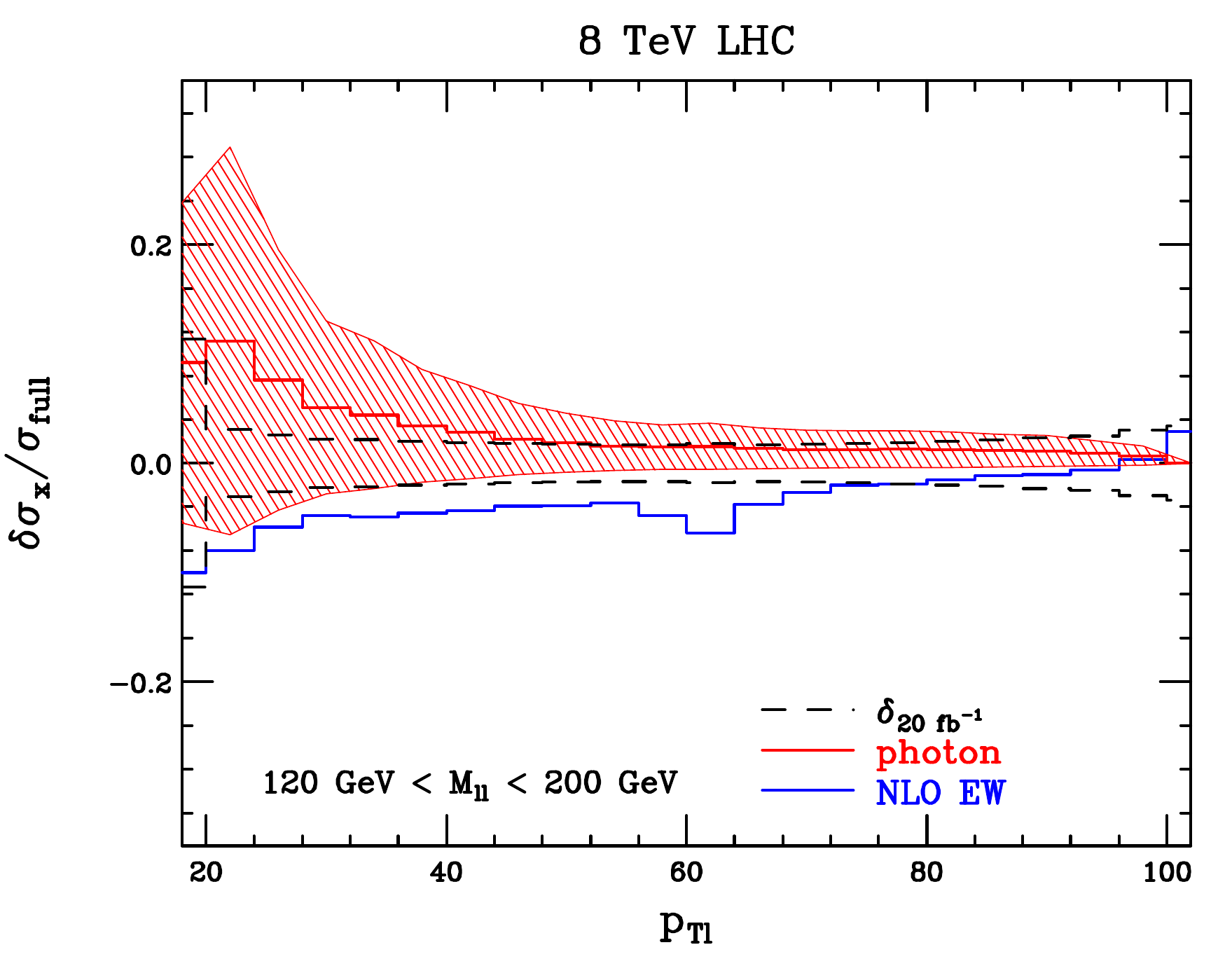}}\quad
\subfigure{\includegraphics[width=3.1in]{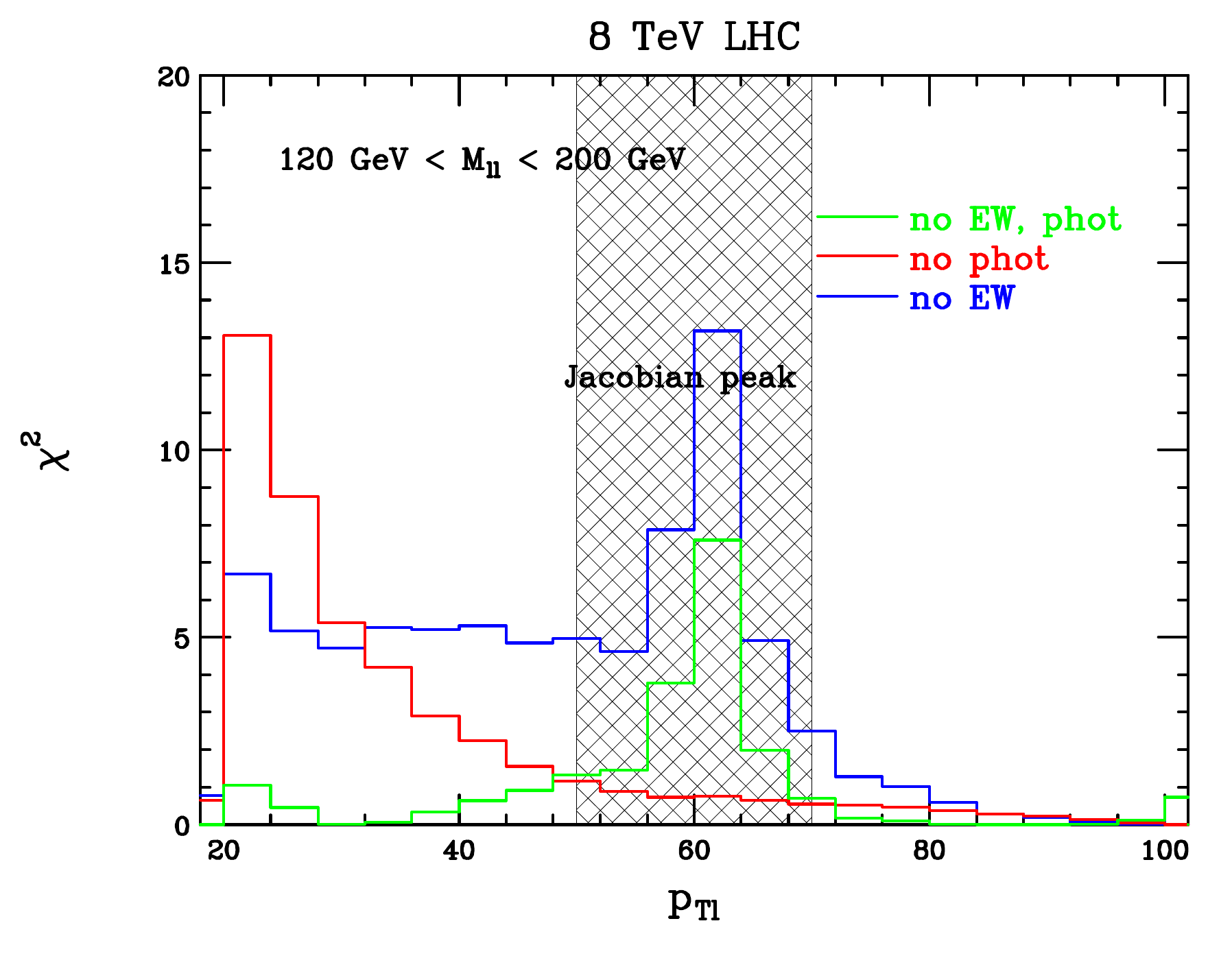}}}
\caption{Shown are the deviations induced by photon-initiated contributions and electroweak corrections to the lepton transverse momentum distribution (left panel) at an 8 TeV LHC, for the invariant mass range $M_{ll} \in [120,200]$ GeV.  The band shows the error coming from the photon distribution function.  The dashed lines show the estimated errors coming from statistics and from uncertainties in the quark and gluon distribution functions.  The right panel shows the $\chi^2$ deviation for each bin assuming 20 fb$^{-1}$ of integrated luminosity.  The region near the Jacobian peak, where fixed-order perturbation theory breaks down, has been shaded out.}  \label{fig:leppT8TeV120_200}
\end{figure}

We next study the lepton $p_{Tl}$ distribution, beginning with the invariant mass region $M_{ll} \in [120,200]$ GeV.  The results are shown in Fig.~\ref{fig:leppT8TeV120_200}.  The structure of this distribution is interesting.  The photon-induced deviations peak at low momentum, while the electroweak corrections are flat over the entire distribution.  The reason for this peak is clear, and has been pointed out previously in the literature~\cite{Dittmaier:2009cr}.  The photon processes proceed via $t$-channel lepton exchange, which have a collinear singularity regulated by the cuts on $p_{Tl}$ and $\eta_l$, leading to an enhancement at low $p_{Tl}$ (the lepton mass would regulate this singularity in the absence of cuts).  The size of the peak is larger than the estimated errors.  The $\chi^2$ distribution is shown in the right panel of Fig.~\ref{fig:leppT8TeV120_200}.  The low-$p_{Tl}$ region is very sensitive to the photon PDF, with a $\chi^2$ value reaching ten per bin.  The importance of controlling the EW corrections is again clear; when both the EW and photon-initiated corrections are turned off, the $\chi^2$ value drops to two.  We note that the region near $p_{Tl} \sim 60$ GeV has been shaded out.  This is the Jacobian peak coming from the kinematic boundary of the leading-order process.  Fixed-order perturbation theory breaks down near this boundary, and the shading is meant to remove it from consideration.

\begin{figure}
\centering
\mbox{\subfigure{\includegraphics[width=3.1in]{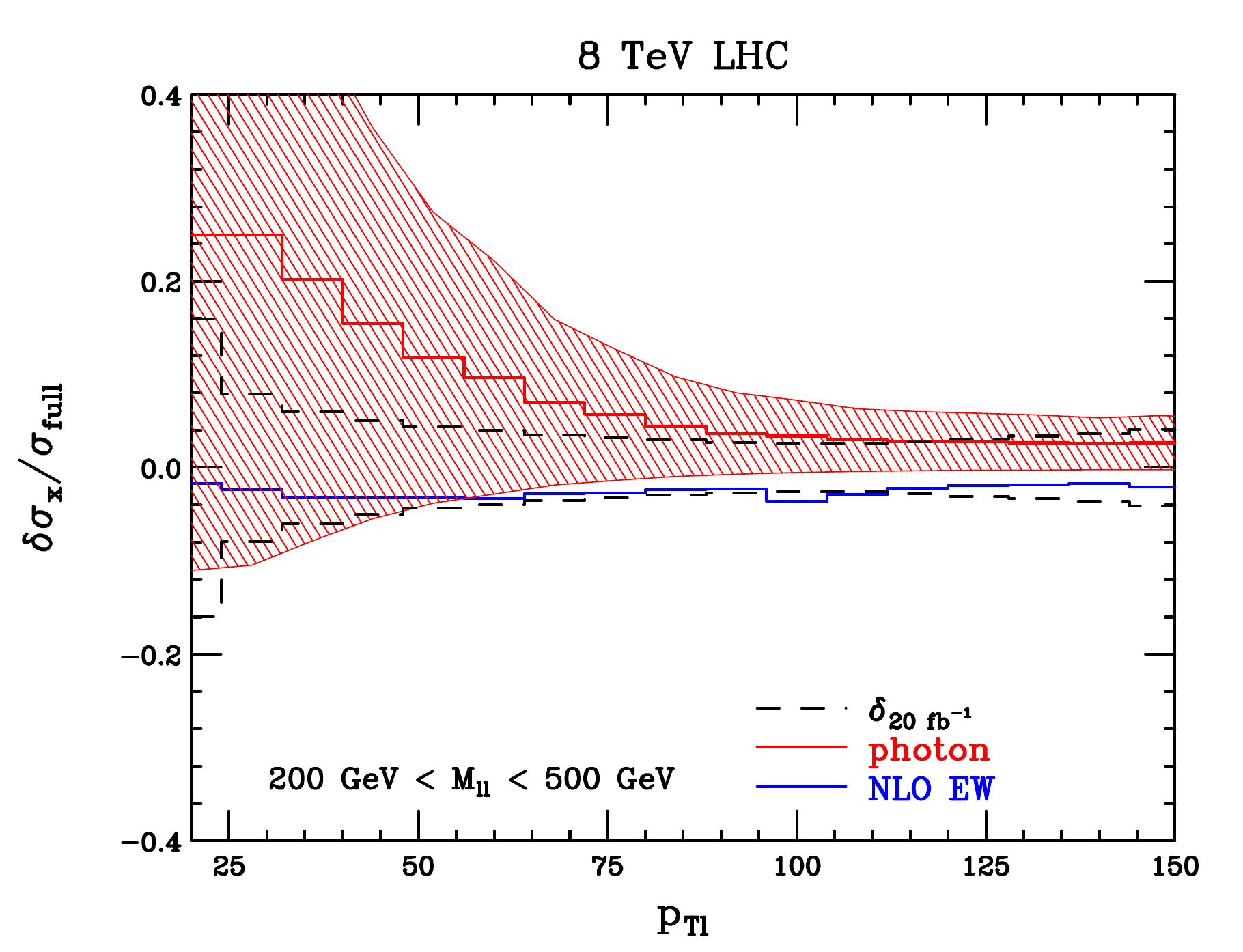}}\quad
\subfigure{\includegraphics[width=3.1in]{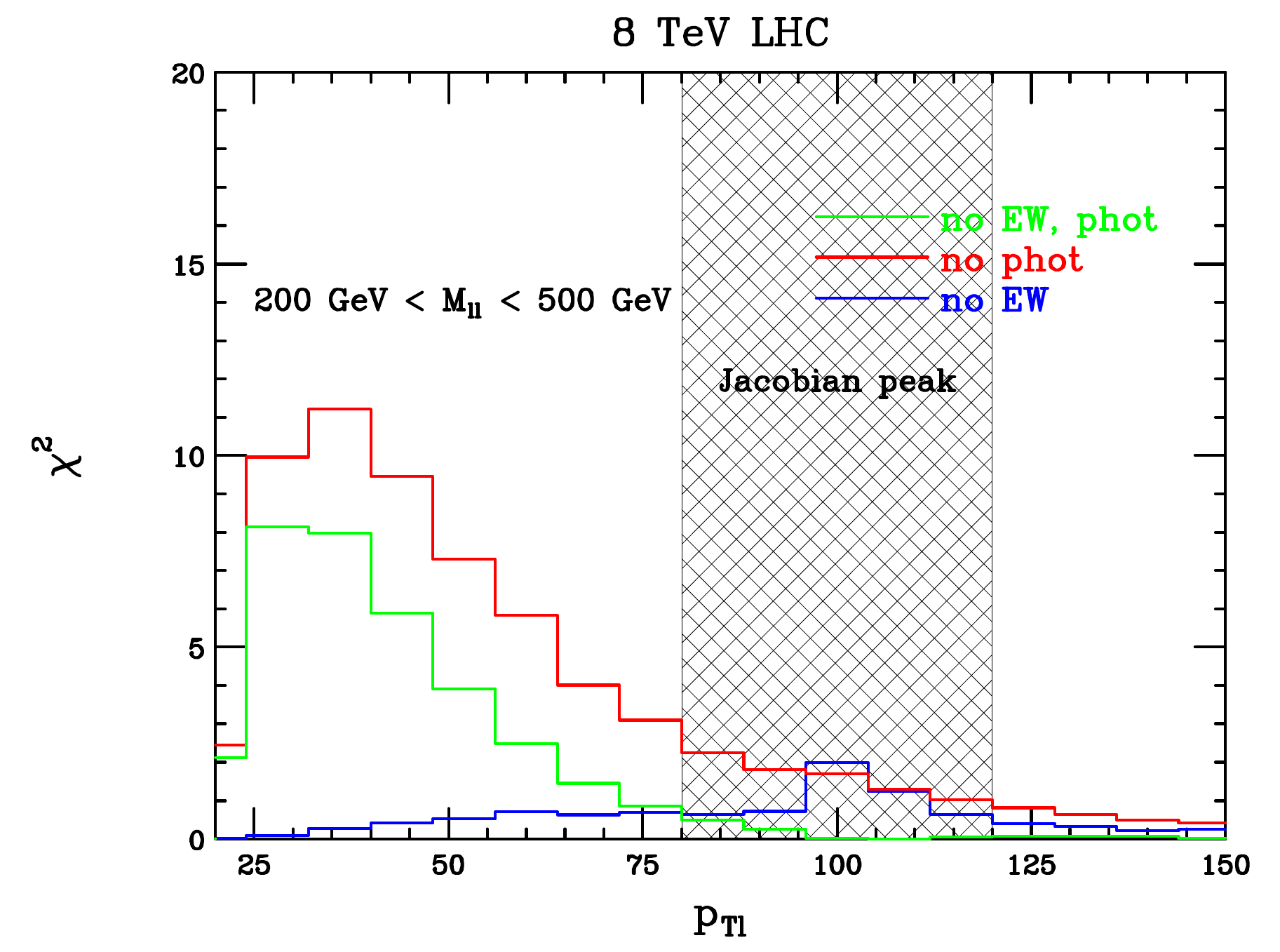}}}
\caption{Shown are the deviations induced by photon-initiated contributions and electroweak corrections to the lepton transverse momentum distribution (left panel) at an 8 TeV LHC, for the invariant mass range $M_{ll} \in [200,500]$ GeV.  The band shows the error coming from the photon distribution function.  The dashed lines show the estimated errors coming from statistics and from uncertainties in the quark and gluon distribution functions.  The right panel shows the $\chi^2$ deviation for each bin assuming 20 fb$^{-1}$ of integrated luminosity.  The region near the Jacobian peak, where fixed-order perturbation theory breaks down, has been shaded out.}  \label{fig:leppT8TeV200_500}
\end{figure}

The same results for the region $M_{ll} \in [200,500]$ GeV are shown in Fig.~\ref{fig:leppT8TeV200_500}.  Athough the estimated error at low $p_{Tl}$ increases in this bin, primarily because of low statistics, the size of the photon deviation increases, so that the $\chi^2$ function indicates as significant a deviation as in the lower invariant mass bin.  The photon-initiated corrections increase quickly with mass, and the relative importance of the electroweak corrections decreases in this higher invariant mass range, making this region a cleaner place from which to extract the photon PDF.  We see from the left panel in Fig.~\ref{fig:leppT8TeV200_500} that the uncertainty from the photon distribution function is far larger than the estimated errors from other sources, indicating that measurement of this distribution would very significantly improve our knowledge of this quantity.  The $\chi^2$ values reach over ten for low $p_{Tl}$, and do not decrease much upon simultaneously neglecting EW corrections, quantitatively demonstrating their reduced importance in this region.  We note that the maximum $\chi^2$ value, indicating the phase-space region with the most sensitivity to the photon PDF, occurs somewhat above the lower bound of $p_{Tl}> 20$ GeV.  This is because the leading-order kinematics of the $\gamma\gamma \to l^+l^-$ process implies that
\begin{equation}
p_{Tl} = \frac{\sqrt{\hat{s}}}{x_2 \text{e}^{\eta}+x_1 \text{e}^{-\eta}},
\label{eq:LOpT}
\end{equation}
where $\hat{s} = x_1 x_s s$ is the standard partonic Mandelstan invariant.  The constraint $|\eta_l|<2.5$ already imposes a constraint on the lepton transverse momentum that is stronger than the explicit cut.  This indicates that lowering the lepton $p_{Tl}$ cut to enhance the effect of photon-induced processes does not help unless the $\eta_l$ cut can simultaneously be relaxed.  We will study the effect of relaxing this cut in a later section.

\begin{figure}
\centering
\mbox{\subfigure{\includegraphics[width=3.1in]{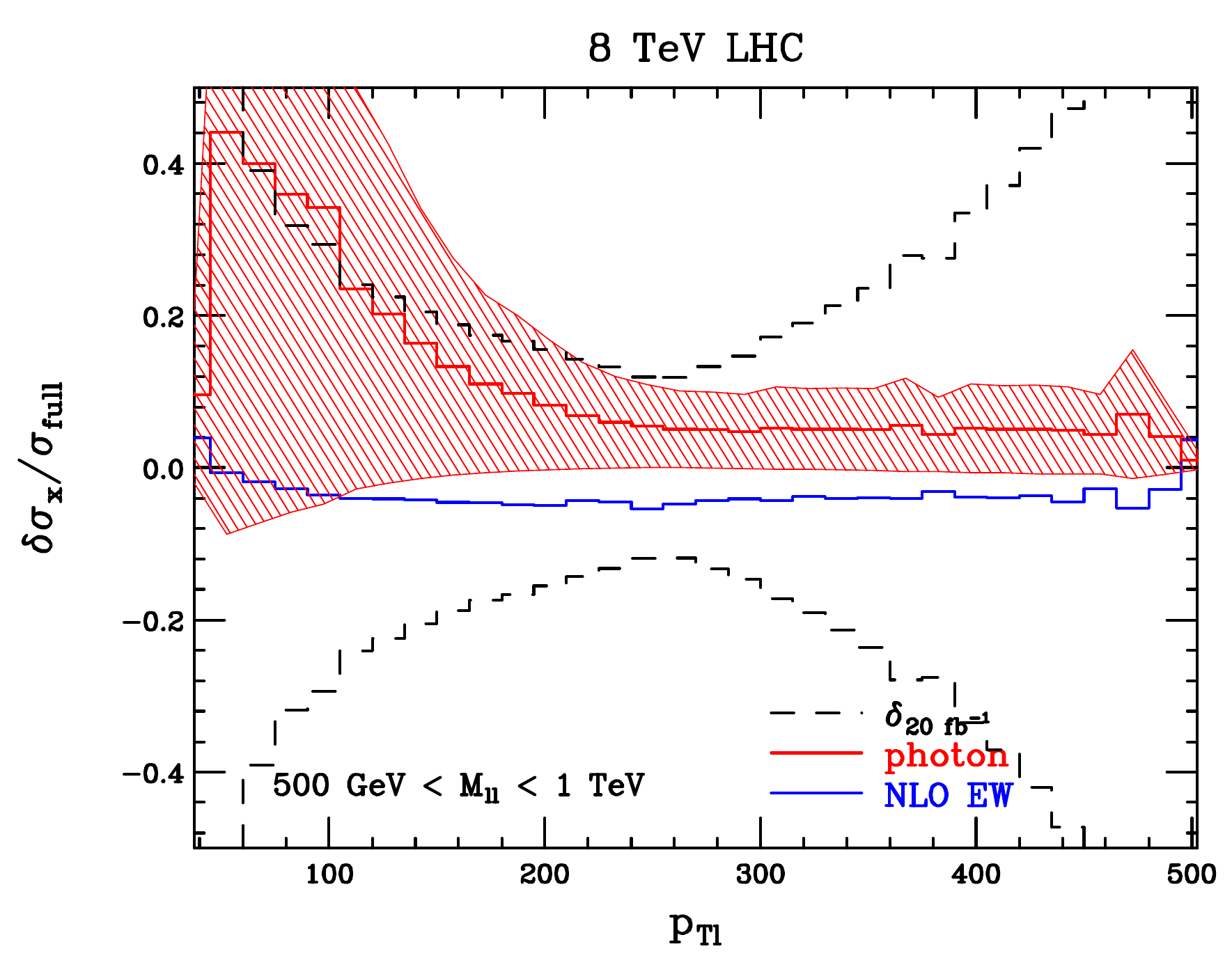}}\quad
\subfigure{\includegraphics[width=3.1in]{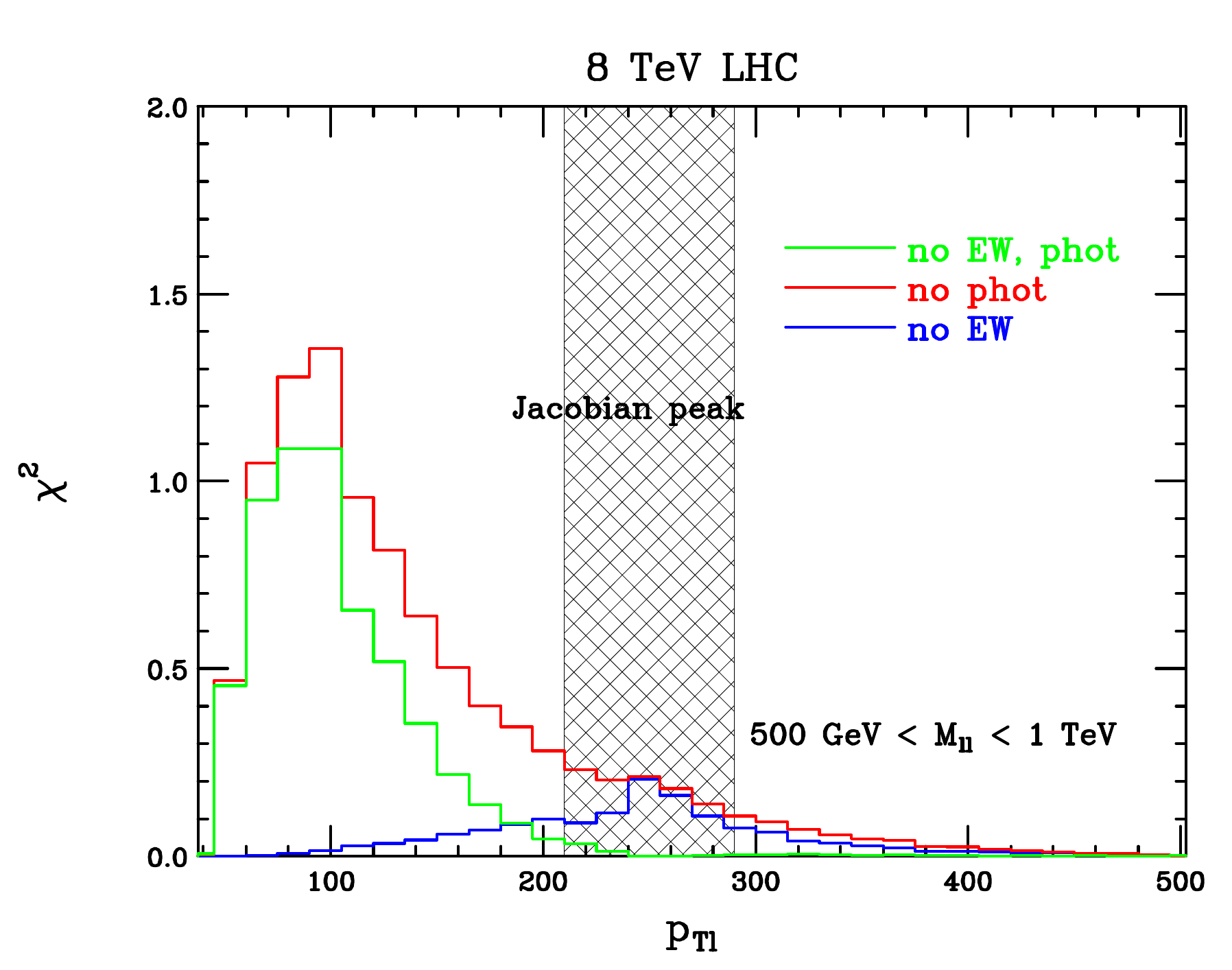}}}
\caption{Shown are the deviations induced by photon-initiated contributions and electroweak corrections to the lepton transverse momentum distribution (left panel) at an 8 TeV LHC, for the invariant mass range $M_{ll} \in [500,1000]$ GeV.  The band shows the error coming from the photon distribution function.  The dashed lines show the estimated errors coming from statistics and from uncertainties in the quark and gluon distribution functions.  The right panel shows the $\chi^2$ deviation for each bin assuming 20 fb$^{-1}$ of integrated luminosity.  The region near the Jacobian peak, where fixed-order perturbation theory breaks down, has been shaded out.}  \label{fig:leppT8TeV500_1000}
\end{figure}

Finally, we show in Fig.~\ref{fig:leppT8TeV500_1000} the invariant mass bin $M_{ll} \in [500,1000]$ GeV.  The event rate is too small to make this bin as sensitive to photon-induced corrections as the lower ones, even though the deviations reach 40\% in the low $p_{Tl}$ region.  However, the $\chi^2$ function indicates that this mass range can still discriminate between different photon distribution functions.  The electroweak corrections are smaller, and constant over the entire kinematics range.  The $\chi^2$ function in the lower $p_{Tl}$ bins does not significantly change if the EW corrections are turned off in addition to the photon contributions.

\begin{figure}
\centering
\mbox{\subfigure{\includegraphics[width=3.1in]{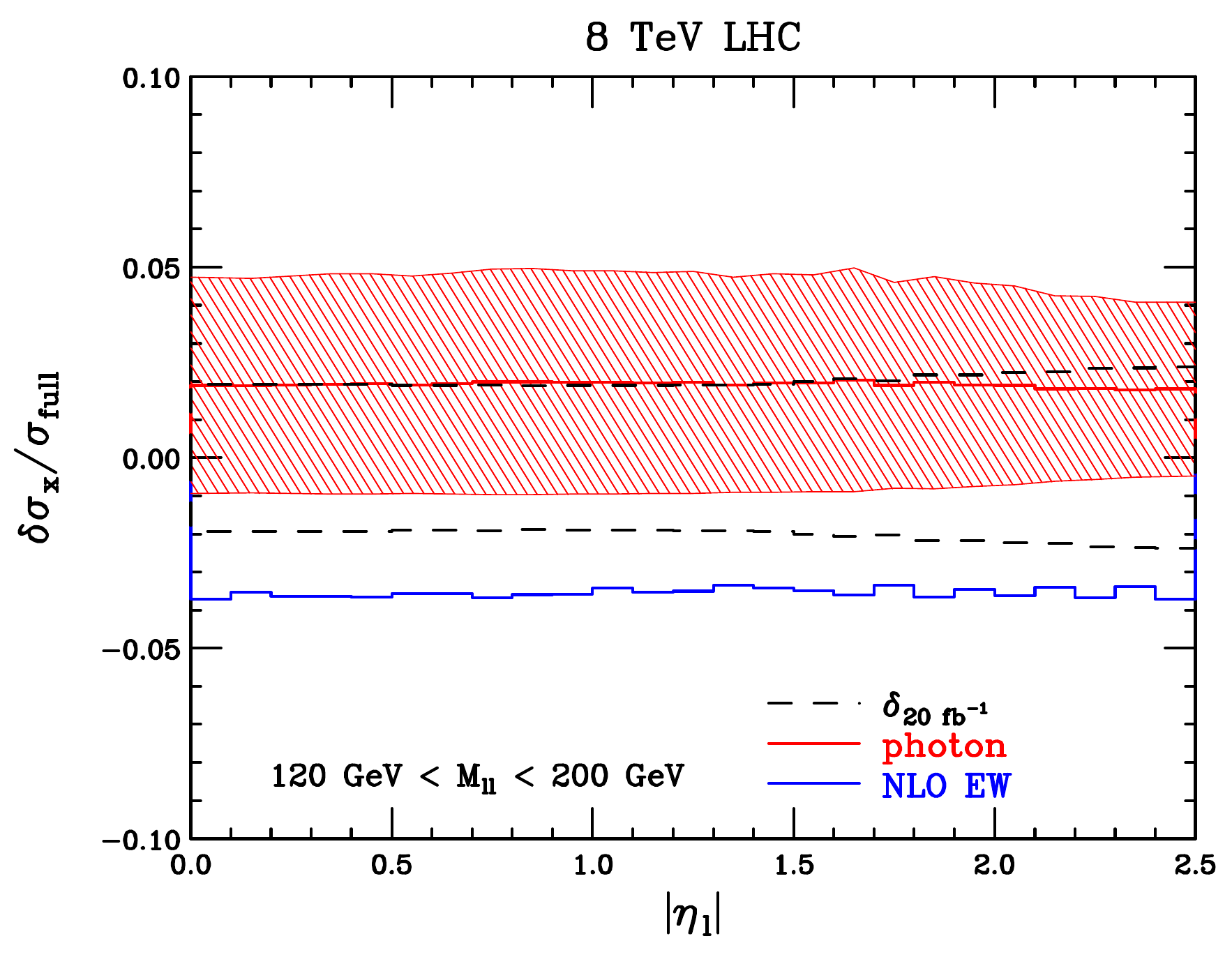}}\quad
\subfigure{\includegraphics[width=3.1in]{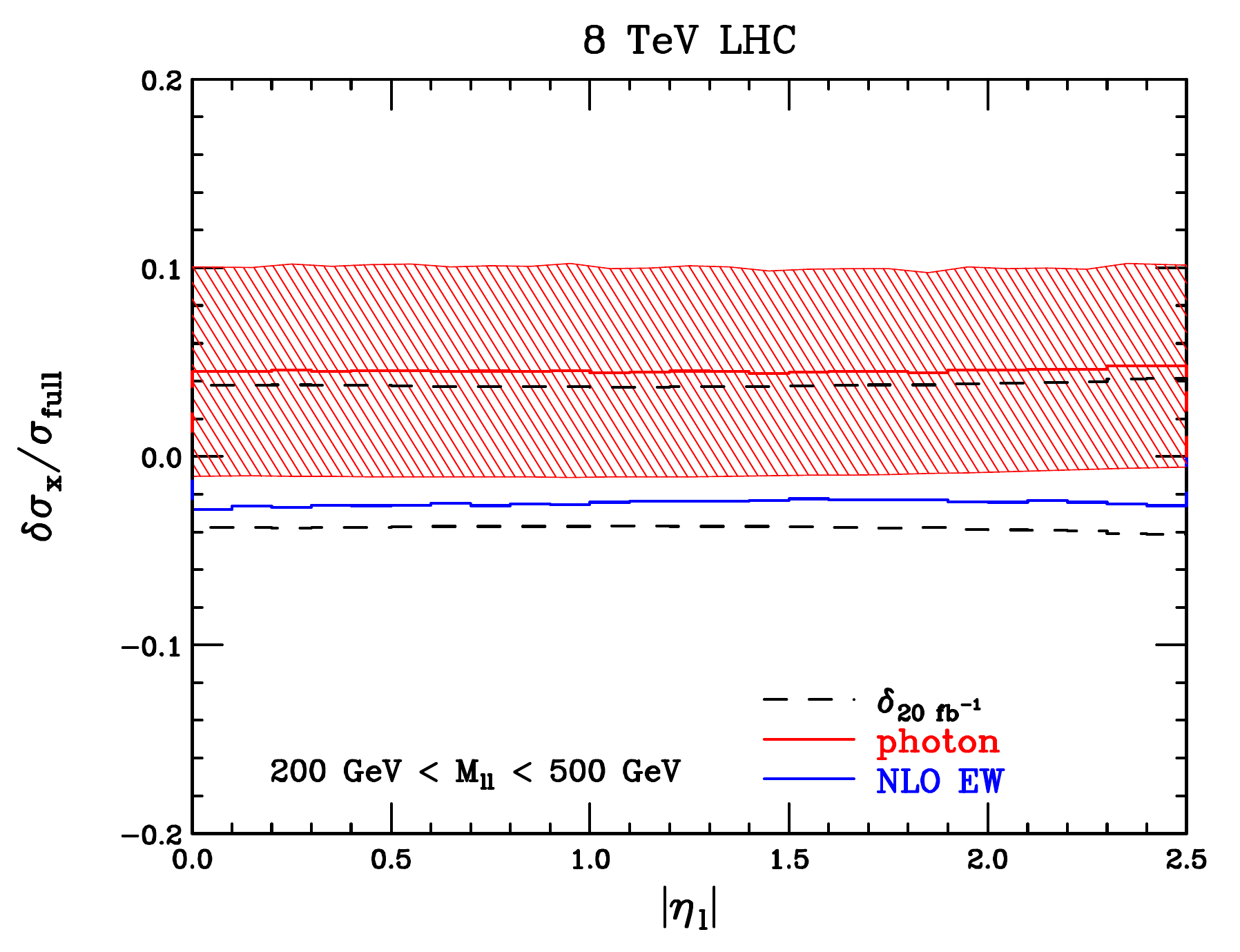}}}
\subfigure{\includegraphics[width=3.1in]{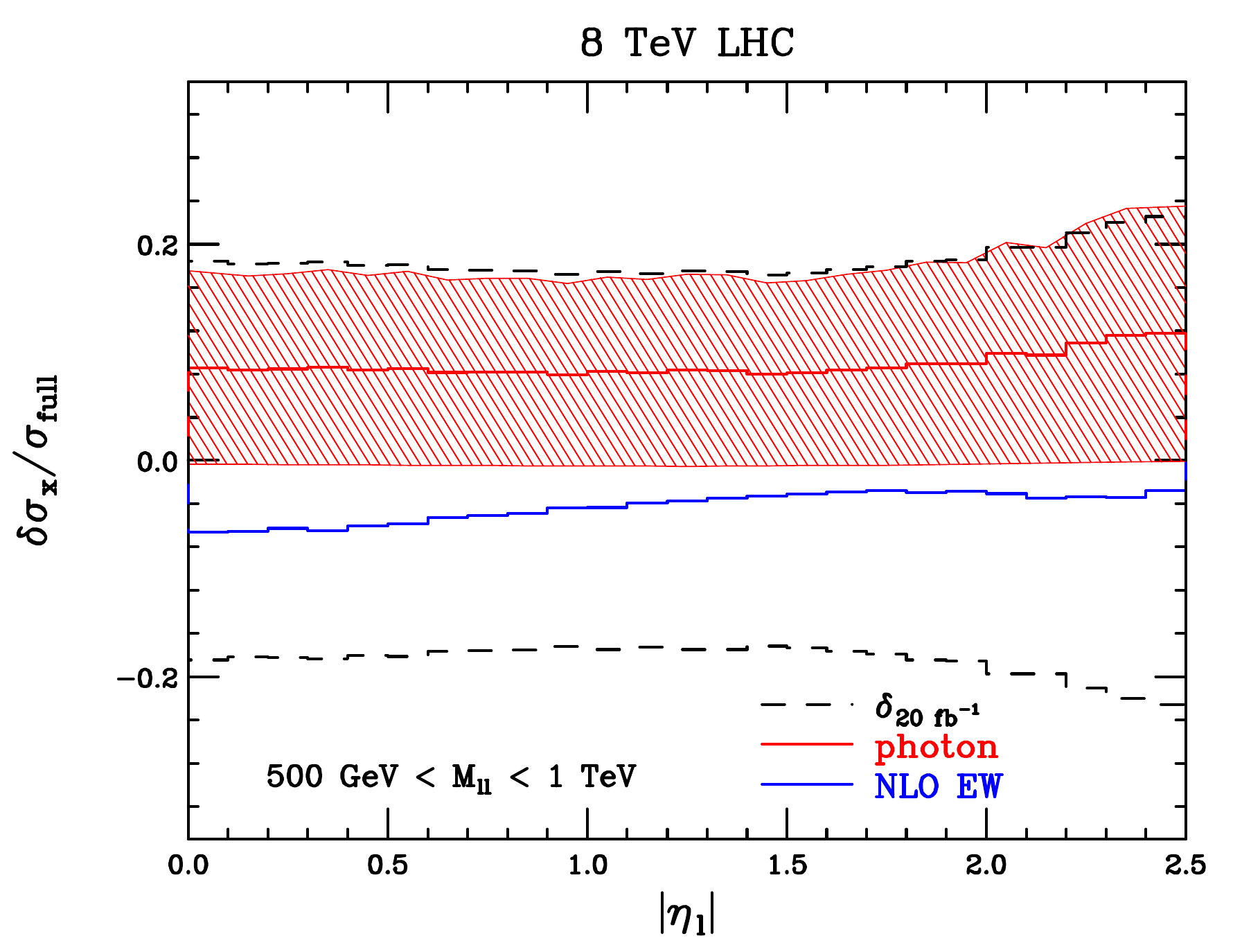}}
\caption{Shown are the deviations induced by photon-initiated contributions and electroweak corrections to the lepton pseudorapidity distribution  at an 8 TeV LHC, for the invariant mass ranges $M_{ll} \in [120,200]$ GeV, $M_{ll} \in [200,500]$ GeV, and $M_{ll} \in [500,1000]$ GeV.  The bands show the errors coming from the photon distribution function.  The dashed lines show the estimated errors coming from statistics and from uncertainties in the quark and gluon distribution functions. }  \label{fig:lepeta8TeV120_1000}
\end{figure}

We next study the lepton pseudorapidity distribution.  Results for the electroweak and photon-induced deviations are shown in Fig.~\ref{fig:lepeta8TeV120_1000} for all three invariant mass bins.  Both corrections are relatively flat over the entire kinematic range, and tend to cancel.  Structure in the distributions only appears in the highest invariant mass bin, but is too small to be observed over the estimated errors.  We will see this structure again when we consider the $|\eta_l|$ distribution at a 14 TeV LHC.  We do not show the $\chi^2$ distributions for this variable, since it is not particularly sensitive to either effects we are interested in extracting.

\begin{figure}
\centering
\mbox{\subfigure{\includegraphics[width=3.1in]{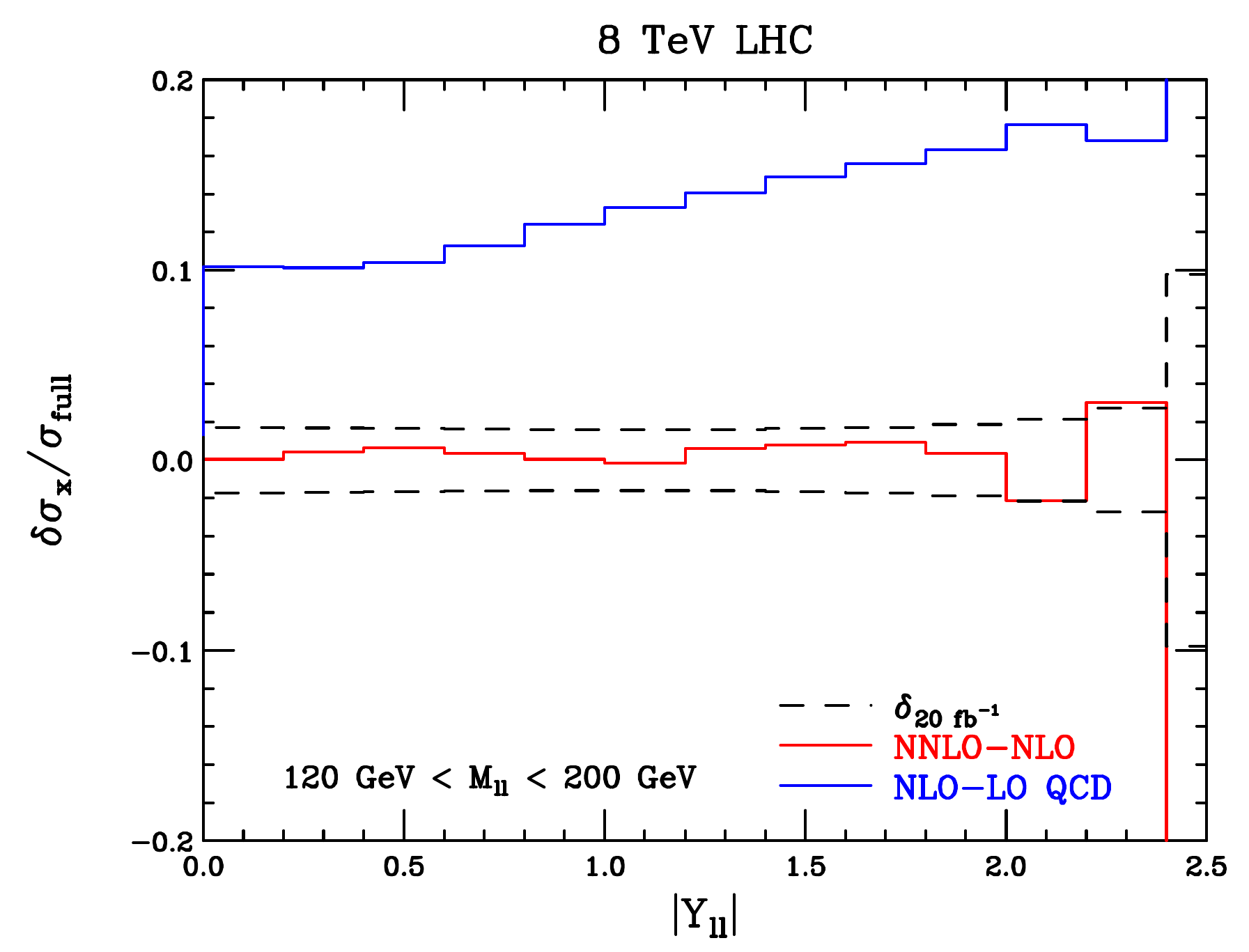}}\quad
\subfigure{\includegraphics[width=3.1in]{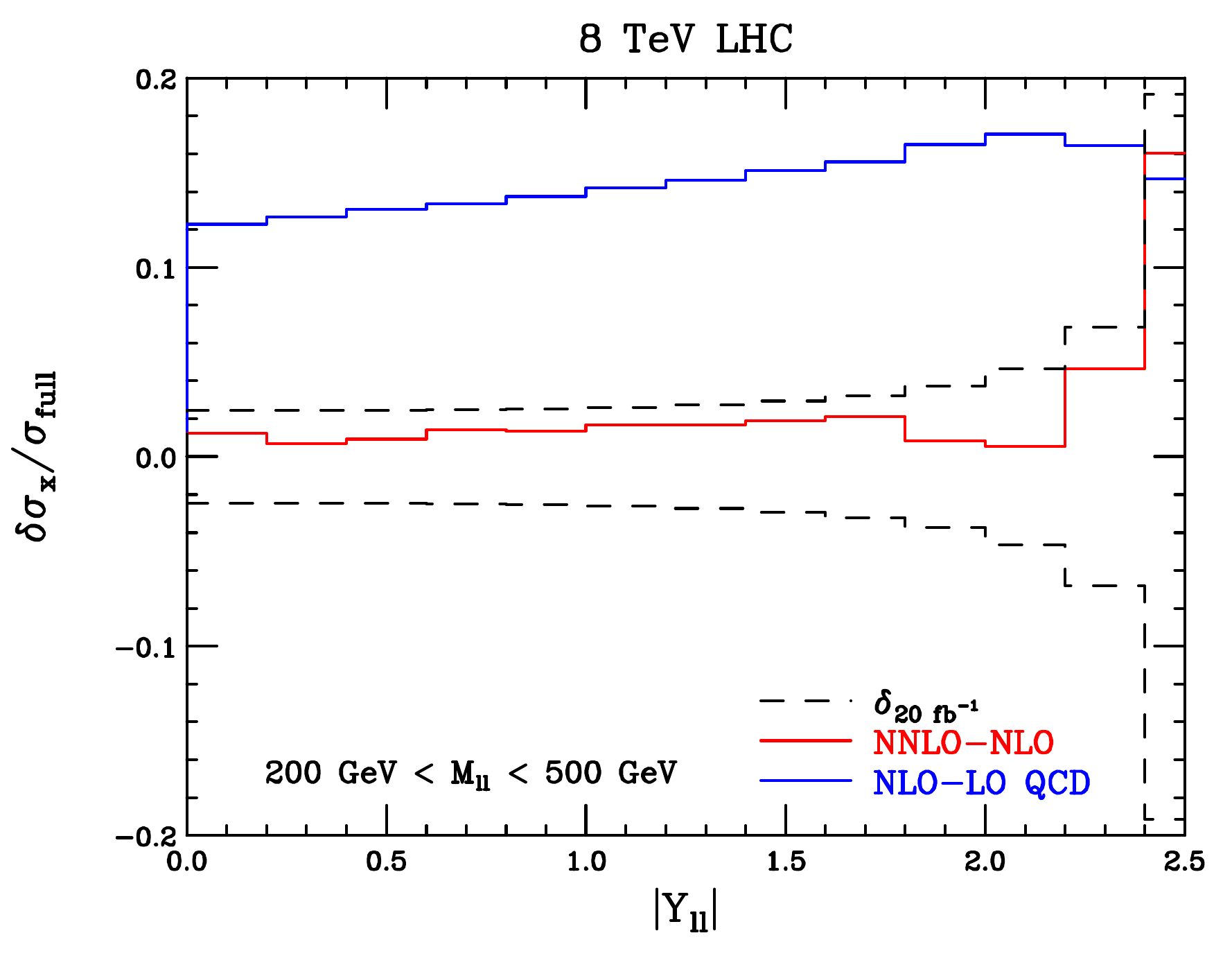}}}
\caption{Shown are the deviations induced by QCD corrections to the dilepton rapidity distribution at an 8 TeV LHC, for the invariant mass ranges $M_{ll} \in [120,200]$ GeV (left panel) and $M_{ll} \in [200,500]$ GeV (left panel).  The two lines indicate the deviation of NLO QCD minus LO relative to the full result, and NNLO minus NLO relative to the full result. The dashed lines show the estimated errors coming from statistics and from uncertainties in the quark and gluon distribution functions. }  \label{fig:ZrapQCD8TeV120_500}
\end{figure}

We conclude this section by considering the impact of higher-order QCD corrections on the distributions studied.  As emphasized in Ref.~\cite{Dittmaier:2009cr}, without sufficient control over QCD, other effects are swamped by its uncertainty.  A crucial aspect of our analysis is the inclusion of the NNLO QCD corrections.  We show in Fig.~\ref{fig:ZrapQCD8TeV120_500} the corrections induced by both NLO QCD and NNLO QCD on the dilepton rapidity distribution for the lowest two invariant mass bins.  While the change from LO to NLO is large, the additional shift in going from NLO to NNLO QCD is small, typically less than or equal to the estimated uncertainty from other sources.  Even adding the difference between NLO and NNLO as an estimate of uncertainty from higher-order corrections to the $\chi^2$ function of Eq.~(\ref{eq:chidef}), which we feel is an overestimate of this error, does not significantly reduce the observability of the photon-initiated or EW terms.  To avoid too large a proliferation of plots we do not show the QCD deviations for other observables and other mass bins, but simply note that the above comments remain true with two exceptions: the $p_{Tl}$ distribution near the Jacobian peak, where we anyway expect fixed-order perturbation theory to break down, and the $p_{Tl}$ distribution right at the lower cut, where the NNLO-NLO QCD corrections are comparable to or slightly larger than the uncertainty from other sources.  Since the most sensitive region to the photon PDF is above the lower $p_{Tl}$ cut as discussed above (due to the $\eta_l$ cut), this does not have a large effect on the presented results, but a careful accounting of QCD effects in this region is important.  Since the $\eta_l$ cut also has a significant effect near this boundary, relaxing it slightly may reduce the impact of higher-order QCD.  We show that this is indeed the case in a later section.

\subsection{Results for a 14 TeV LHC}

We next proceed to study the $|Y_{ll}|$, $|\eta_l|$, and $p_{Tl}$ distributions at a 14 TeV LHC.  We now additionally consider the invariant mass bin $M_{ll} \in [1000,3000]$ GeV.  When estimating the statistical errors, we assume 30 fb$^{-1}$ for the lower three bins, an amount expected after roughly one year of LHC operation.  For the new bin we assume 100 fb$^{-1}$, consistent with roughly two years of LHC run time, since the event rate for this bin is lower than the others.  Since the errors for this bin are dominated by statistics and not PDF errors, uncertainty estimates for different values of integrated luminosity can be approximated by a simple rescaling of the presented results.

\begin{figure}
\centering
\mbox{\subfigure{\includegraphics[width=3.1in]{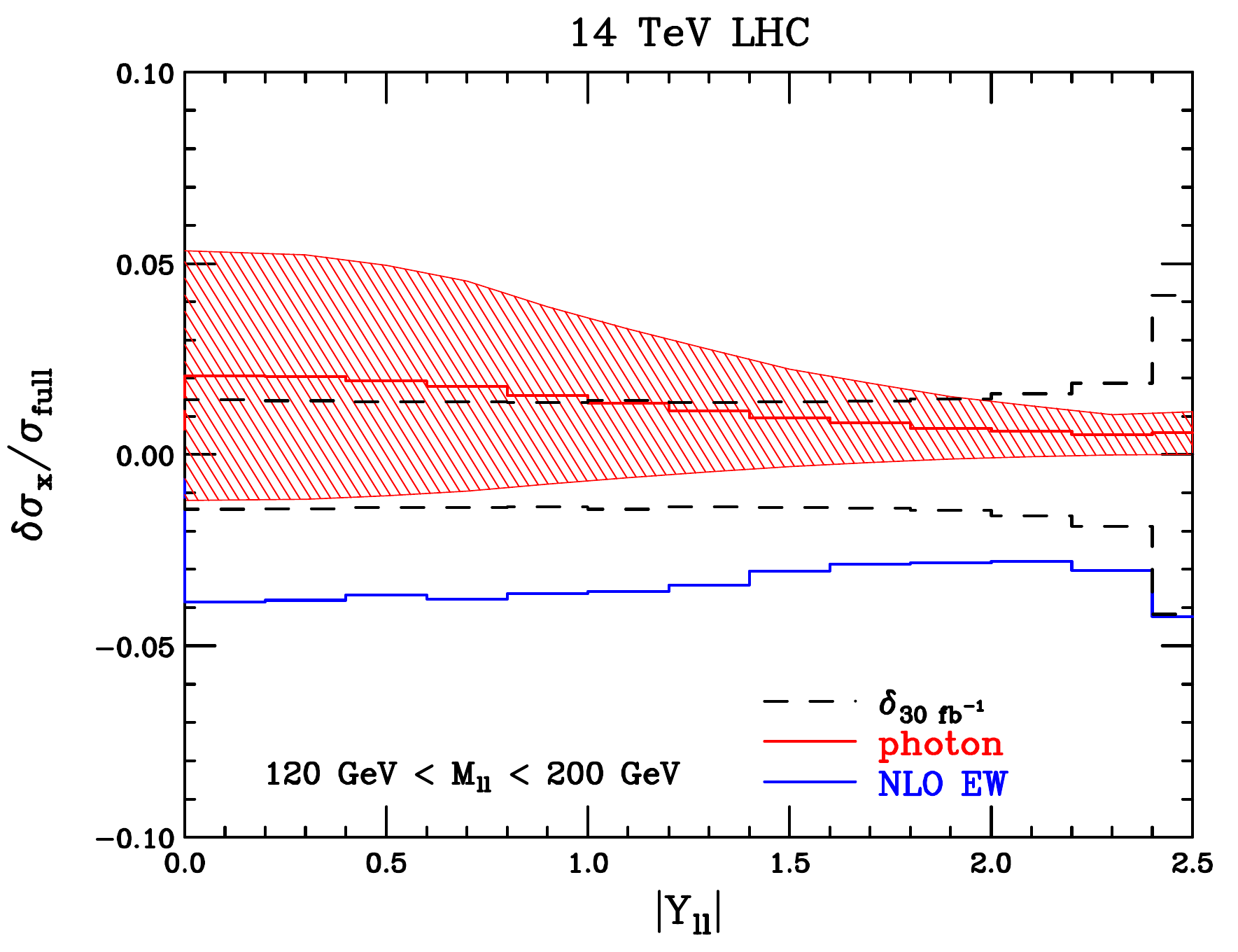}}\quad
\subfigure{\includegraphics[width=3.1in]{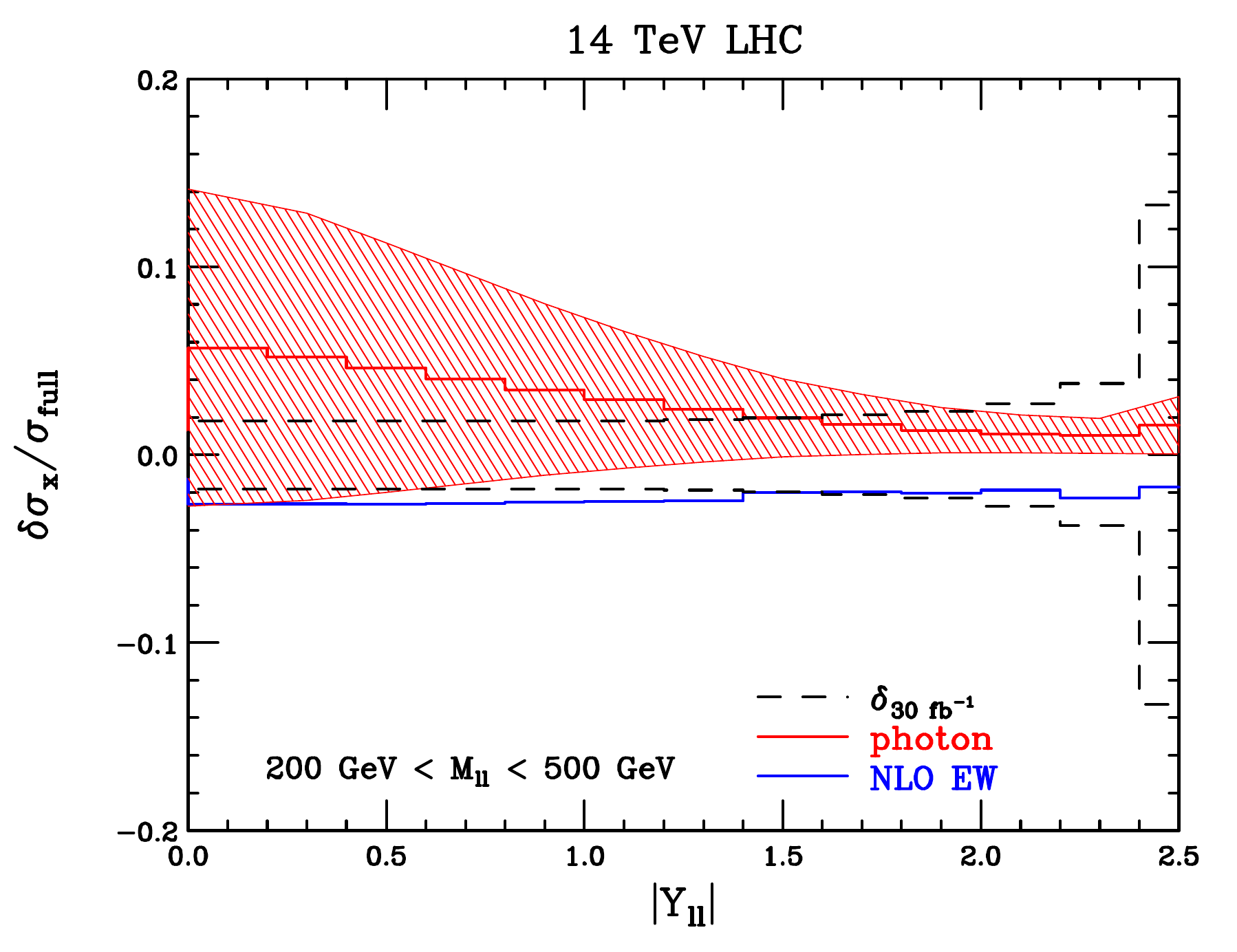}}}
\mbox{\subfigure{\includegraphics[width=3.1in]{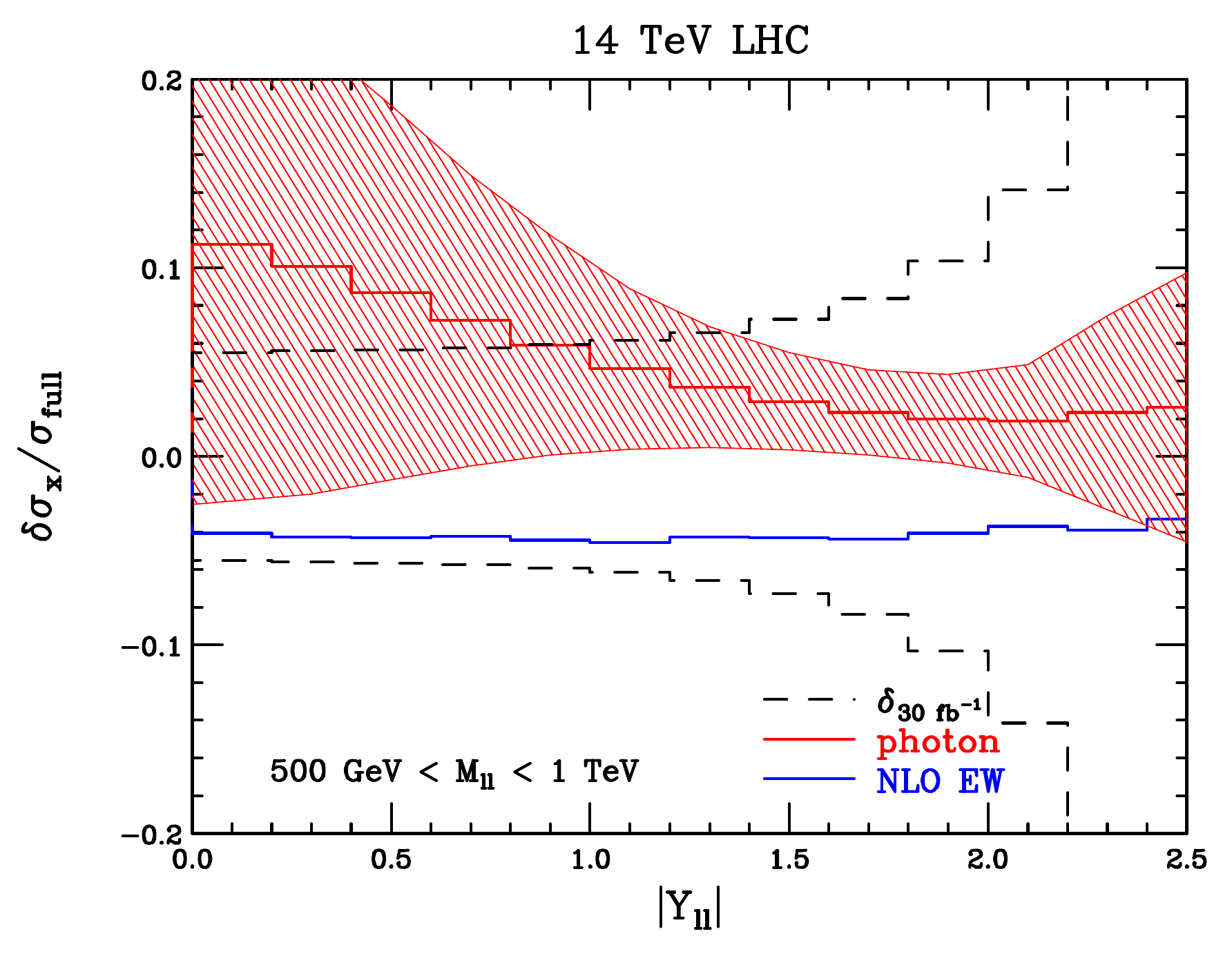}}\quad
\subfigure{\includegraphics[width=3.1in]{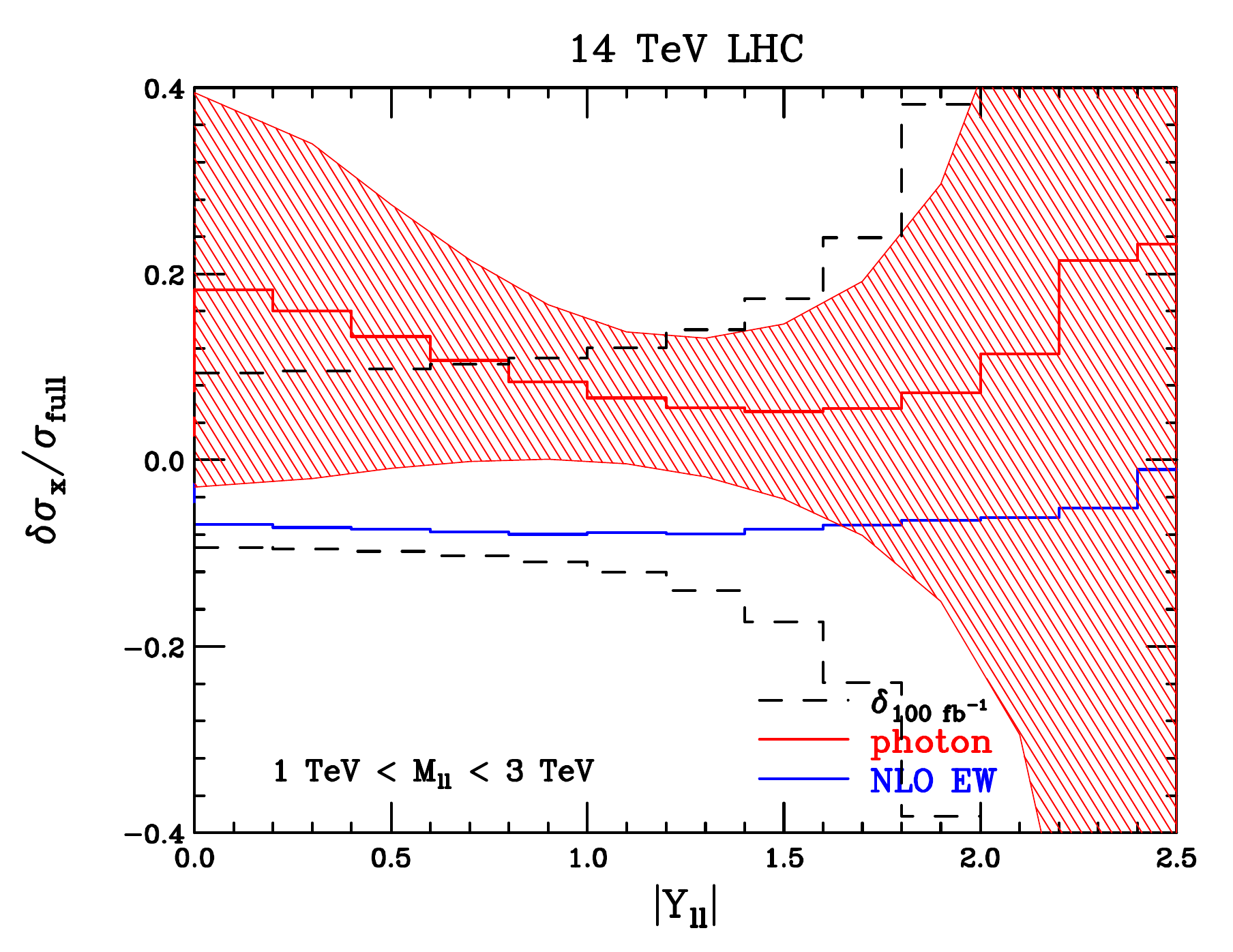}}}
\caption{Shown are the deviations induced by photon-initiated contributions and electroweak corrections to the dilepton rapidity distribution at a 14 TeV LHC.  Clockwise from the top left, the plots show the following invariant mass ranges: $M_{ll} \in [120,200]$ GeV, $M_{ll} \in [200,500]$ GeV, $M_{ll} \in [500,1000]$ GeV and $M_{ll} \in [1000,3000]$ GeV.  The bands show the errors coming from the photon distribution function.  The dashed lines show the estimated errors coming from statistics and from uncertainties in the quark and gluon distribution functions. For the first three invariant mass bins, 30 fb$^{-1}$ are assumed, while 100 fb$^{-1}$ are assumed for $M_{ll} \in [1000,3000]$ GeV.}  \label{fig:Zrap14TeV120_3000}
\end{figure}

\begin{figure}
\centering
\mbox{\subfigure{\includegraphics[width=3.1in]{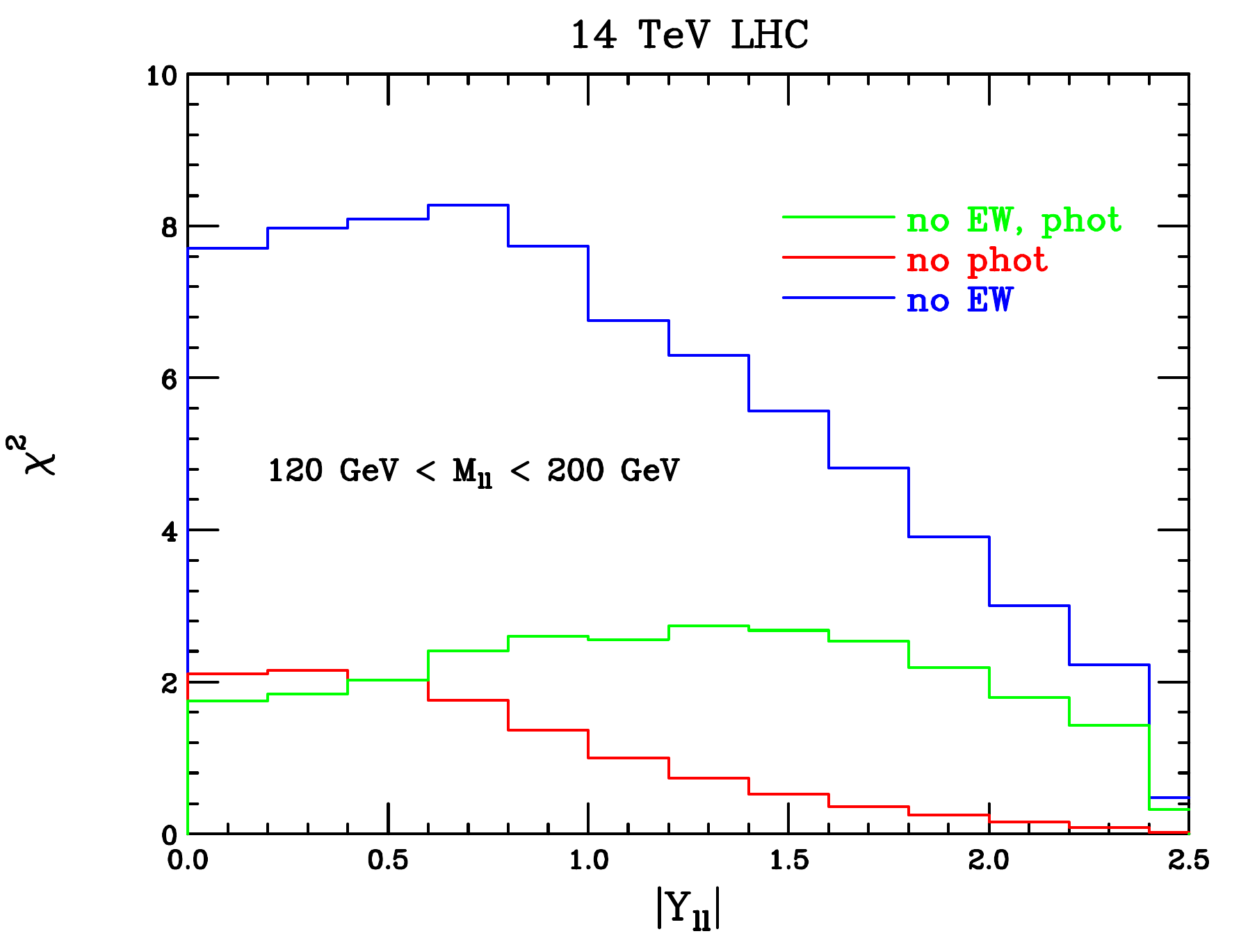}}\quad
\subfigure{\includegraphics[width=3.1in]{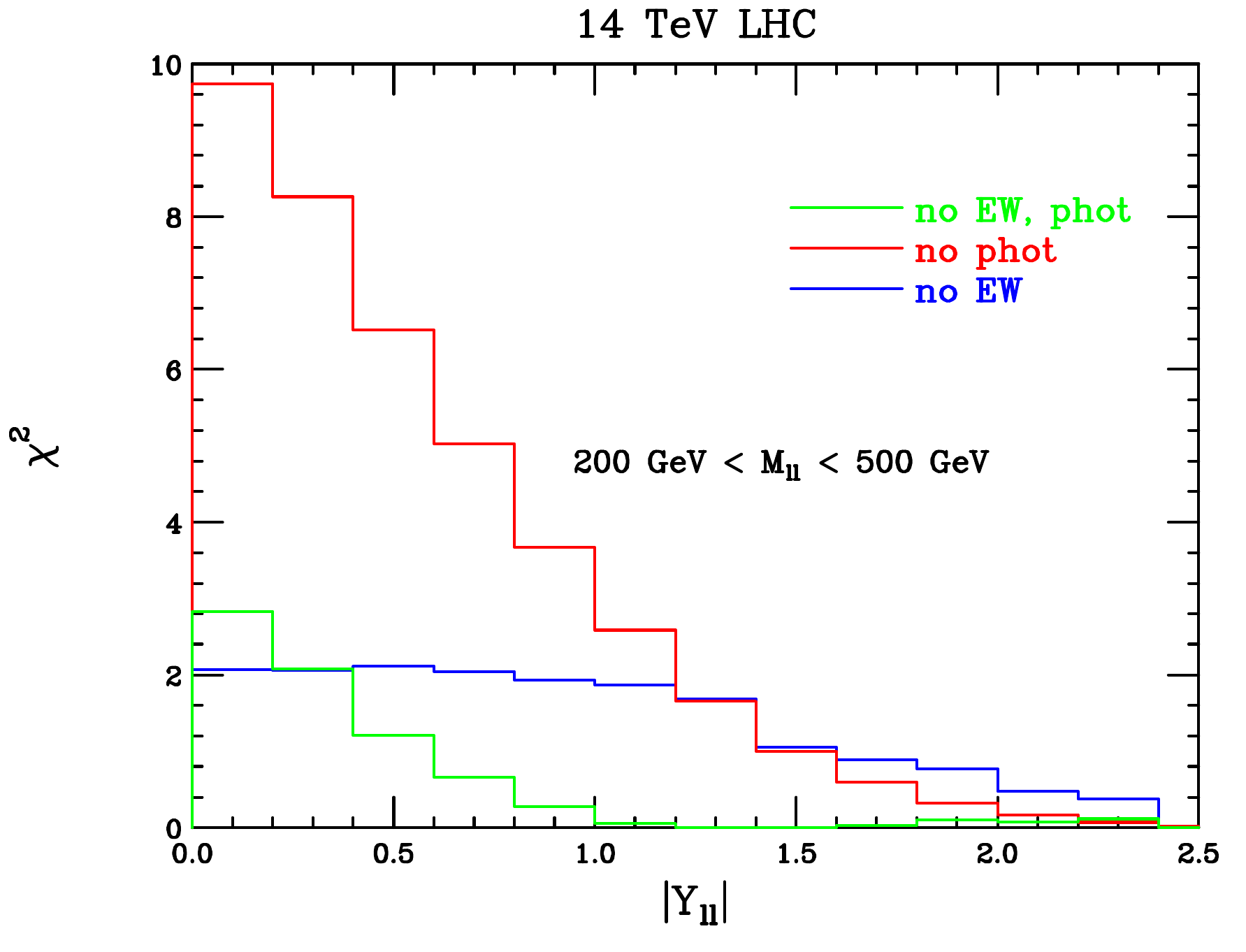}}}
\mbox{\subfigure{\includegraphics[width=3.1in]{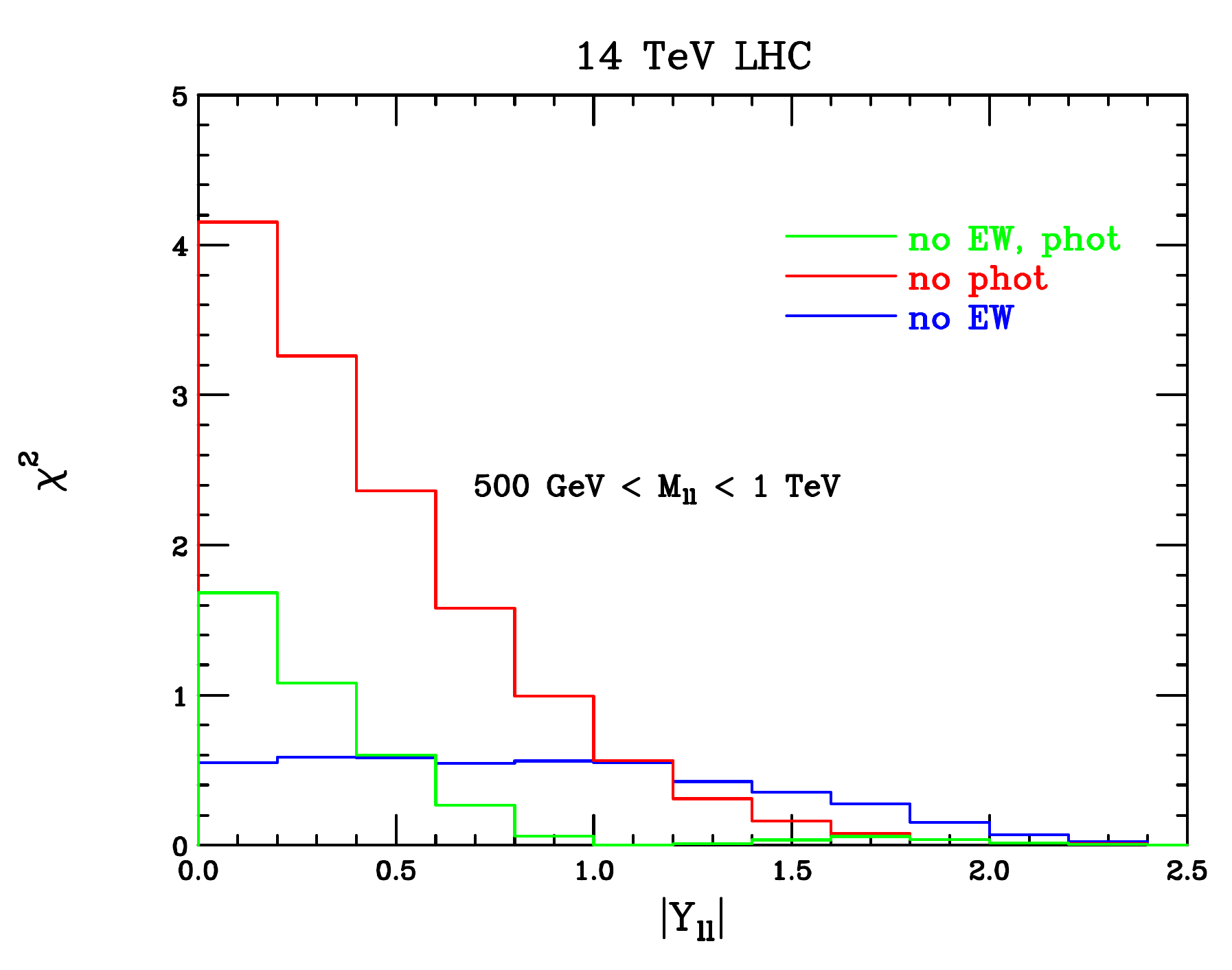}}\quad
\subfigure{\includegraphics[width=3.1in]{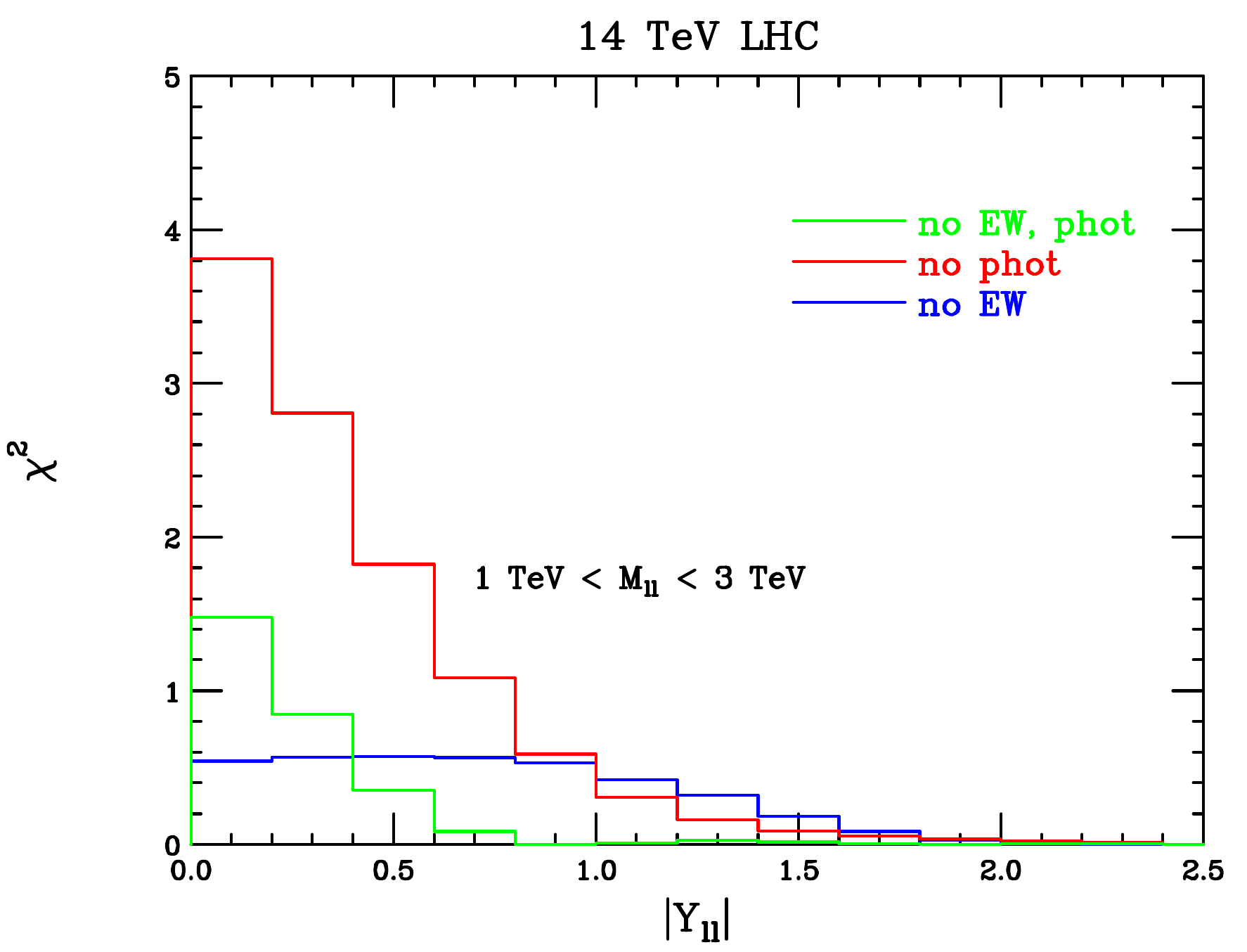}}}
\caption{Shown are the $\chi^2$ distributions for the dilepton rapidity distribution.  Clockwise from the top left, the plots show the following invariant mass ranges: $M_{ll} \in [120,200]$ GeV, $M_{ll} \in [200,500]$ GeV, $M_{ll} \in [500,1000]$ GeV and $M_{ll} \in [1000,3000]$ GeV. For the first three invariant mass bins, 30 fb$^{-1}$ are assumed, while 100 fb$^{-1}$ are assumed for $M_{ll} \in [1000,3000]$ GeV.}  \label{fig:Zrap14TeV120_3000_chi2}
\end{figure}

We start with the dilepton rapidity distribution.  The deviations coming from electroweak corrections and photon-induced processes for all four invariant mass bins are shown in Fig.~\ref{fig:Zrap14TeV120_3000}.  Several trends are apparent from the plot.  Just like in 8 TeV collisions, the photon contributions are peaked toward central rapidity, while the electroweak corrections are flat.  The electroweak corrections grow slightly as the invariant mass is increased, changing from $-4\%$ to $-7\%$ at central rapidity when going from the lowest bin to the highest bin.  The photon terms grow more quickly, increasing from $+5\%$ to $+20\%$ when going from the lowest bin to the highest bin.  They are larger than the estimated statistical+PDF uncertainty for central rapidities in all bins.  The error on the photon PDF is much larger than the other considered sources of uncertainty, especially in the higher-invariant mass bins.  This has already been noted in the literature~\cite{Ball:2013hta}.  Improved control over this quantity will be critical to enable high-mass searches at Run II of the LHC.  The corresponding $\chi^2$ distributions are shown in Fig.~\ref{fig:Zrap14TeV120_3000_chi2}.  The most sensitive bin is the $M_{ll} \in [200,500]$ GeV one, although all four show a strong sensitivity to the photon PDF.  The cancellations between the negative EW corrections and the positive photon-initiated ones are important for all four invariant mass bins.

\begin{figure}
\centering
\mbox{\subfigure{\includegraphics[width=3.1in]{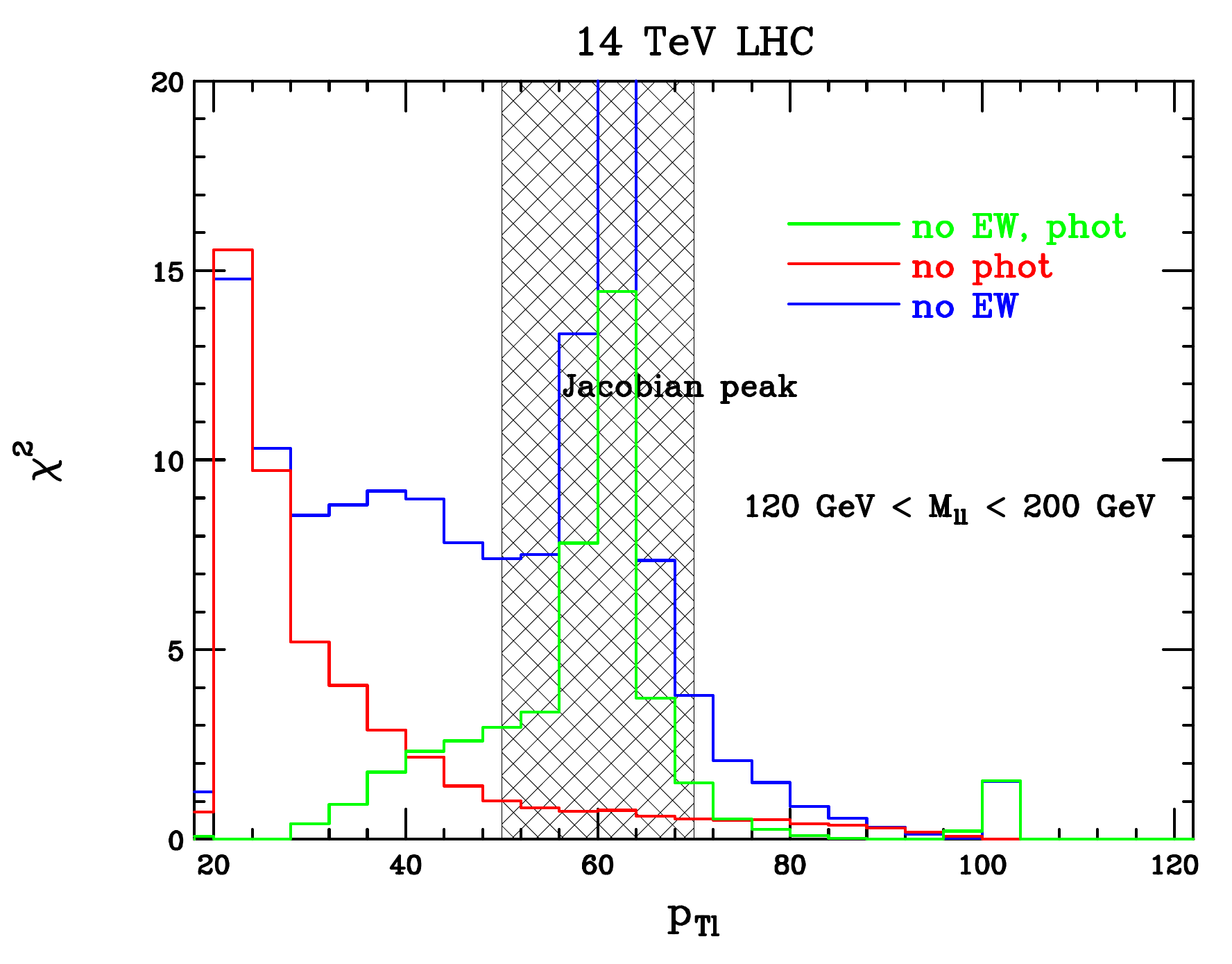}}\quad
\subfigure{\includegraphics[width=3.1in]{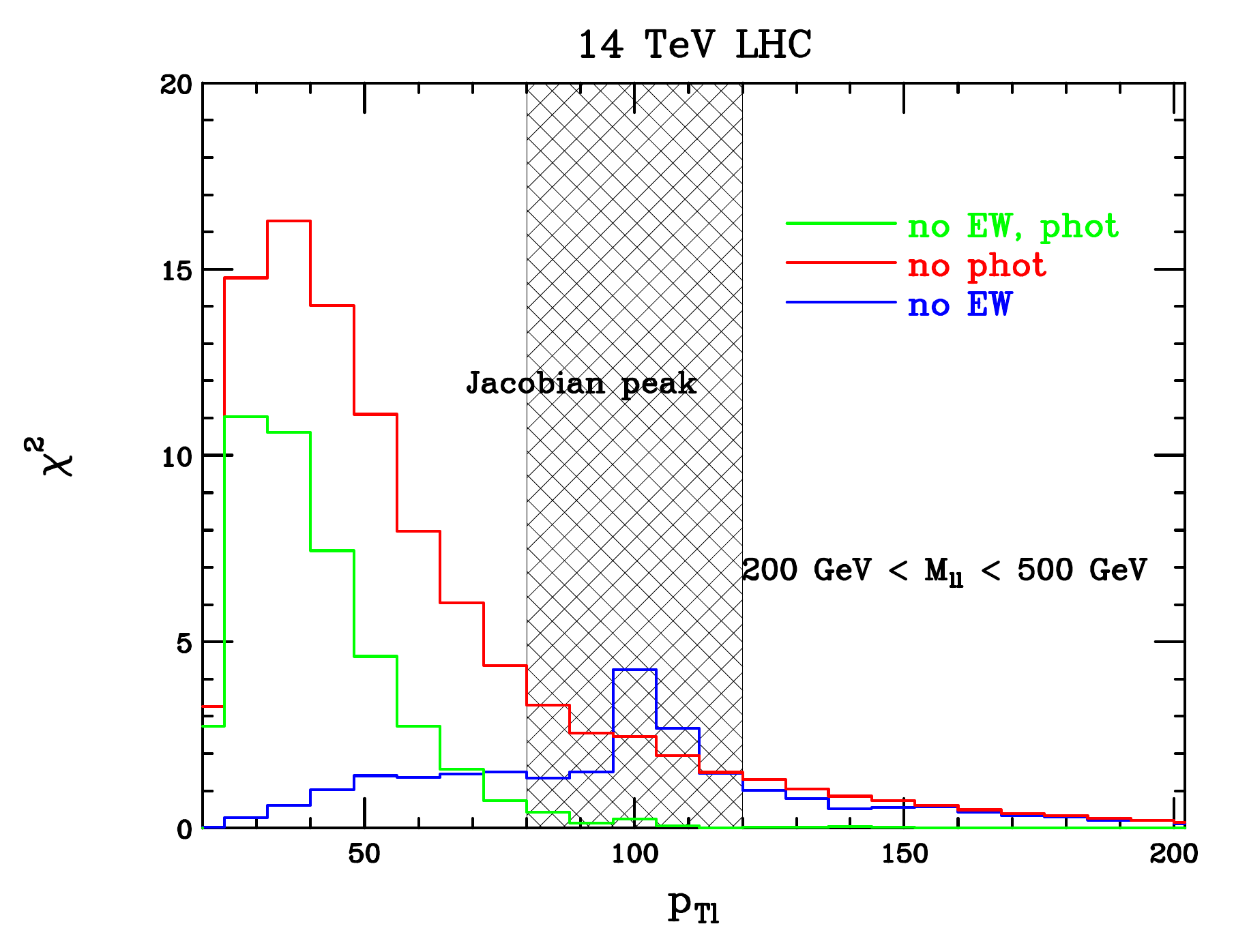}}}
\mbox{\subfigure{\includegraphics[width=3.1in]{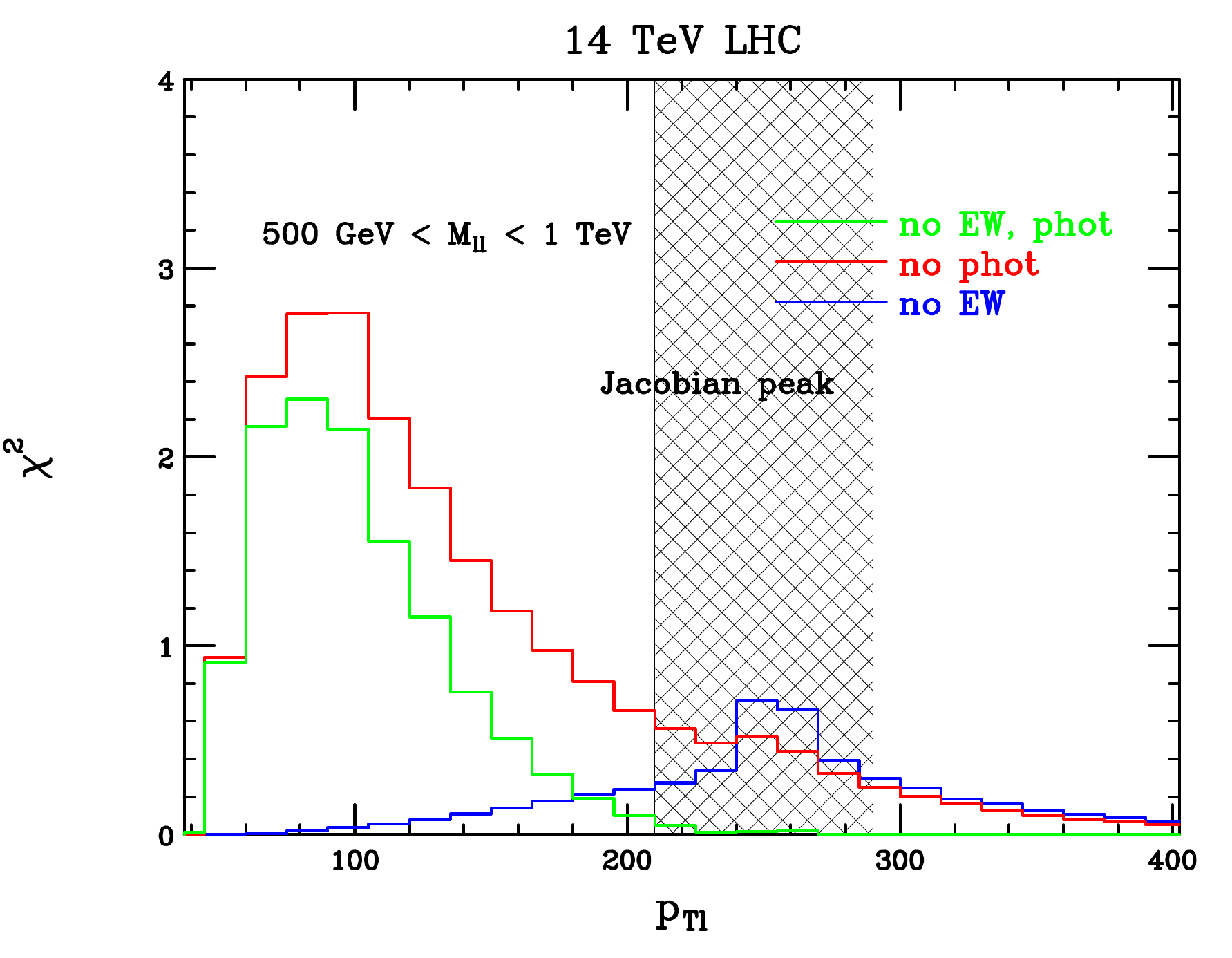}}\quad
\subfigure{\includegraphics[width=3.1in]{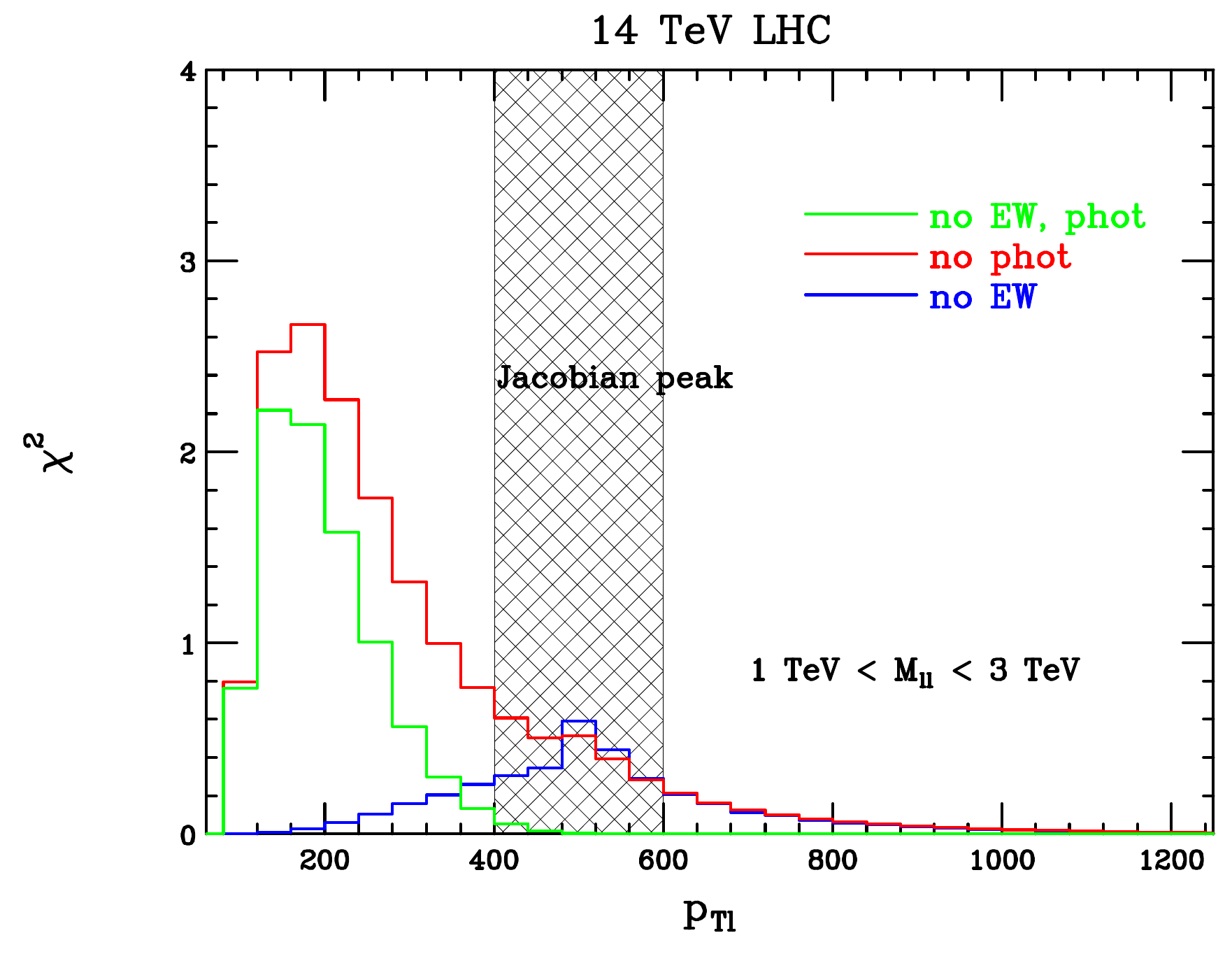}}}
\caption{Shown are the $\chi^2$ distributions for the lepton transverse momentum distribution.  Clockwise from the top left, the plots show the following invariant mass ranges: $M_{ll} \in [120,200]$ GeV, $M_{ll} \in [200,500]$ GeV, $M_{ll} \in [500,1000]$ GeV and $M_{ll} \in [1000,3000]$ GeV. For the first three invariant mass bins, 30 fb$^{-1}$ are assumed, while 100 fb$^{-1}$ are assumed for $M_{ll} \in [1000,3000]$ GeV.  The region near the Jacobian peak, where fixed-order perturbation theory breaks down, has been shaded out.}  \label{fig:leppT14TeV120_3000_chi2}
\end{figure}

We now consider the lepton transverse momentum distribution.  Results for the four invariant mass bins of Eq.~(\ref{eq:inv14}) are shown below in Fig.~\ref{fig:leppT14TeV120_3000}.  The photon deviations are large in all four bins, larger than the estimated statistical+PDF errors.  They strongly peak toward low $p_{Tl}$, as discussed in the presentation of 8 TeV results.  They grow from 8\% in the $M_{ll} \in [120,200]$ GeV bin to over 40\% in the $M_{ll} \in [1,3]$ TeV; photon-initiated scattering is expected to be an extremely large component of the total Drell-Yan cross section at Run II of the LHC.  The electroweak corrections grow much more mildly with invariant mass, reaching only $-7\%$ in the highest invariant mass bin.  They are flat as a function of lepton $p_{Tl}$.  The $\chi^2$ distributions for $p_{Tl}$ are shown in Fig.~\ref{fig:leppT14TeV120_3000_chi2} for each invariant mass bin.  The $\chi^2$ values are in general larger than those for the dilepton rapidity ones for a given invariant mass bin, indicating that the $p_{Tl}$ distribution is particularly sensitive to the photon PDF.  In the $M_{ll} \in [120,200]$ GeV bin it is crucial to have control over the electroweak contributions in order to properly extract the photon-initiated contributions.  Above this first bin the photon-initiated corrections increase quickly enough in size that the electroweak contributions become relatively less important.  The $p_{Tl}$ distribution is a good place to extract the photon PDF while lessening the effect of EW corrections.  As mentioned in the introduction, another good reason to use the $p_{Tl}$ distribution to extract the photon PDF is that potential physics beyond the Standard Model is expected to populate the high $p_{Tl}$ region, and the chance of confusing these two effects is therefore reduced.  We note that as in 8 TeV collisions, we have shaded out the region near the Jacobian peak in order to indicate the breakdown of fixed-order perturbation theory.

\begin{figure}
\centering
\mbox{\subfigure{\includegraphics[width=3.1in]{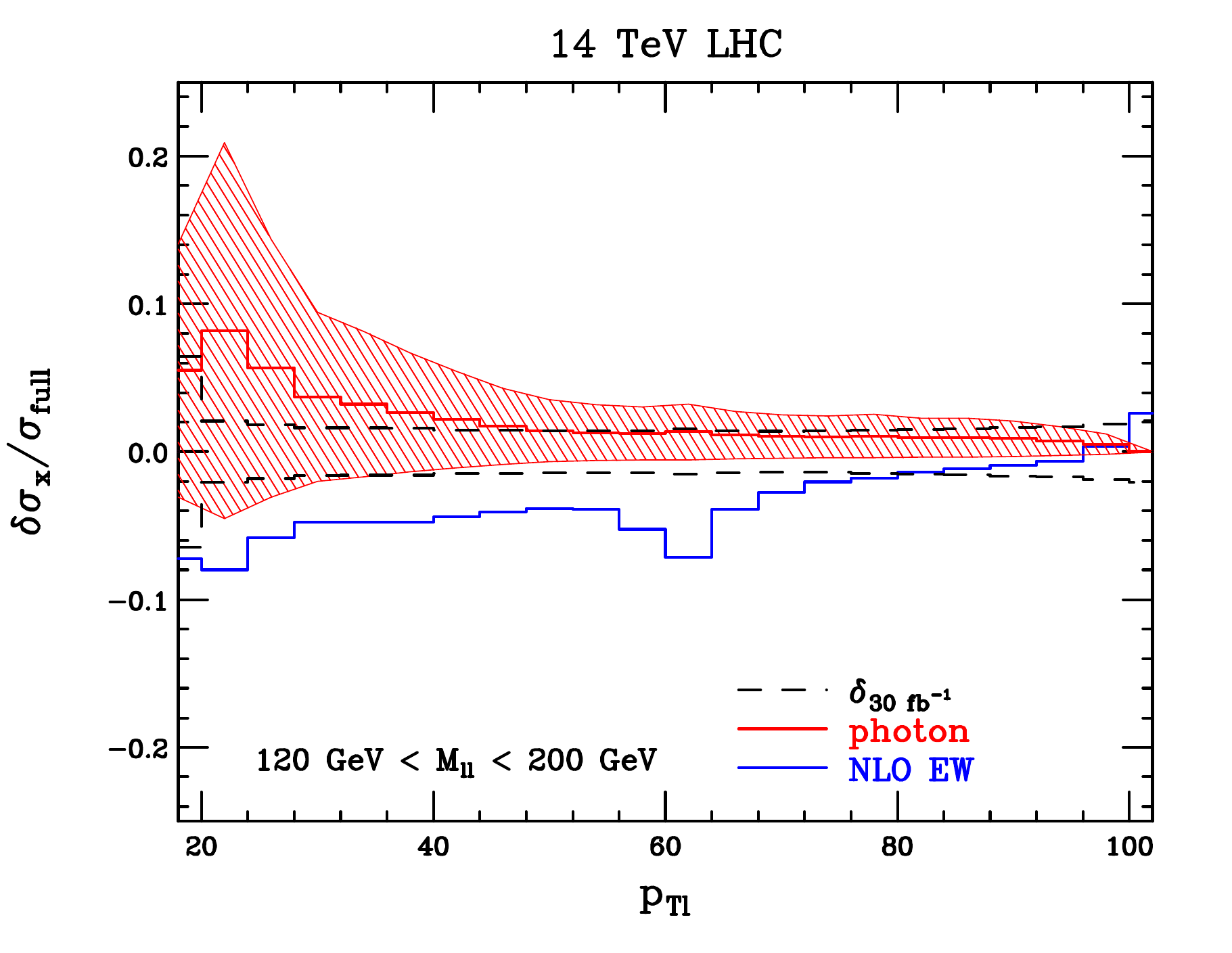}}\quad
\subfigure{\includegraphics[width=3.1in]{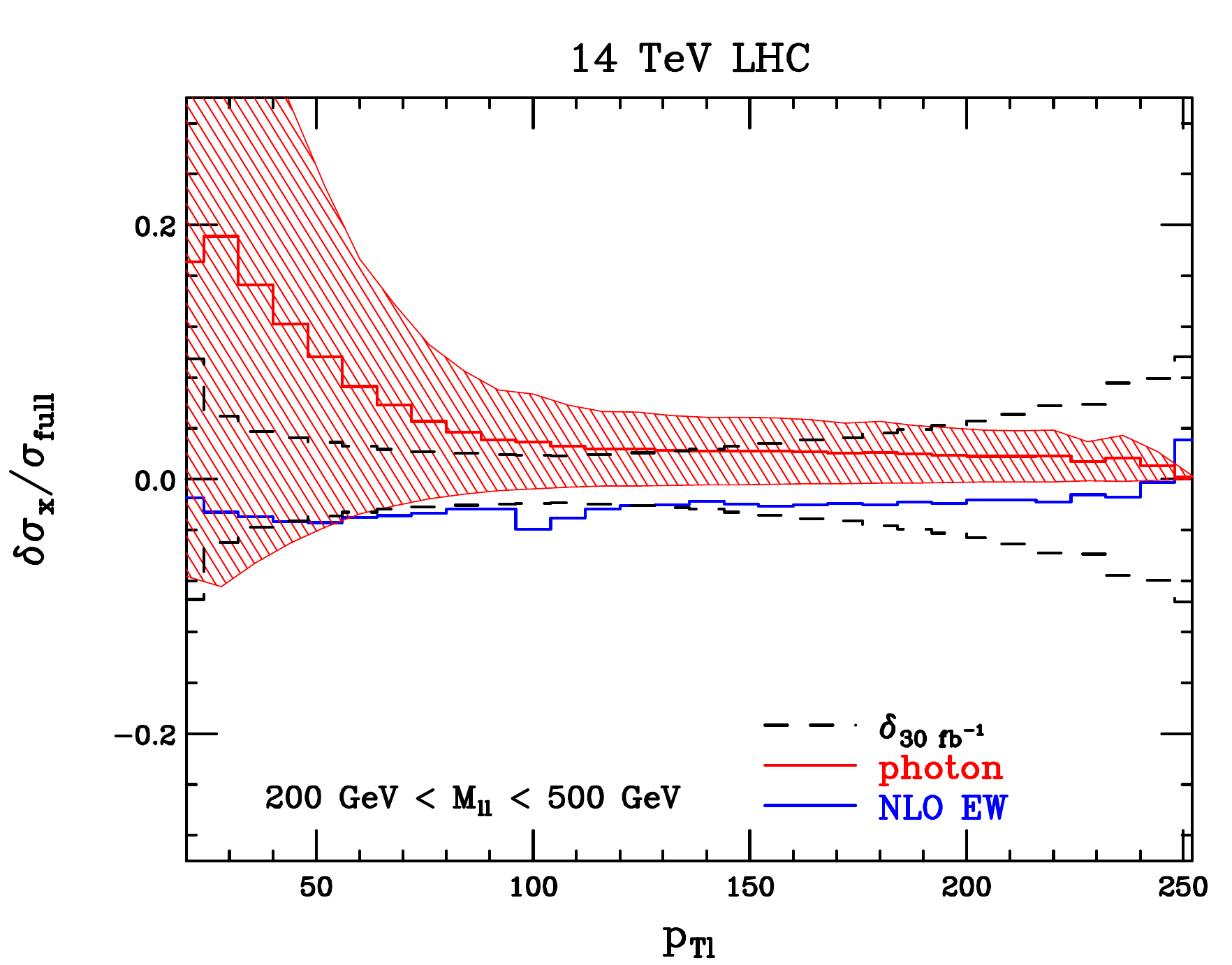}}}
\mbox{\subfigure{\includegraphics[width=3.1in]{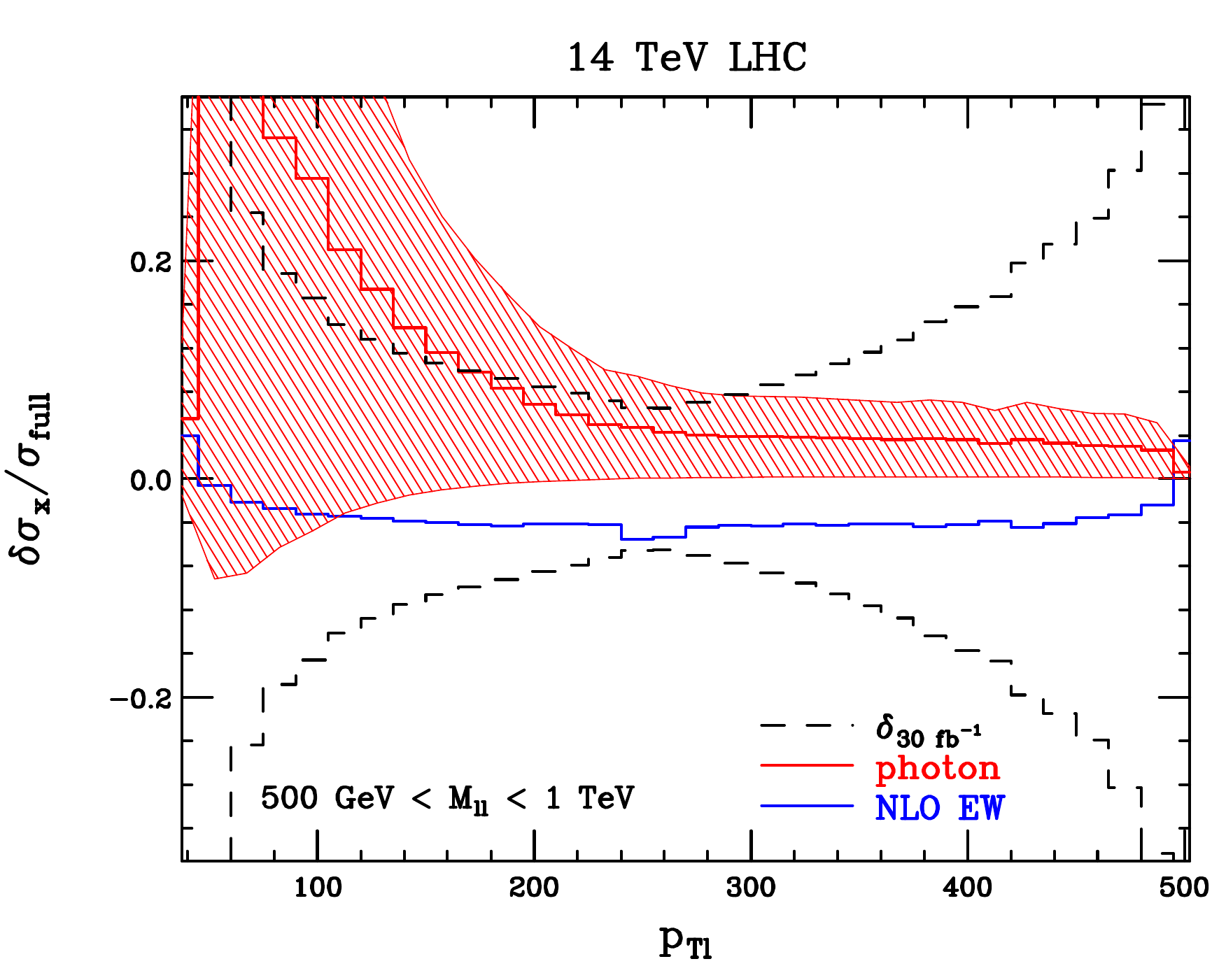}}\quad
\subfigure{\includegraphics[width=3.1in]{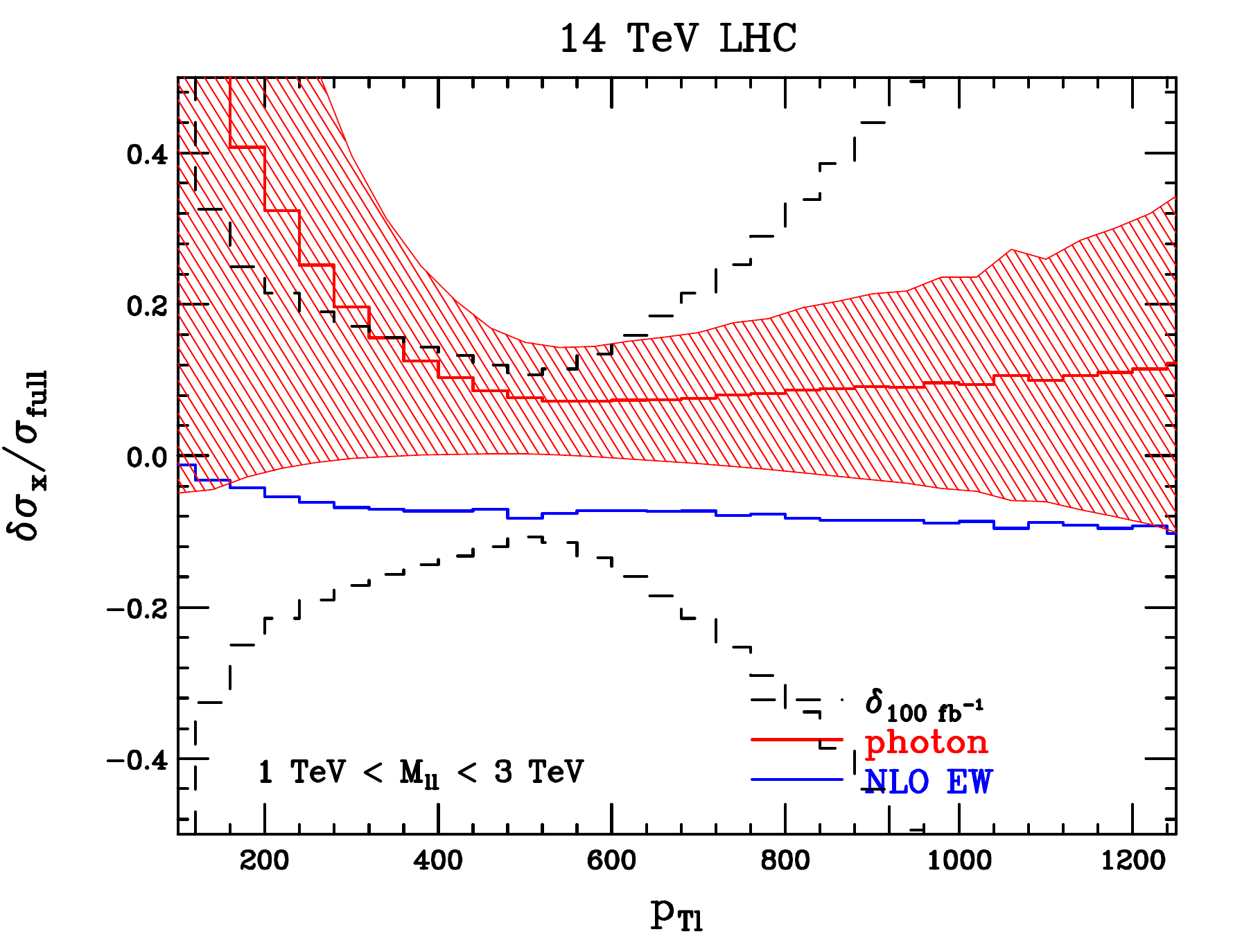}}}
\caption{Shown are the deviations induced by photon-initiated contributions and electroweak corrections to the $p_{Tl}$ distribution at a 14 TeV LHC.  Clockwise from the top left, the plots show the following invariant mass ranges: $M_{ll} \in [120,200]$ GeV, $M_{ll} \in [200,500]$ GeV, $M_{ll} \in [500,1000]$ GeV and $M_{ll} \in [1000,3000]$ GeV.  The bands show the errors coming from the photon distribution function.  The dashed lines show the estimated errors coming from statistics and from uncertainties in the quark and gluon distribution functions. For the first three invariant mass bins, 30 fb$^{-1}$ are assumed, while 100 fb$^{-1}$ are assumed for $M_{ll} \in [1000,3000]$ GeV.}  \label{fig:leppT14TeV120_3000}
\end{figure}

\begin{figure}
\centering
\mbox{\subfigure{\includegraphics[width=3.1in]{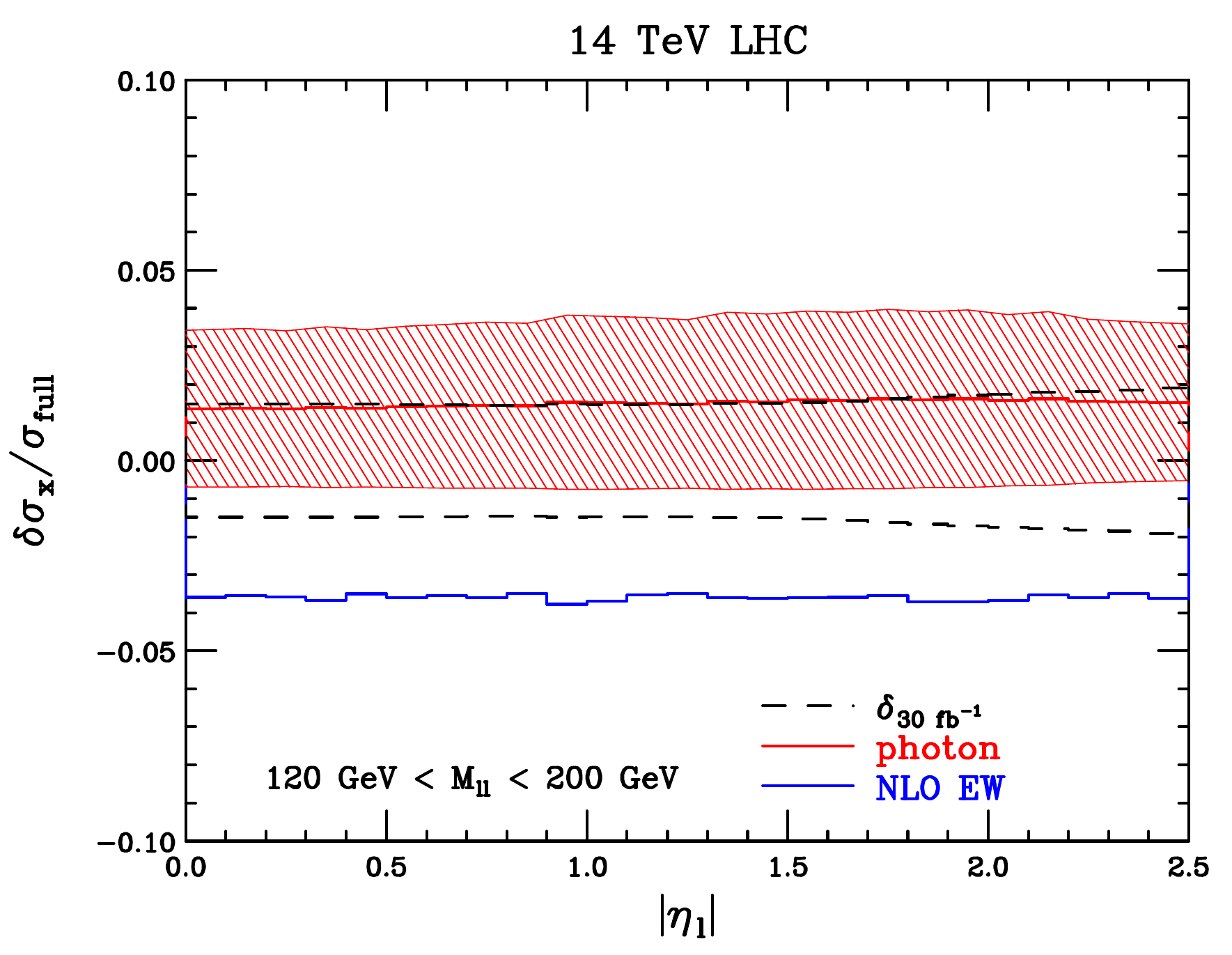}}\quad
\subfigure{\includegraphics[width=3.1in]{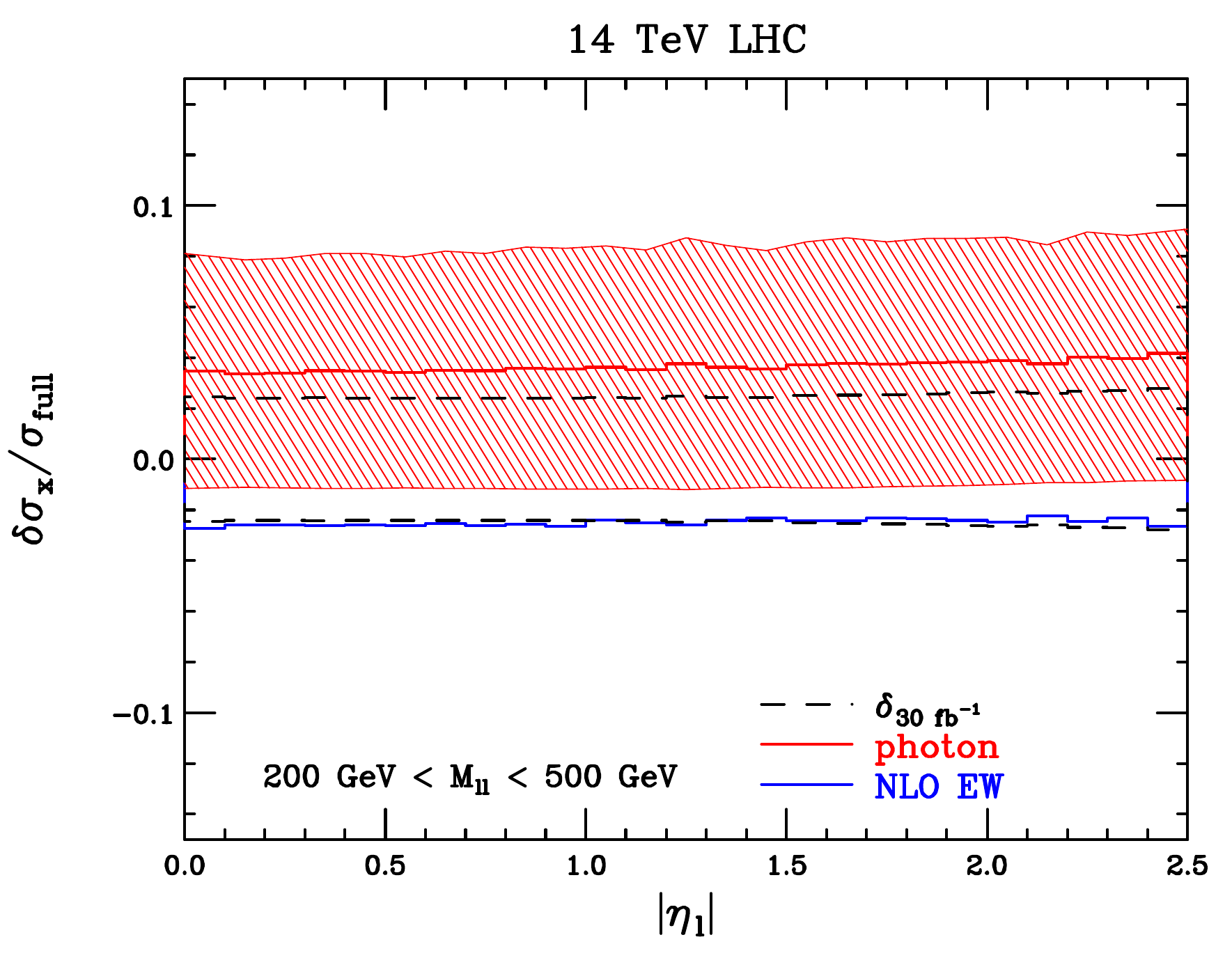}}}
\mbox{\subfigure{\includegraphics[width=3.1in]{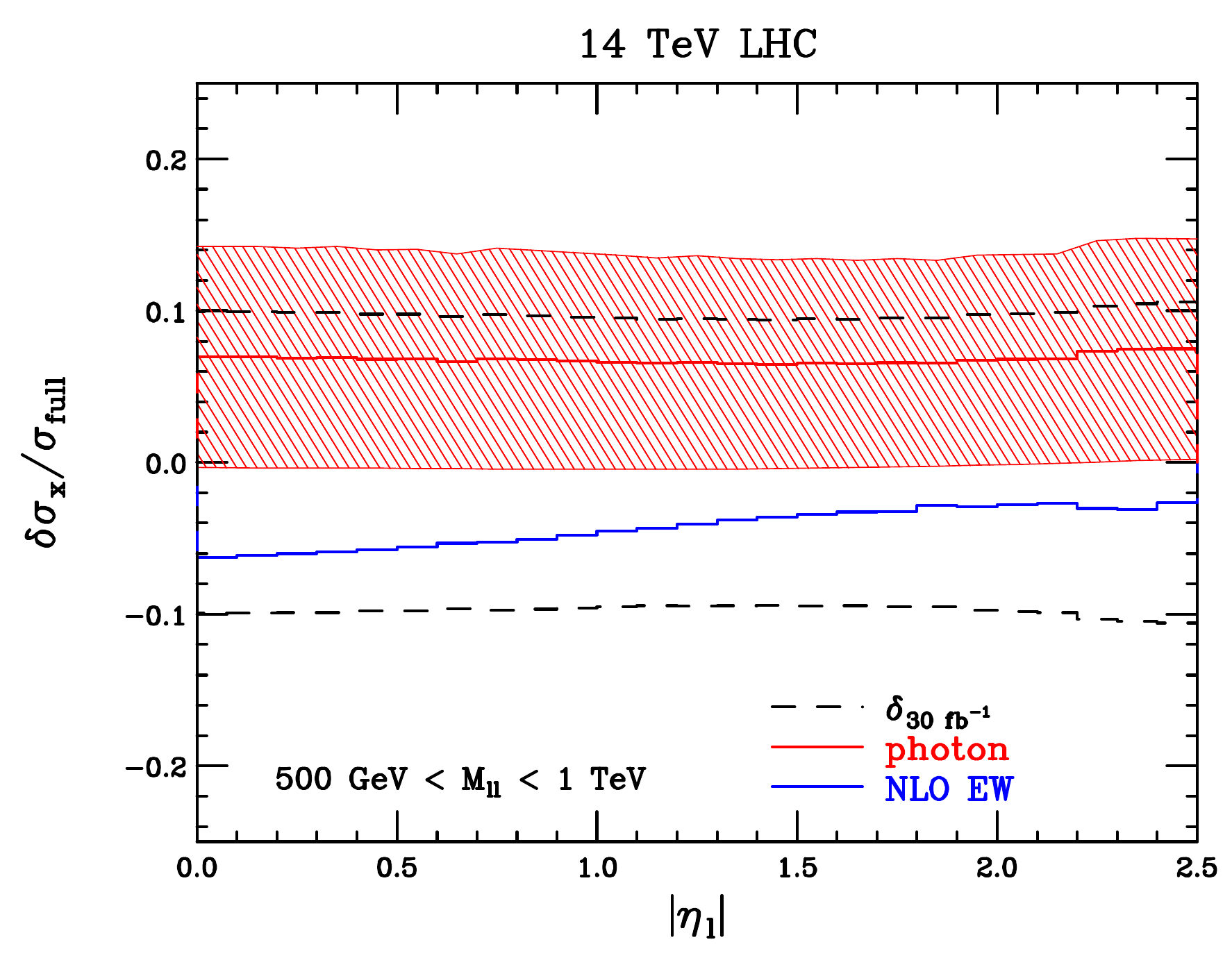}}\quad
\subfigure{\includegraphics[width=3.1in]{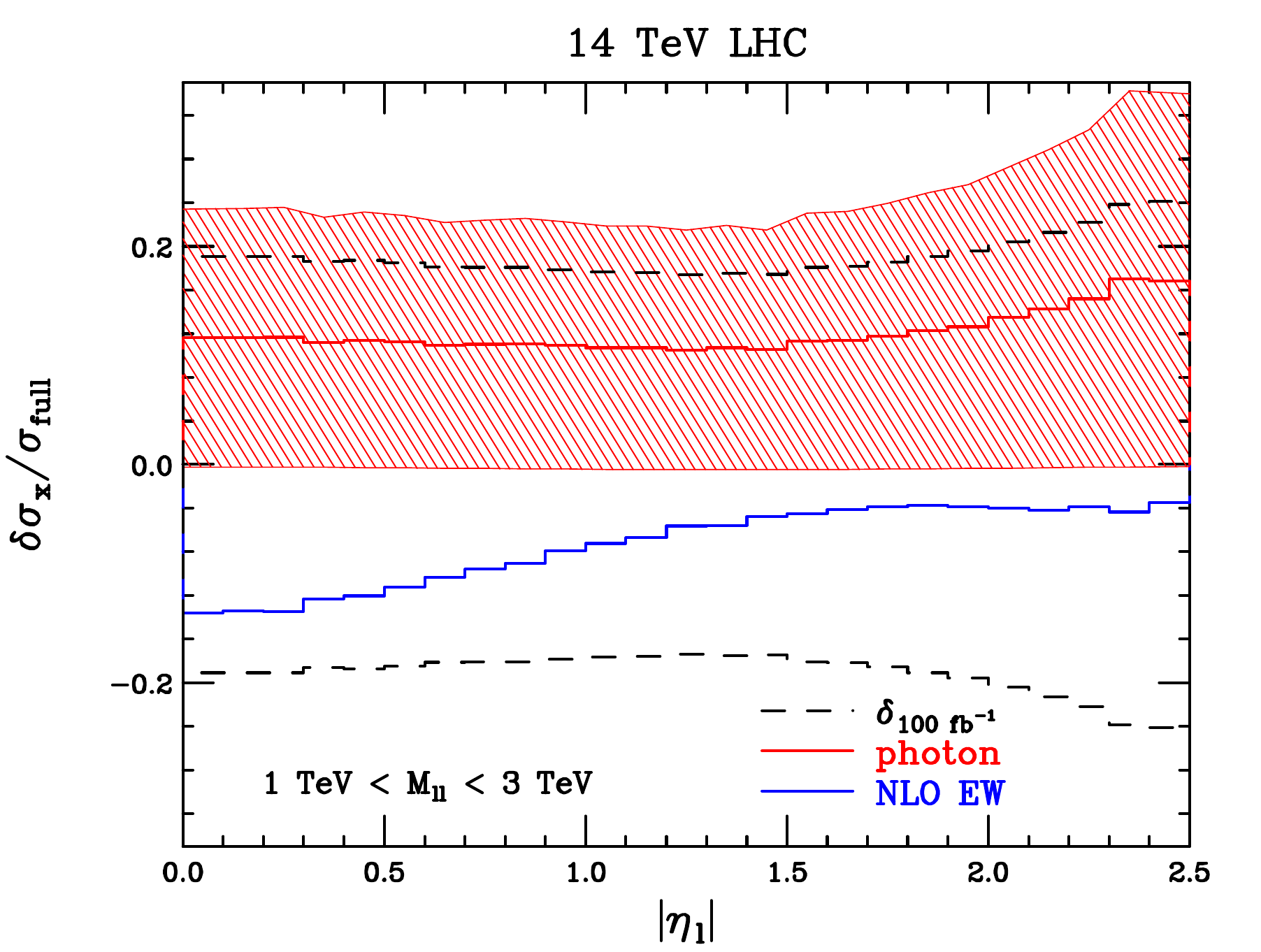}}}
\caption{Shown are the deviations induced by photon-initiated contributions and electroweak corrections to the $|\eta_l|$ distribution at a 14 TeV LHC.  Clockwise from the top left, the plots show the following invariant mass ranges: $M_{ll} \in [120,200]$ GeV, $M_{ll} \in [200,500]$ GeV, $M_{ll} \in [500,1000]$ GeV and $M_{ll} \in [1000,3000]$ GeV.  The bands show the errors coming from the photon distribution function.  The dashed lines show the estimated errors coming from statistics and from uncertainties in the quark and gluon distribution functions. For the first three invariant mass bins, 30 fb$^{-1}$ are assumed, while 100 fb$^{-1}$ are assumed for $M_{ll} \in [1000,3000]$ GeV.}  \label{fig:lepeta14TeV120_3000}
\end{figure}

Finally, results for the lepton $|\eta_l|$ distributions are shown in Fig.~\ref{fig:lepeta14TeV120_3000}.  For the lower two invariant mass bins, the deviations from both photon-initiated processes and from electroweak corrections are relatively flat over the entire kinematic range.  The size of the deviations is generally at or below the expected uncertainties arising from statistics and imperfect knowledge of quark and gluon PDFs.  Structure begins to appear in the electroweak corrections in the $M_{ll} \in [500,1000]$ GeV bin, and becomes more pronounced in the highest bin, $M_{ll} \in [1000,3000]$ GeV.  The electroweak corrections reach $-15\%$ and are peaked toward central pseudorapidity.  This arises because of the strong angular dependence of the Sudakov logarithms which dominate the electroweak corrections at high invariant masses.  A detailed discussion of the angular dependence of Sudakov logarithms is given in Ref.~\cite{Chiu:2008vv}.  For example, it is shown there that in the partonic center-of-mass frame, the EW Sudakov logarithms are largest for the scattering angle $\text{cos}(\theta_{CM}) \approx 0$ for the $d \bar{d} \to \mu^+ \mu^-$ channel, while they are peaked toward forward scattering for the $u \bar{u} \to \mu^+ \mu^-$ channel.  A combination of these underlying processes leads to the effect seen in Fig.~\ref{fig:lepeta14TeV120_3000}.  Distributions such as the dilepton rapidity are relatively insensitive to this dependence.  When forming this observable the four-momenta of the two leptons are added, removing the dependence on the CM-frame scattering angle and therefore removing the angular structure of these corrections.  The $p_{Tl}$ distribution is more sensitive to the $t$-channel enhancement of the underlying matrix elements.  The lepton $|\eta_l|$ distribution offers a window onto the underlying structure of the Sudakov logarithms.  We see from Fig.~\ref{fig:lepeta14TeV120_3000} that while large, the deviations caused by the electroweak corrections are small compared to the estimated errors, at least with the relatively fine binning chosen.  More luminosity or a coarser binning is needed to uncover the underlying structure of the Sudakov corrections.  The need to control the photon PDF using other distributions is clear; in agreement with the trend observed so far, the two types of corrections tend to cancel.

\begin{figure}
\centering
\mbox{\subfigure{\includegraphics[width=3.1in]{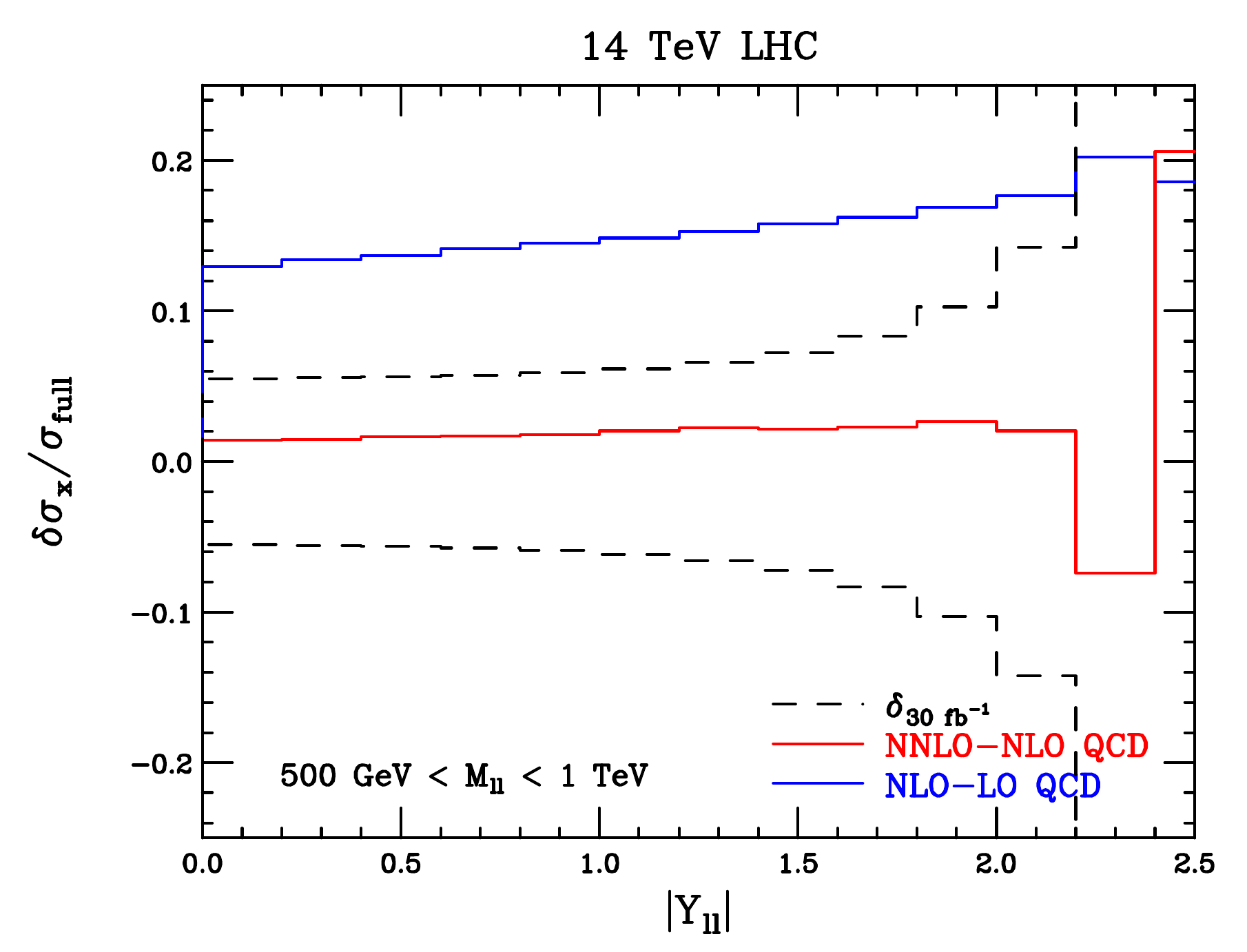}}\quad
\subfigure{\includegraphics[width=3.1in]{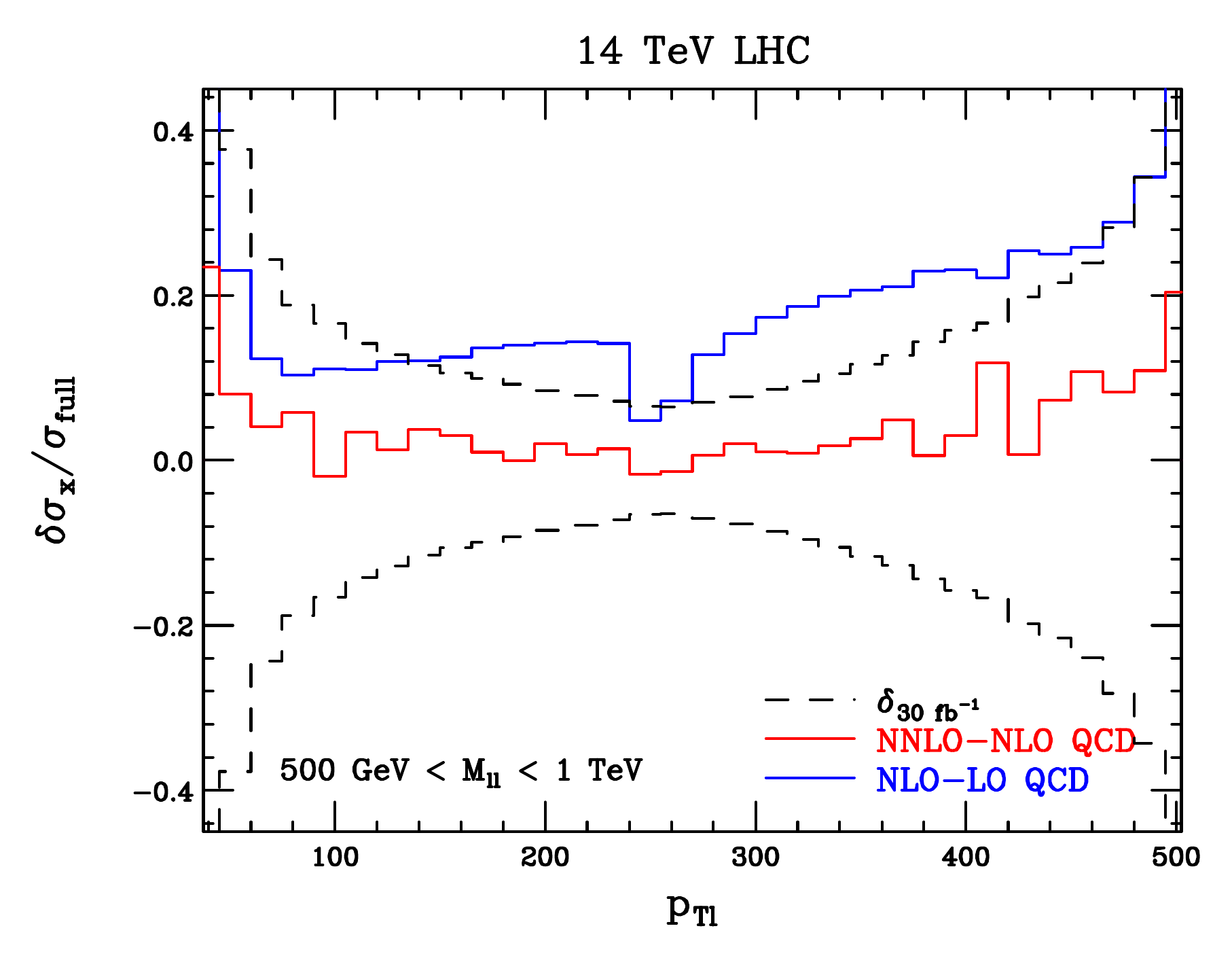}}}
\caption{Shown are the deviations induced by QCD corrections to the dilepton rapidity distribution (left panel) and lepton $p_{Tl}$ distribution (right panel) at a 14 TeV LHC, for the invariant mass range $M_{ll} \in [500,1000]$ GeV.  The two lines indicate the deviation of NLO QCD minus LO relative to the full result, and NNLO minus NLO relative to the full result. The dashed lines show the estimated errors coming from statistics and from uncertainties in the quark and gluon distribution functions. }  \label{fig:14TeVQCD}
\end{figure}

We again conclude this section by considering the impact of higher-order QCD corrections on the distributions studied in Fig.~\ref{fig:14TeVQCD}.  We show the dilepton rapidity and lepton $p_{Tl}$ distribution in the invariant mass region $M_{ll} \in [500,1000]$ GeV.  The shift in going from NLO to NNLO QCD is smaller than the uncertainty from other sources over the entire dilepton rapidity range.  This is also true for the $p_{Tl}$ distribution, but the corrections do become large near the lower kinematic boundary.  Relaxing the $\eta_l$ cut reduces the size of these corrections, as we now show.

\subsection{Effects of modified acceptance cuts}

An additional issue that we wish to address in this study is what can be gained by modifying the acceptance cuts away from the values in Eq.~(\ref{eq:acccuts}), which were used so far in our study.  These might be tightened by the experimental collaborations due to trigger requirements during higher luminosity running, or they may be loosened because of improved analysis techniques that allow for additional kinematic regions to be accessed.  To probe such possibilities, we will consider in this section two potential changes in the acceptance cuts on the leptons.
\begin{itemize}

\item We will study the effect of loosening the pseudorapidity cut on the leptons to $|\eta_l|<4.0$.  Although this is an aggressive relaxation of this bound, a similar loosening may be possible in the electron channel by using calorimetric information in the forward region\footnote{We thank S.~Stoynev for discussions on this topic.}.

\item We consider the effect of a staggered transverse momentum cut on the leptons; we demand that the harder lepton satisfy $p_T >40$ GeV, while the softer one satisfies $p_T >20$ GeV.

\end{itemize}
For simplicity, we restrict this study to the invariant mass bin $M_{ll} \in [200,500]$ GeV.

\begin{figure}
\centering
\mbox{\subfigure{\includegraphics[width=3.1in]{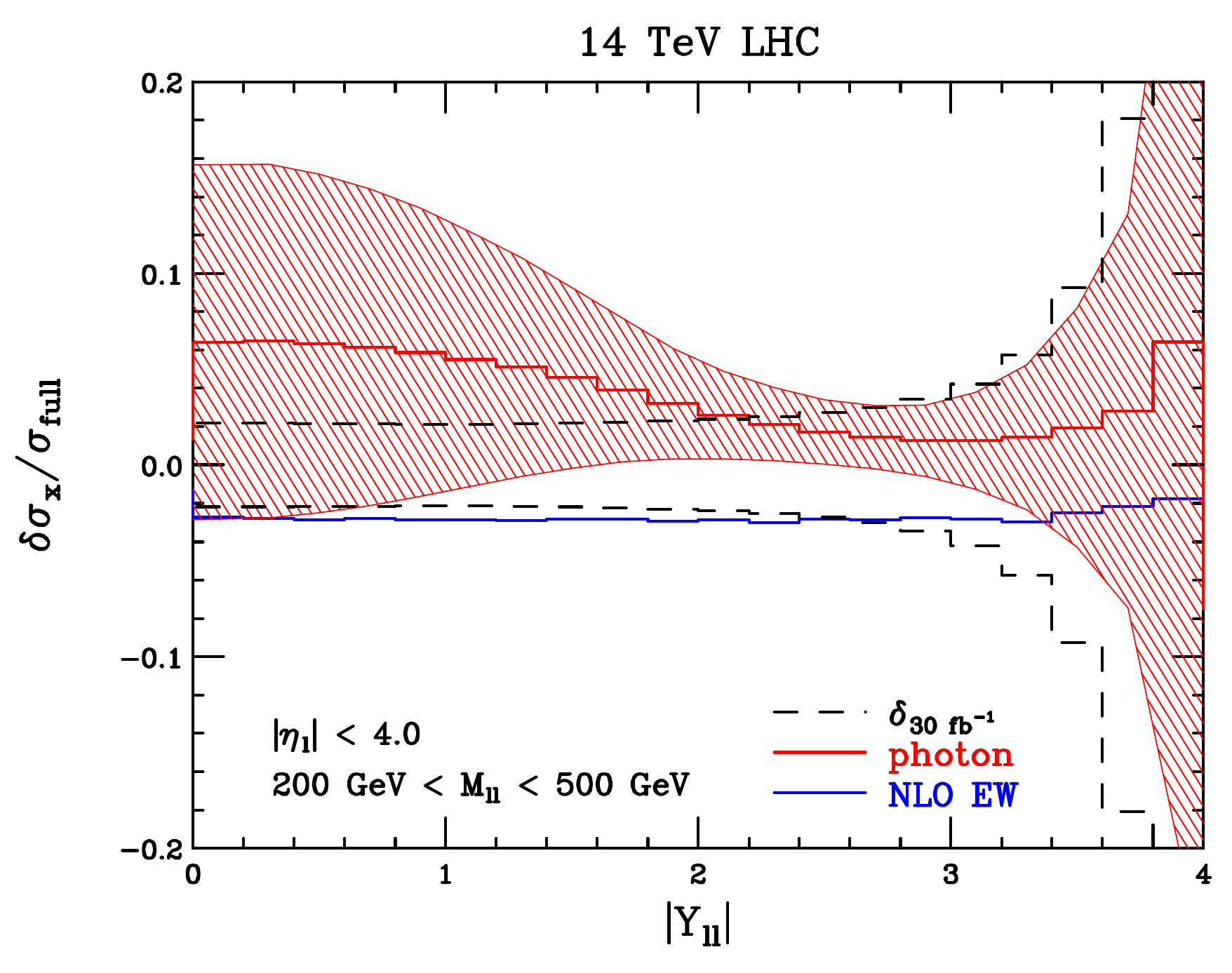}}\quad
\subfigure{\includegraphics[width=3.1in]{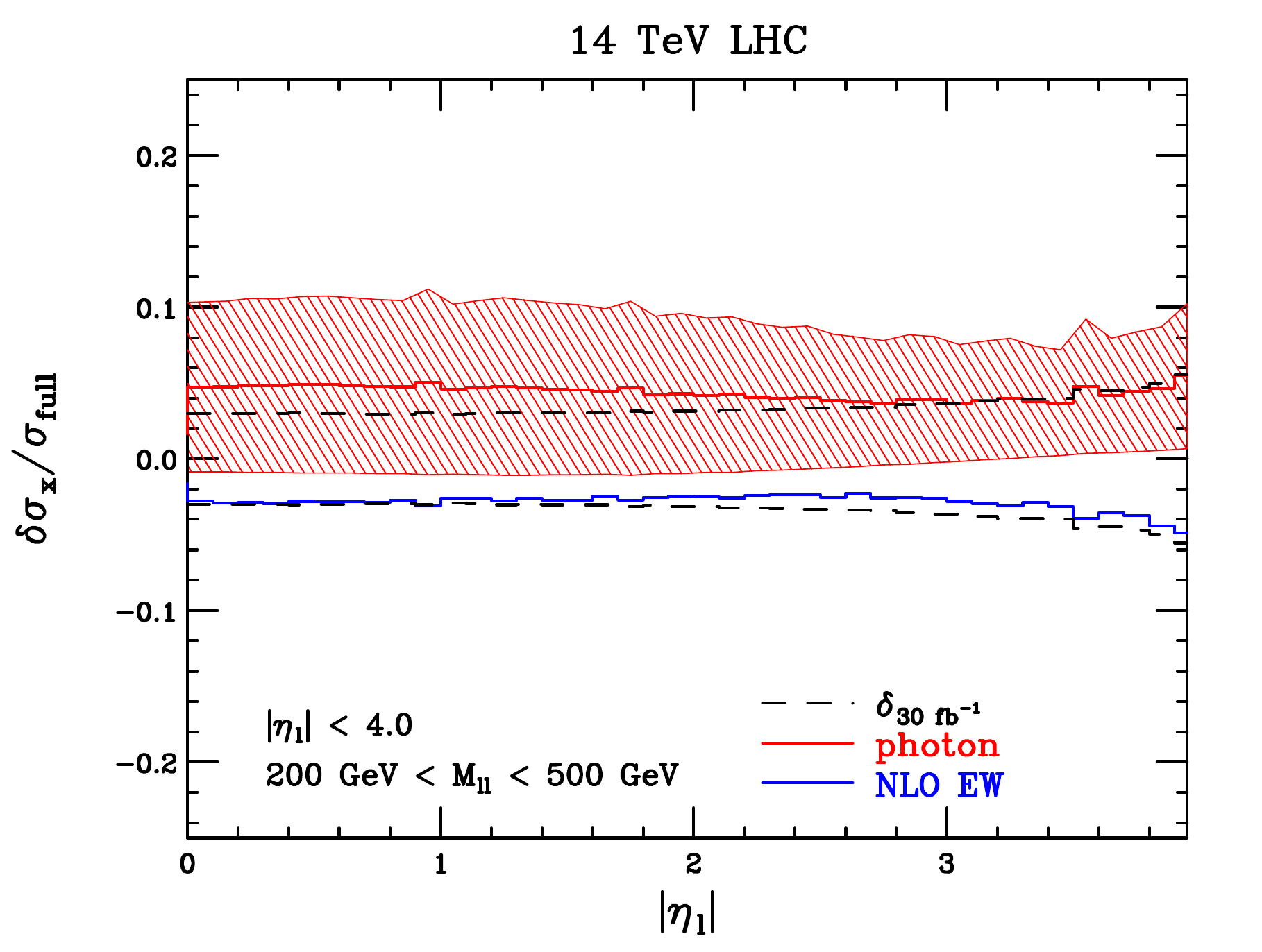}}}
\caption{Shown are the deviations induced by photon-initiated contributions and electroweak corrections to the dilepton rapidity distribution (left panel) and the lepton pseudorapidity distribution (right panel), for the invariant mass range $M_{ll} \in [200,500]$ GeV.  The lepton pseudorapidity cut has been extended to $|\eta_l|>4$.  The bands show the errors coming from the photon distribution function.  The dashed lines show the estimated errors coming from statistics and from uncertainties in the quark and gluon distribution functions. }  \label{fig:eta4_200_500}
\end{figure}

\begin{figure}
\centering
\includegraphics[width=3.5in]{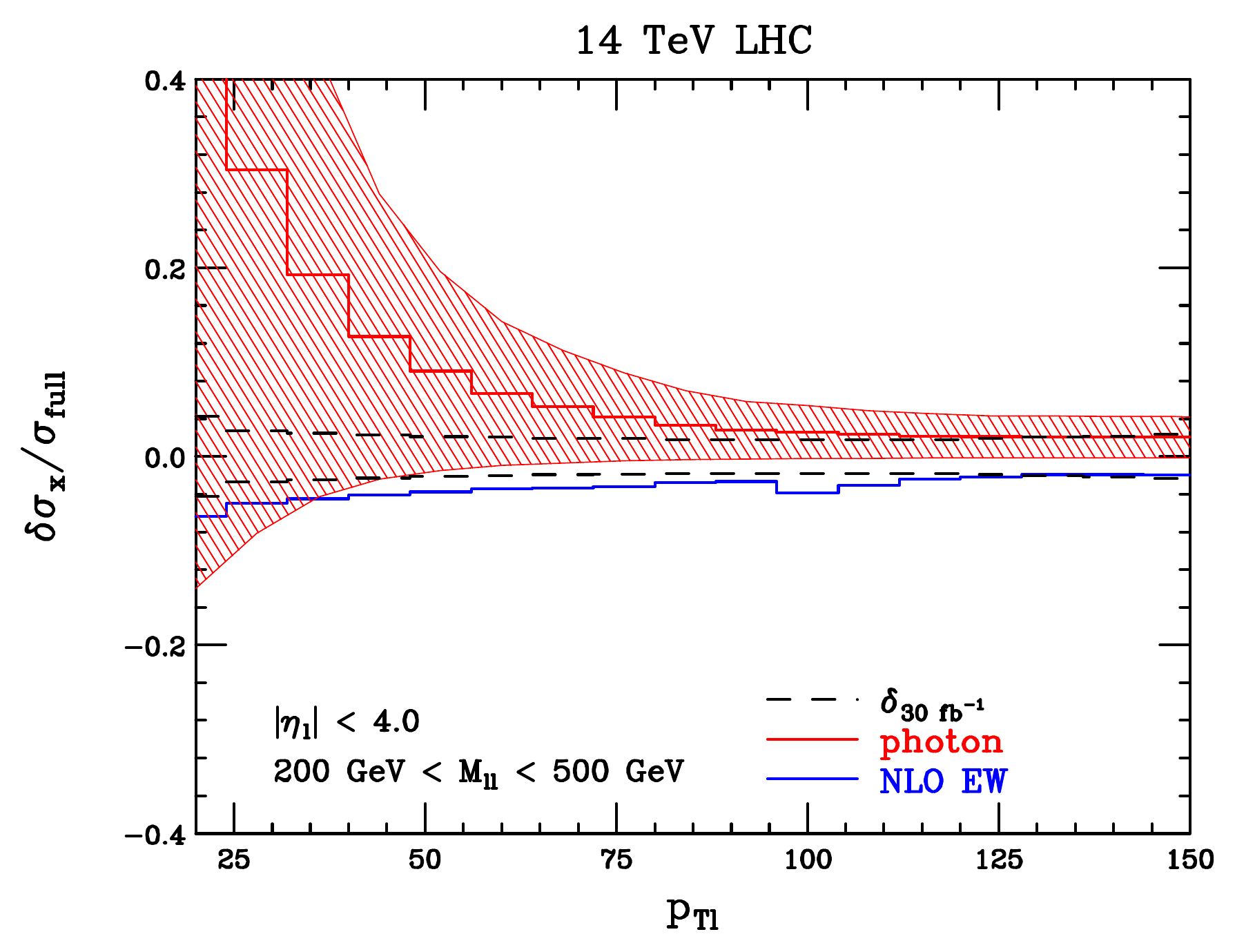}
\caption{Shown are the deviations induced by photon-initiated contributions and electroweak corrections to the lepton transverse momentum distribution for the invariant mass range $M_{ll} \in [200,500]$ GeV.  The lepton pseudorapidity cut has been extended to $|\eta_l|<4$.  The bands show the errors coming from the photon distribution function.  The dashed lines show the estimated errors coming from statistics and from uncertainties in the quark and gluon distribution functions. }  \label{fig:leppTeta4_200_500}
\end{figure}

\begin{figure}
\centering
\mbox{\subfigure{\includegraphics[width=3.1in]{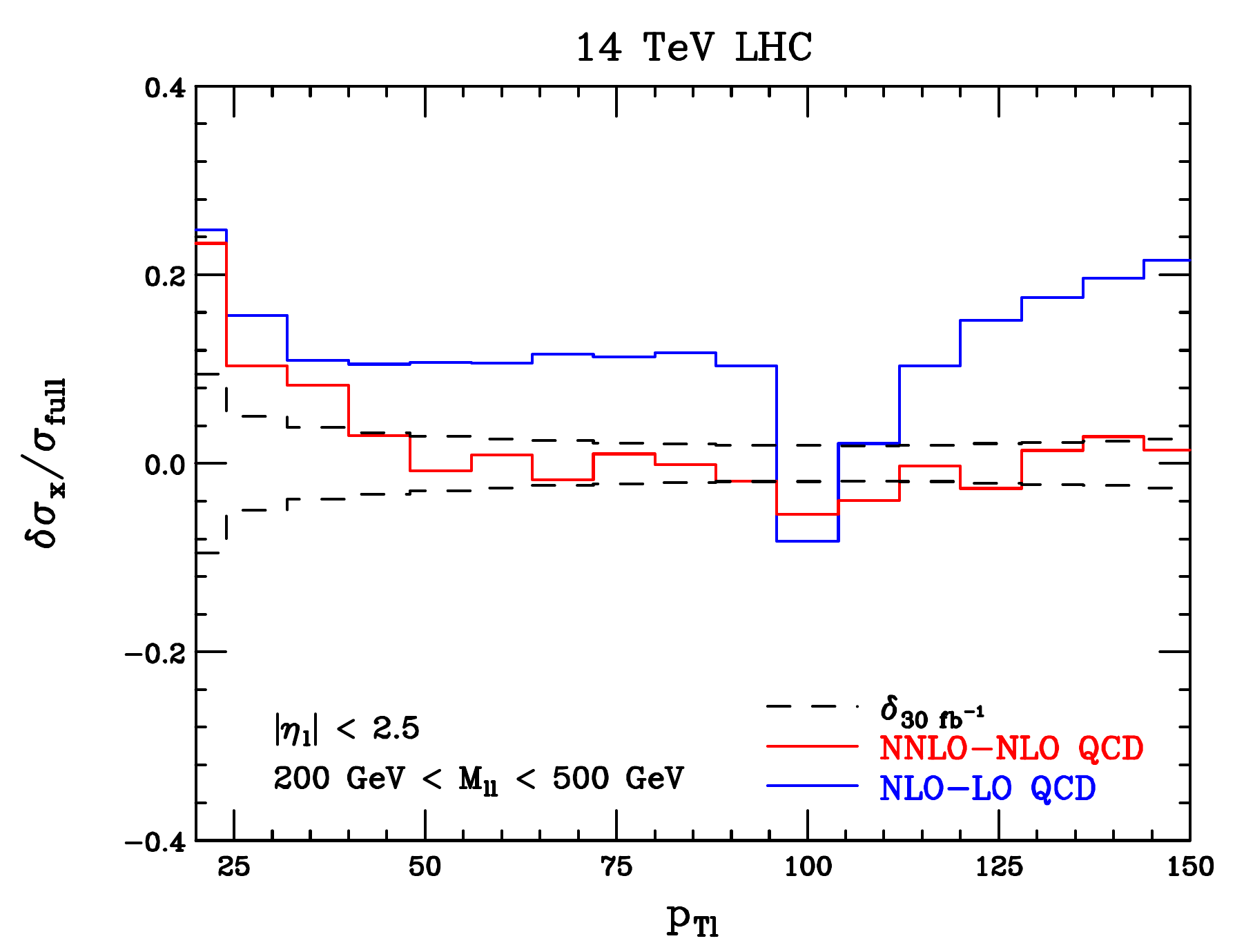}}\quad
\subfigure{\includegraphics[width=3.1in]{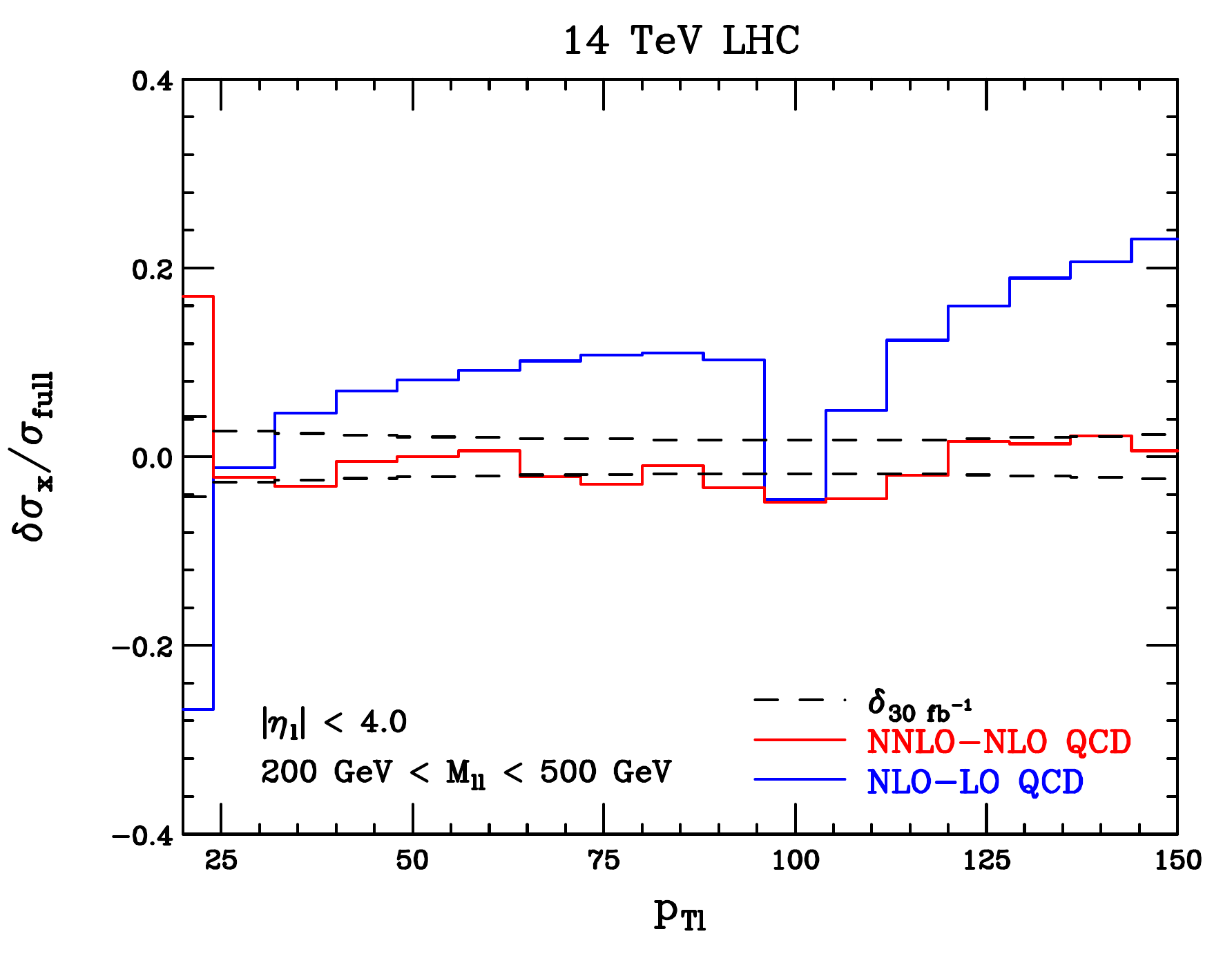}}}
\caption{Shown are the deviations induced by QCD corrections to the lepton $p_{Tl}$ distribution for the invariant mass range $M_{ll} \in [200,500]$ GeV.  The lepton pseudorapidity cut has been extended to $|\eta_l|<4$ in the right panel; int he left panel it has been kept at $|\eta_l|<2.5$.  The two lines indicate the deviation of NLO QCD minus LO relative to the full result, and NNLO minus NLO relative to the full result. }  \label{fig:leppTeta_200_500}
\end{figure}

We begin by considering the loosening of the pseudorapidity cut to $|\eta_l|<4.0$.  Results for the dilepton rapidity and lepton pseudorapidity distributions are shown in Fig.~\ref{fig:eta4_200_500}.  Both of these observables have an increased phase space upon changing the $|\eta_l|$ cut.  However,  the expected errors are large in the new regions of phase space, and not much is gained from these distributions for this change in the cut.  Something more interesting occurs in the lepton transverse momentum distribution, shown in Fig.~\ref{fig:leppTeta4_200_500}.  This plot should be contrasted with the result shown with $|\eta_l|<2.5$ in Fig.~\ref{fig:leppT14TeV120_3000}.  The photon-induced deviation continues to grow at low $p_T$, instead of decreasing near the lower $p_T$ boundary.  The reason for this was explained below Eq.~(\ref{eq:LOpT}).  The previous pseudorapidity cut was restricting the low $p_T$ region due to the connection between these variables shown in Eq.~(\ref{eq:LOpT}).  This constraint is now lifted.  This has an added benefit.  With the previous pseudorapidity cut, additional low $p_T$ phase space regions were opened beyond LO in QCD, increasing the size of the QCD corrections there and potentially reducing the power of this region in constraining the photon PDF.  With the relaxed cuts, all of the available phase space is open already at LO, reducing the impact of QCD corrections.  This can be seen in Fig.~\ref{fig:leppTeta_200_500}, where the impact of QCD corrections on $p_{Tl}$ at NLO and NNLO is shown.  When $|\eta_l|<2.5$, the shift from NLO QCD to NNLO QCD is larger than the estimated errors from statistics and PDFs when $p_{Tl}<40$ GeV, complicating the use of this phase space region in extracting the photon PDF.  This is not the case when $|\eta_l|<4.0$.  Only right at the lower boundary of $p_{Tl}=20$ GeV are QCD corrections large.  The exact numerical values for the $p_{Tl}$ and $\eta_l$ cuts used in this study are meant to be illustrative only, but the importance of carefully considering the impact of phase-space restrictions on QCD radiation in future measurements should be clear from this example.

\begin{figure}
\centering
\mbox{\subfigure{\includegraphics[width=3.1in]{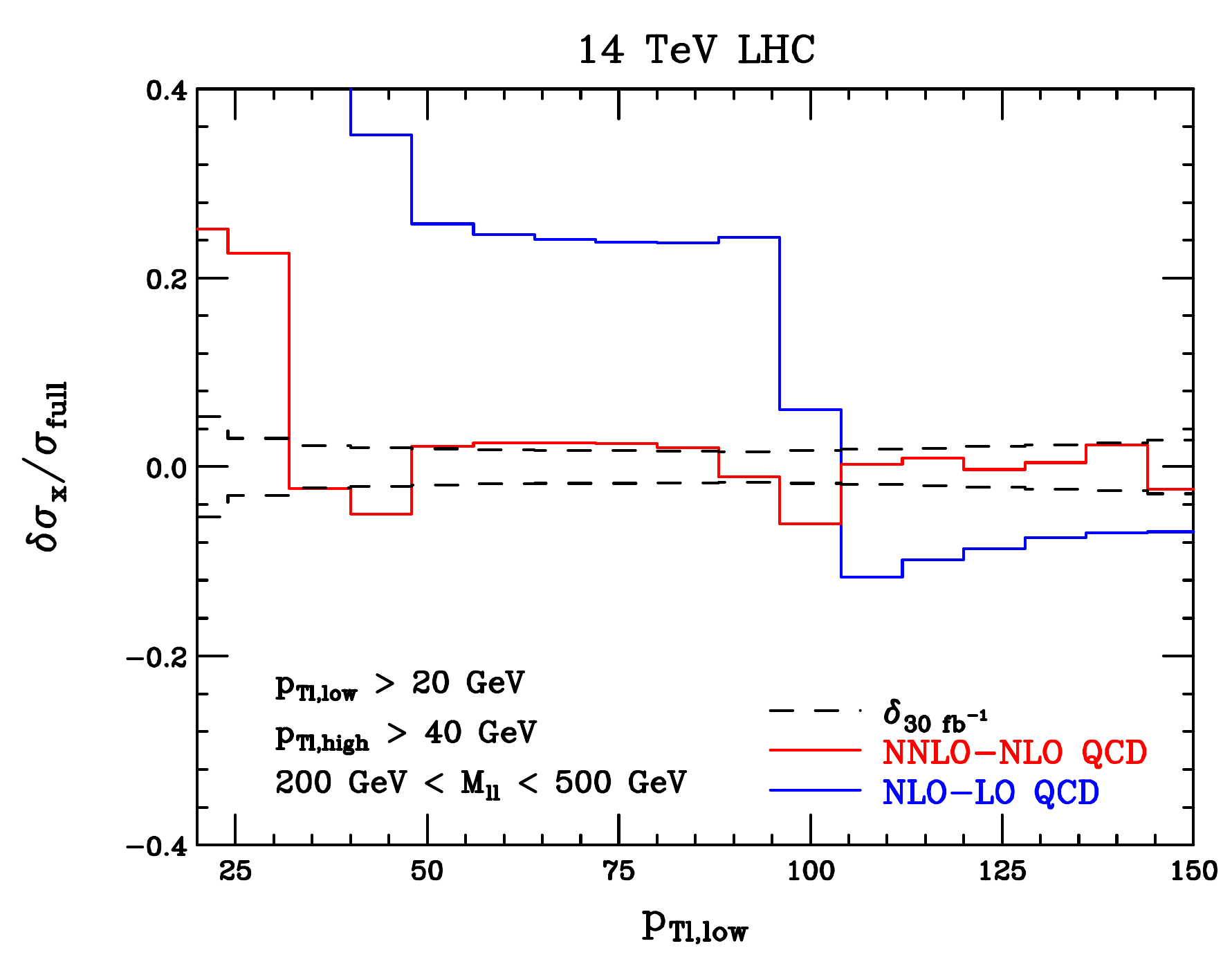}}\quad
\subfigure{\includegraphics[width=3.1in]{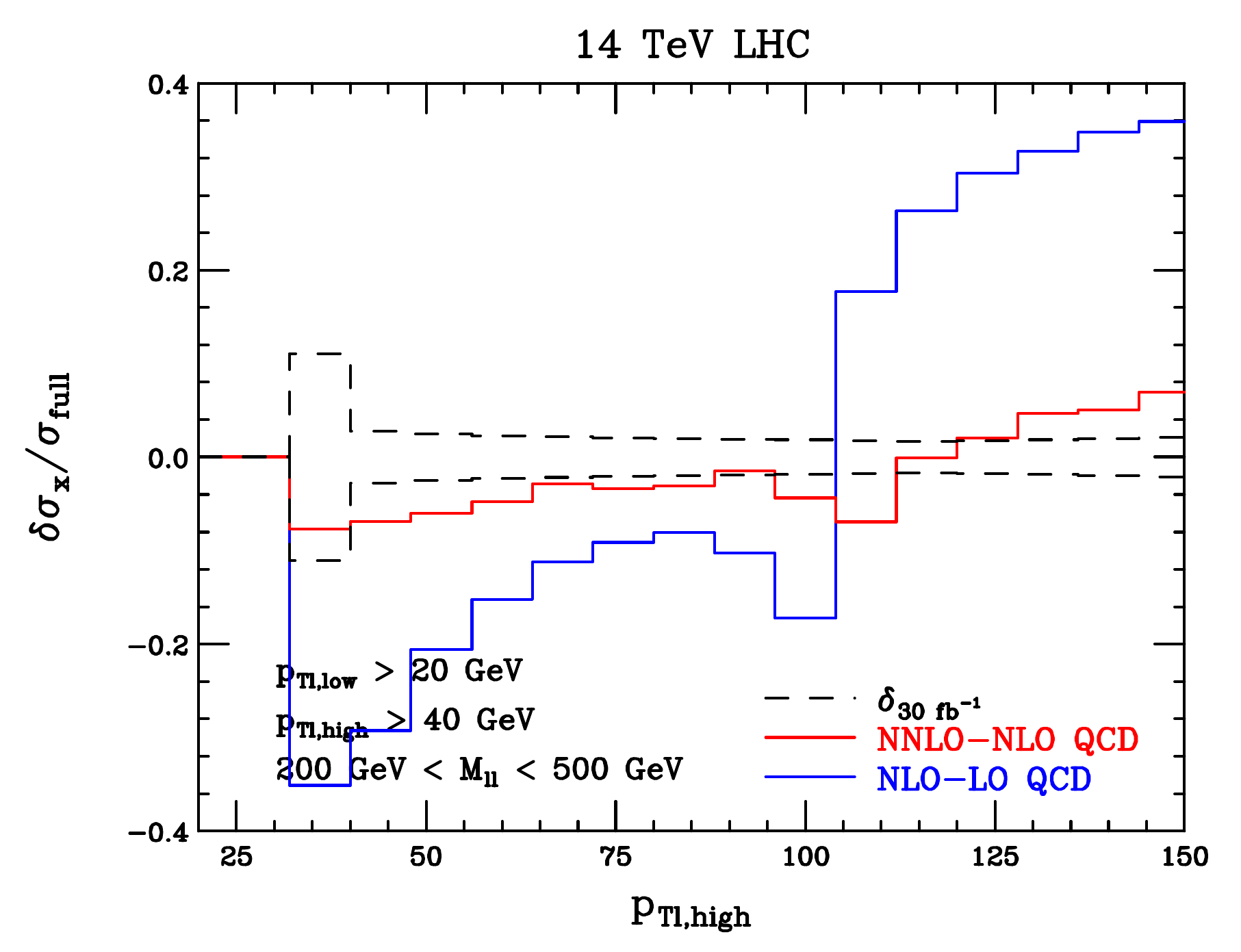}}}
\caption{Shown are the deviations induced by QCD corrections to the harder (right panel) and softer (left panel) lepton $p_{Tl}$ distributions for the invariant mass range $M_{ll} \in [200,500]$ GeV.  The lepton pseudorapidity cut has been extended to $|\eta_l|<4$ in both plots.  The two lines indicate the deviation of NLO QCD minus LO relative to the full result, and NNLO minus NLO relative to the full result. }  \label{fig:leppTstag_200_500}
\end{figure}

We now demand that the harder of the two leptons satisfies $p_{T} >40$ GeV while the softer has $p_{T} >20$ GeV, while keeping $|\eta_l|<4$.  The major effect of this staggered cut arises from the fact that at Born level, the two leptons are both forced to have $p_{T} >40$ GeV, since they are back-to-back in the transverse plane.  The region where the softer lepton is in the range $p_T \in [20,40]$ GeV only opens up at NLO in QCD when there is additional radiation for the leptons to recoil against.  This enhances the impact of QCD corrections in exactly the phase-space region where there is the most sensitivity to the photon PDF.  The deviations induced by higher-order QCD corrections to the softer and harder lepton transverse momentum distributions are shown in Fig.~\ref{fig:leppTstag_200_500}.  The shift when going from NLO to NNLO in QCD reaches over 20\% for the softer lepton in the region below 40 GeV.  This shows that higher-order QCD can potentially mask other effects appearing in the low $p_{Tl}$ region if the cuts enhance their effect.  One should appreciate this effect in future experimental analyses.

\section{Summary and Conclusions}
\label{sec:conc}

In this note we have mapped the structure of radiative corrections affecting high-mass Drell-Yan production in both 8 TeV and 14 TeV LHC collisions.  We have carefully studied the effect of photon-induced processes and electroweak corrections in all relevant kinematic variables, for a host of invariant mass regions.  We have estimated the observability of these effects given expected statistical errors, and uncertainties coming from quark and gluon PDFs.  The considered effects are larger than the expected errors over a large kinematic range, making high-mass Drell-Yan production an ideal place to understand perturbative corrections in the Standard Model, and to better determine the proton structure.  Our analysis helps guide future measurements by showing what corrections must be accounted for in which distributions, and also indicates how to individually extract each effect.  By doing this one can determine whether such effects as electroweak Sudakov logarithms are under theoretical control, and can therefore be applied in other processes.  Our study also helps inform future studies of proton structure.  We have performed this study using the most up-to-date theoretical tools: NNLO QCD corrections combined with NLO electroweak effects, together with NNPDF photon PDFs for the leading photon-initiated processes.

The main physics conclusions of our study have already been presented in the Introduction, so we conclude with several general comments on our results.  
\begin{itemize}

\item Not surprisingly, our results show the importance of simultaneously controlling all sources of radiative corrections.  While carefully chosen observables can reduce the size of EW corrections or photon-initiated effects, in general attempting to determine one without accounting for the other will lead to incorrect results.

\item In our view the experimental collaborations should attempt to measure all possible kinematic distributions.  They are all interesting for different reasons.  The dilepton rapidity shows sensitivity to the photon PDF at central values.  The lepton transverse momentum distribution is especially sensitive to the photon PDF.  The lepton $\eta_l$ distribution allows the angular structure of EW Sudakov logarithms to be probed.  They all provide a different window into the structure of the Standard Model.  

\item The interplay between QCD corrections and experimental cuts, and particularly the opening up of new phase space regions at higher orders, should be carefully studied.  If not, large QCD uncertainties may mask the other effects one wishes to measure.

\end{itemize}
We look forward to the continued precision study of the Drell-Yan process during Run II of the LHC.

\section{Acknowledgments}
\label{sec:acks}

We are grateful to U.~Klein, A.~Kubik, M.~Schmitt, and S.~Stoynev for many helpful discussions.

The work of R.~B. was supported by the U.S. Department of Energy, Division of High Energy Physics, under contract DE-AC02-06CH11357.  The work of Y.~L. was supported by the U.S. Department of Energy under contract DEÐAC02Ð76SF00515.  The work of F.~P. was supported by the U.S. Department of Energy, Division of High Energy Physics, under contract DE-AC02-06CH11357 and the grants DE-FG02-95ER40896 and DE-FG02-08ER4153.

This research used resources of the National Energy Research Scientific Computing Center, which is supported by the Office of Science of the U.S. Department of Energy under Contract No. DE-AC02-05CH11231.

\end{document}